\documentclass[aps,showpacs,showkeys,preprint,preprintnumbers,amsmath,amssymb,nofootinbib]{revtex4-1}
\usepackage{graphicx}
\usepackage{dcolumn}
\usepackage{bm}
\usepackage{siunitx}
\usepackage{slashed} 
\usepackage{subfigure}
\usepackage{natbib}
\newcommand{\bea}{\begin{eqnarray}}
\newcommand{\eea}{\end{eqnarray}}
\newcommand{\ud}{\mathrm{d}}
\def\missE{\slashed E} 
\def\missP{\slashed P} 

\begin{document}

\title{ Discovery potential of Higgs boson pair production through 4$\ell$+$\slashed{E}$ final states at a 100 TeV collider}
\author{Xiaoran Zhao$^{1,2}$\footnote{Correspondence Author:zhaoxiaoran13@mails.ucas.ac.cn}, Qiang Li$^{3}$, Zhao Li$^{2}$, Qi-Shu Yan$^{1,4}$
\\$^{1}$ School of Physics Sciences, University of Chinese Academy of Sciences, Beijing 100039, P.R. China
\\$^{2}$ Institute of High Energy Physics, Chinese Academy of Sciences, Beijing 100039, P.R. China
\\$^{3}$ Department of Physics and State Key Laboratory of Nuclear Physics and Technology, Peking University, Beijing, 100871, China
\\$^{4}$ Center for future high energy physics, CAS, Beijing 100039, P. R. China
}
\begin{abstract}
We explore the discovery potential of Higgs pair production at a 100 TeV collider via full leptonic mode. The same mode can be explored at the LHC when Higgs pair production is enhanced by new physics. We examine two types of fully leptonic final states and propose a partial reconstruction method. The reconstruction method can reconstruct some kinematic observables. It is found that the $m_{T2}$ variable determined by this reconstruction method and the reconstructed visible Higgs mass are important and crucial to discriminate the signal and background events. It is also noticed that a new variable, denoted as $\Delta m$ which is defined as the mass difference of two possible combinations, is very useful as a discriminant. We also investigate the interplay between the direct measurements of $t\bar{t} h$ couplings and other related couplings and trilinear Higgs coupling at hadron colliders and electron-positron colliders. 
\end{abstract}



\maketitle

\section{Introduction}


The discovery of Higgs boson at the LHC has motivated the high energy community to think of the next generation p p colliders. A 100 TeV collider can offer us a huge potential to probe various new physics \cite{Arkani-Hamed:2015vfh}. For example, new vector boson $W^\prime$ and $Z^\prime$ can be discovered up to 25-35 TeV\cite{Rizzo:2014xma}. A heavy Higgs bosons of the two Higgs doublet model can be probed up to 20 TeV or so via single associate production \cite{Hajer:2015gka}. In the simplified model, the superpartners of top quark and the gluino can be probed up to 5 TeV and 10 TeV, respectively \cite{Gershtein:2013iqa}, which can make decisive evidences on the fate of the electroweak supersymmetry models. Dark matter candidate can be probed up to 10 TeV or higher \cite{Bramante:2014tba,Low:2014cba,Acharya:2014pua,Xiang:2015lfa}. A 100 TeV collider could also perform high precision measurements on Higgs properties \cite{Baglio:2015wcg}, top quark properties, EW physics, and so on.

Among various new physics candidates, the shape of Higgs potential plays a very special role. As we know, the shape of Higgs potential is determined by Higgs fields and Higgs self-couplings, new Higgs fields and self-couplings exist in most of extensions of the standard model (SM). Therefore, it is well-known \cite{Barger:1988jk,Ilyin:1995iy,Djouadi:1999gv,Ferrera:2007sp,Asakawa:2010xj,Baglio:2012np} that  to probe Higgs self-couplings at colliders can offer us a way to understand the nature of Higgs bosons, to reconstruct the shape of Higgs potentials, to understand the mystery of electroweak symmetry breaking. These couplings could play crucial parts in the EW baryogenesis scenarios \cite{Trodden:1998ym,Riotto:1999yt}. For example, they are crucial to determine whether CP violation is strong enough to produce a large enough matter anti-matter asymmetry, which is needed in terms of the one of three Sakharov criteria. In the two Higgs doublet model, it is possible to introduce meaningful complex Higgs self couplings which can induce a large enough CP violation which is needed for the EW baryogensis scenarios. These couplings are also important to determine whether the strong first-order phase transition could occur for a realistic EW baryogenesis scenarios. These couplings can affect the gravitational wave radiation in the process of bubble collisions \cite{Kosowsky:1992vn,Grojean:2006bp,Huber:2008hg}, while the gravitational wave can induced  a B mode which is detectable from the cosmological microwave background \cite{Kakizaki:2015wua,Jiang:2015qor,Huang:2016odd}. So it is well-motived to explore the shape of Higgs potential in our world.

Compared to the standard model, new physics could modify either effective trilinear (or cubic) or quartic couplings or both, either in $10-20\%$ via loop corrections \cite{Noble:2007kk} or more than $100\% - 300\%$ via tree-level corrections (say adding a dimension-6 operator \cite{Zhang:1992fs,Grojean:2004xa,Huang:2015izx} or many higher dimensional operators \cite{Huang:2015tdv}). The Lorentz structure of triple Higgs boson can even be modified in the Higgs-Gravity model \cite{Xianyu:2013rya,Ren:2014sya}, which could lead to energetic Higgs bosons in the final states \cite{He:2015spf}. Using the discovered Higgs boson as a probe, to measure the self-couplings of Higgs boson could help us to further understand the nature of the Higgs bosons and to extract the information of the shape of Higgs potentials which encodes the electroweak symmetry breaking. Therefore to measure trilinear couplings and quartic couplings \cite{Papaefstathiou:2015paa,Chen:2015gva,Fuks:2015hna,Dicus:2016rpf} will be of great importance and could be one of the prime targets for both the LHC high luminosity runs and future collider projects. 

The study on the di-Higgs boson final states in the SM and new physics models at hadronic colliders has been being a hot topic recently, various production processes and final states have been explored in literatures, $b \bar{b} \gamma \gamma$ \cite{Dolan:2012rv,He:2015spf}, $W W  b \bar{b}$ \cite{Papaefstathiou:2012qe}, $W W  \gamma \gamma$ \cite{Lu:2015qqa}, $b \bar{b} \tau \tau$ \cite{Dolan:2012rv}, and rare decay final states $3\ell 2 j$ \cite{3l2j} and others \cite{Papaefstathiou:2015iba}. Both ATLAS and CMS collaborations of the LHC had performed realistic simulation and analysis on di-Higgs boson final states \cite{ATL-PHYS-PUB-2014-019,CMS-PAS-FTR-15-002}. 

In this work, we extend our study in \cite{3l2j} to the pure leptonic mode, i.e. $p p \to h h  \to 4 \ell + \missE $ in a 100 TeV collider. To our best knowledge, this mode has not been carefully studied in literatures due to its tiny production rate in the SM at the collision energy of the LHC. But for some new physics models, the production rate of di-Higgs can be enhanced by a factor from 10 to 100, then this mode could be accessible even at the LHC. For a 100 TeV collision, the production rate of this mode in the SM itself is large enough and is accessible. Meanwhile, since it is pure leptonic final states, this mode can be searched by experimental groups relatively easy. Therefore, it is meaningful to perform a careful analysis on this mode either for the LHC runs or for a future 100 TeV collider project.

In order to determine the Higgs self-couplings at future hadron colliders, a precision measurement of top Yukawa coupling at the LHC and future colliders is very crucial.  The top quark Yukawa coupling plays a remarkable role to help us to probe the properties of Higgs boson. It is the strongest Yukawa coupling, which almost saturates the perturbation bounds; it can affect the vacuum stability \cite{Degrassi:2012ry} much seriously than any other Yukawa couplings in the SM; it determines the multi-Higgs production at hadron colliders and affects the decay of Higgs boson to gluon pair, di-photon  and Z$\gamma$ final states much larger than the other fermions in the SM; it affects the Higgs self-coupling measurements at hadron colliders, both trilinear and quartic coupling measurements. 

Therefore, in this work, to examine how top quark Yukawa coupling can affect the measurement of trilinear Higgs coupling, we take the following effective Lagrangian
\bea
\label{efl} {\cal L}_1 = Y_t \, (a \, \bar{t} t   + i \, b \,\bar{t} \gamma_5 t ) \, h + \lambda_3 \, \lambda_{SM} \, v \, h \, h \, h + \cdots \,,
\eea
where the term $Y_t = \sqrt{2} m_t/v$ is the Yukawa couplings of top quark in the SM,  and both $a$ and $b$ are dimensionless parameters. The parameter $b$ is related with the CP violation. In the standard model, $a =1$ and $b=0$. In the two Higgs doublet model (2HDM) with no CP violation, $a=\textrm{ctg} \beta$ and $b=0$. If there is CP violation in the 2HDM, the CP even and CP odd neutral scalars could mix which leads to a non-vanishing $b$. Early efforts to probe this coupling at hadron colliders could be found in \cite{Gunion:1996vv}. The study of measurement of these couplings at linear colliders can be found \cite{BhupalDev:2007ftb}. A recent study at the LHC and future hadronic colliders how to measure these two free parameters can be found in \cite{He:2014xla,Boudjema:2015nda}, where a different but equivalent parametrisation was used. Theoretical calculation of loop corrections from $t\bar{t} h$ can be found at \cite{Huang:2001ns}. A recent study on the CP properties of the 2HDM could be found in \cite{Mao:2014oya}.  A systematic analysis on the constraints from the Higgs precision measurement to either $a$ or $b$, interested readers can refer \cite{Cheung:2013kla,Chang:2014rfa,Lu:2015jza} for such a detailed study.

The term $\lambda_{SM}=m^2_h/2v^2\approx 0.13 $, while $\lambda_3$ is a free dimensionless parameter. In various new physics models, this coupling can vary in a large range. For example, in the framework of an effective operator, the strong first order electroweak phase transition has been explored in \cite{Huang:2015izx}, where $\lambda_3$ can be in the range [5/3, 3]. In the model with a singlet + SM, the trilinear coupling could be larger than the value of the SM by more than $20\%$ to $200\%$ \cite{Curtin:2014jma}, and this deviation is dependent upon the mass of the singlet. 

There are mainly two-fold aims for this work: 1) we explore the sensitivity of the pure leptonic mode $p p \to h h  \to 4 \ell + \missE $ at a 100 TeV collider,  2) we examine the complementarity of the direct measurement of $\bar{t} t h$ and the direct measurement of $\lambda_3$ in the future colliders. 

The work is organised as follows.  In section II, we study the cross section of the process $g g \to h h$. In section III, we analyse the sensitivity of two types of the same sign leptonic final states $gg \to h h \to 4 \ell + \missE$ in a 100 TeV collider. In section IV, we examine the issue how $t \bar{t} h$ measurement can affect the determination of the trilinear Higgs couplings. In section V, we examine the complimentarily to determine related Higgs couplings at hadronic colliders and electron-positron colliders. We end this work with a few discussions on the detector issues to probe pure leptonic modes in a 100 TeV collider. We provide an appendix to describe the quasi-Monte Carlo method implemented in our code "wat" which has been used in the work to evaluate cross section and to  generate unweighted signal events.

\section{The cross section of Higgs pair production at Hadron colliders \label{xsec}}
We implemented the effective Lagrangian described in Eq. (\ref{efl}) as a new model file by modifying the
one-loop SM model file in MadGraph5/aMC@NLO\cite{mg5amc}. The parameters $a$, $b$, and $\lambda_3$ and the
corresponding tree level vertices are added by following the UFO protocol \cite{ufo}. According to the OPP
method\cite{opp}, we add two new R2 terms at one-loop level which are related with the process $g  g \to h h$
by following the information given in the two-Higgs doublet model \cite{r2auto}, shown as below:

\begin{align}
  \includegraphics[width=0.2\textwidth]{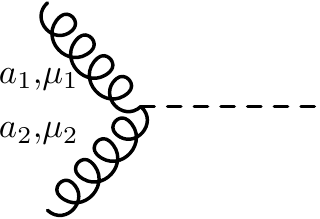}
 &=-i\frac{\sqrt{2}g_s^2m_taY_t\delta_{a_1a_2}g_{\mu_1\mu_2}}{16\pi^2} \,\\
\includegraphics[width=0.2\textwidth]{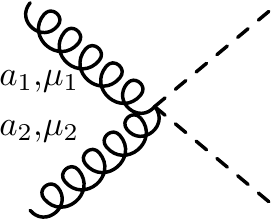}
 &=-i\frac{g_s^2Y_t^2\delta_{a_1a_2}g_{\mu_1\mu_2}}{16\pi^2}(a^2+b^2)\,,
\end{align}
where $a_1,a_2$ are the color indice, and $\mu_1,\mu_2$ are the Lorentz indice, and $g_s$ is the QCD coupling constant.

Then we interface the loop matrix element produced by MadGraph5/aMC@NLO\cite{mg5amc} to our integration and event generation code to obtain the leading order cross section and unweighted events of signal.

At the leading order, cross sections in hadronic colliders can be parametrised as the function of theoretical free parameters$a$, $b$, $\lambda_3$ in the following form 
\bea
\sigma( gg \to h h ) = G_1 \, a^4 + G_2 \, b^4 + G_3 \, a^2 \,  b^2 
+ (G_4 \,  a^3  + G_5 \, a \, b^2  ) \,  \lambda_3  + + ( G_6 \, a^2 + G_7 \,  b^2 ) \,  \lambda_3^2 \,.
\label{xsgghh}
\eea
It is noticed that this cross section is sensitive to the signs of $a$ and $\lambda_3$, respectively, but is insensitive to the sign of $b$.  There are two types of diagrams contributing to the process $gg \to hh$: the box diagrams and the triangle diagrams. The terms independent of $\lambda_3$ come from the squared amplitudes of box diagrams. The terms proportional to $\lambda_3^2$ are from the squared amplitudes of triangle diagrams, and the terms proportional to $\lambda_3$ is from the interference between the box and triangle diagrams.

We use the numerical approach to fit the coefficients $G_1-G_7$ for the LHC at 14 TeV, 33 TeV and a 100 TeV collider. We generate more than 100 points in (a,b,$\lambda_3)$ space for each collision energies. The coefficients of cross sections at the 14 TeV, 33 TeV and 100 TeV are presented in Table \ref{ce14tev}.
\begin{table}[th]
\begin{center}
\begin{tabular}{|c|c|c|c|c|c|c|c|}
\hline
&$ G_1$ (fb) & $G_2$  (fb)& $G_3$  (fb)& $G_4$  (fb)& $G_5 $  (fb)& $G_6 $  (fb)& $G_7 $  (fb)\\ \hline
14 TeV &$34.5$  & $3.37$ & $267.6$&$-23.1$& $-118.7$ & $4.82$ &$15.1$\\ \hline
\hline
33 TeV& $2.27 \times 10^2$ & $23.0$& $1.67 \times 10^3$ & $-143.4$ & $-722.6$  &  $28.7$ & $89.0$ \\
$R^{33}$ & $6.6$ & $6.8$ & $6.2$ & $6.2$ & $6.2$ & $5.9$ & $5.9$ \\\hline \hline
100 TeV &$1.71 \times 10^3 $  & $1.86 \times 10^2$ & $1.20 \times 10^5$&$-1.03 \times 10^3$& $-5.08 \times 10^3$ & $1.97 \times 10^2$ &$6.09 \times 10^2$\\ 
$R^{100}$ & $49.6$ & $55.2$ & $44.9$ & $44.6$ & $42.8$ & $41.2$ & $40.4$ \\ \hline 
\end{tabular}
\end{center}
\caption{The fitting coefficients for the LHC 14 TeV and a 100 TeV collider are tabulated, where the superscript denotes the collision energy 14 TeV and 100 TeV, respectively. $R^{33}$($R^{100}$) is defined as $K^{33}$/$K^{14}$($K^{100}$/$K^{14}$), where K denotes $G_1-G_7$. \label{ce14tev}}%
\end{table}

Due to the enhancement of gluon flux at a 100 TeV, it is noticed that all coefficients $G_i^{100}$ are enhanced compared with $G_i^{14}$. Due to the difference in the form factors, $G_1^{100}$ and $G_2^{100}$ are around 50 times larger than their counterparts $G_1^{14}$ and $G_2^{14}$. The coefficient $G_3^{100}$ is enhanced by a factor almost $45$. The squared triangle diagram coefficients $G_4^{100}$ and $G_5^{100}$ are 40 times larger than $G_4^{14}$ and $G_5^{14}$, this enhancement factor is smaller than that of box diagrams due to the s-channel suppression for energetic gluon fluxes; and the interference coefficients  $G_6^{100}$ and $G_7^{100}$ are also 40  times larger than $G_6^{14}$ and $G_7^{14}$. 

Another interesting observation is that the coefficient $G_3$ is $7$ times larger than the coefficient $G_1$. Typically, when $b$ is much smaller than $1$, the contribution of $G_3$ term can not be large. But if $b$ can be of order one, then the contribution of $G_3$ can be sizeable. In the works \cite{Azatov:2015oxa,Lu:2015jza}, more operators have been taken into account. We have compared our results presented there and found agreement.

\section{Full leptonic modes of signal and background events}
There are quite a few advantages of the full leptonic mode of $g g \to h h \to 4 \ell + \missE$ for a 100 TeV collision. On the first hand, it is relatively efficient to be searched by experiments. Targeted objects in the final state are leptons and missing energy. They can be reconstructed efficiently by subdetector systems of LHC detectors. The lepton reconstruction efficiency with $P_t > 5$ GeV is more than $90\%$ and particle identification can be made at detector level. On the second hand, the relatively clean signal and the signal is robust against the contamination of pileups and underlying events, since primary collision vertices of the signal events can be reconstructed. 

As explained in Sec. \ref{xsec}, we interface loop matrix element from Madgraph5/aMC@NLO\cite{mg5amc} with our code based on QMC to perform phase-space integration and event generation. The generated events are further showered by PYTHIA6\cite{py6} and then used to perform physical analysis. We have not taken into account the detector simulation in this work.

We use Madgraph5 to generate background events in a collision energy $\sqrt{s} = 100$ TeV and use PYTHIA6 to perform showering and decay simulations. Background processes without Z bosons in the final states are generated by using on-shell approximation for top quark, W boson and Higgs boson. While for background processes with single Z boson in the final state which decays into two leptons, like $t\bar{t}Z$ , $ZW^{+}W^{-}$ and $Z h$ background processes, we have included the effects of off-shell Z and $\gamma$ and their interferences.
\begin{center}
\begin{table}
  \begin{center}
\begin{tabular}{|c|c|c|c|c|}
\hline
 & $\sigma\times Br $ & Expected number of events  & Number of events &\\
 & ( fb ) & at \SI{3000}{fb^{-1}}&  generated & K-factors\\
\hline \hline
HH & 0.18 & \SI{5.7e2}{} & 500,000 &  1.6\\
\hline \hline
ZZ & \SI{4.8e2}{} & \SI{1.4e6}{} & - & -\\
\hline \hline
Z h & 5.56 & \SI{1.67e4}{} & {} 500,000 &0.97 \\

Z$W^{+}W^{-}$ & 6.34& \SI{1.90e4}{} & 500,000  & 2.8 \\

$Z t\bar{t}$ & \SI{1.97e2}{} & \SI{5.91e5}{} & 5,000,000  & 1.1 \\
\hline \hline
$t\bar{t}$ h & \SI{1.41e1}{} & \SI{4.22e4}{} & 1,000,000  & 1.2\\

$t\bar{t}t\bar{t}$ & 5.48 & \SI{1.65e4}{} & 400,000  & 1.3\\

$t\bar{t}W^{+}W^{-}$ & \SI{1.78}{} & \SI{5.35e3}{} & 200,000 & 1.3 \\
\hline \hline
h$W^{+}W^{-}$ & \SI{6.02e-2}{} & \SI{1.81e2}{} & 50,000 & 1.4 \\

$W^{+}W^{-}W^{+}W^{-}$ & \SI{2.74e-2}{} & \SI{8.23e1}{} & 10,000 & 2.8\\
\hline
\end{tabular}
\end{center}
\caption{\label{evgn}The expected number of events with 3 ab$^{-1}$ integrated luminosity at $\sqrt{s}=100$ TeV and the generated events for all processes are displayed.}
\end{table}
\end{center}
For both signal and backgrounds, higher order corrections are taken into account by normalizing the total cross section to their (N)NLO results, which is indicated by K factor($K={\sigma_{(N)NLO}}/{\sigma_{LO}}$) in Table \ref{lptn}. We adopt the NNLO result for signal in Ref. \cite{hhnnlo}, and K factor for processes $hZ$, $t\bar{t}Z$, $ZW^+W^-$,$t\bar{t} h$, $h W^+ W^-$ are obtained by MadGraph5/aMC@NLO\cite{mg5amc} under on-shell approximation. The K factor for $t\bar{t}t\bar{t}$ and $t\bar{t}W^+ W^-$ at \SI{100}{TeV} collider is unknown, and we adopt a value $K=1.3$, which is the K factor for $t\bar{t}t\bar{t}$ at \SI{14}{TeV} LHC. The K factor for the process $p p \to W^{+}W^{-}W^{+}W^{-}$ is also unknown, and we use the K factor for $ZW^{+}W^{-}$ since both of them have the same initial states at hadron colliders.

\begin{figure}[htbp]
  \centering
  \subfigure{
  \label{Fig1.sub.1}\thesubfigure
  \includegraphics[width=0.4\textwidth]{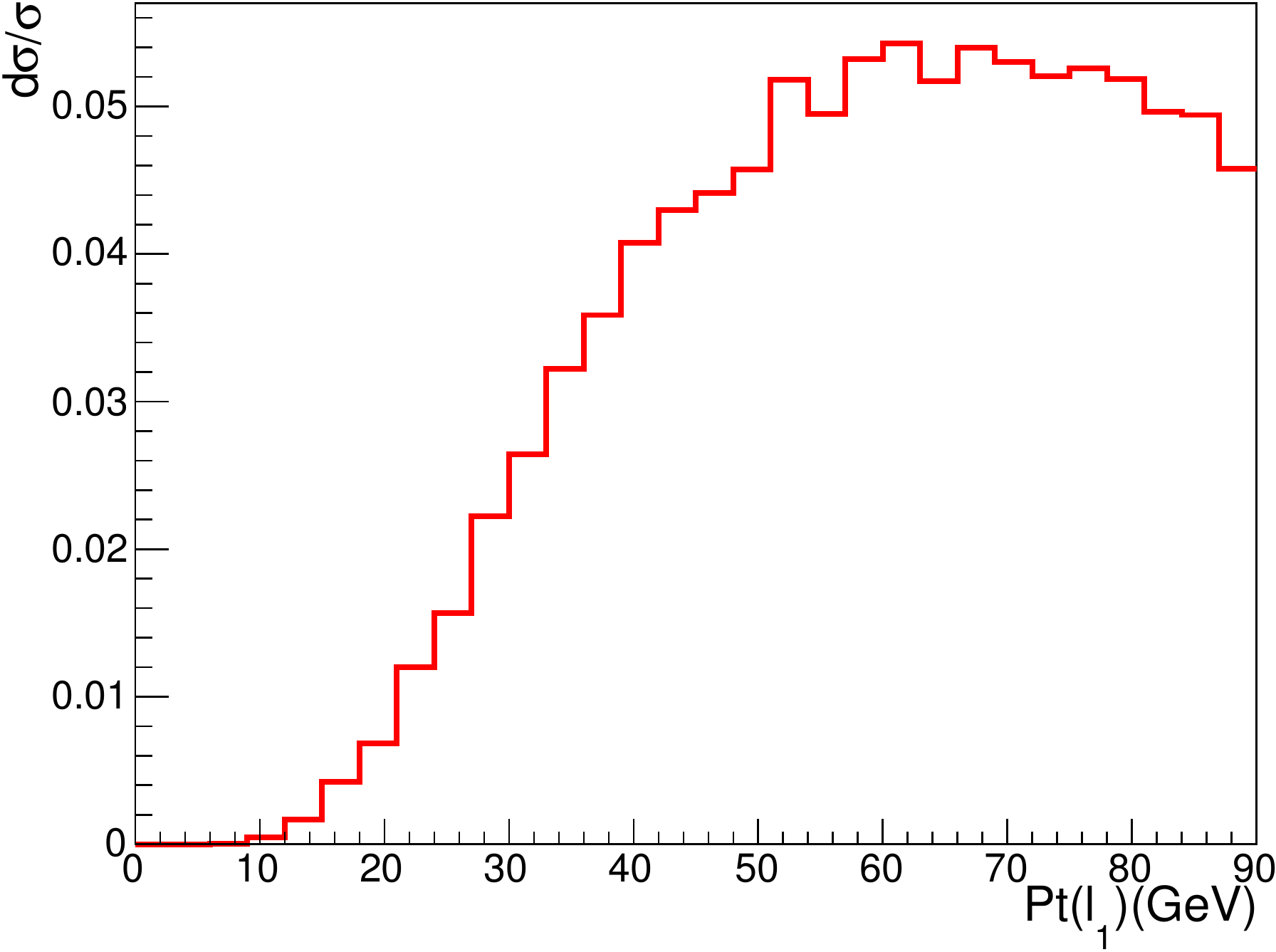}}
  \subfigure{
  \label{Fig1.sub.2}\thesubfigure
  \includegraphics[width=0.4\textwidth]{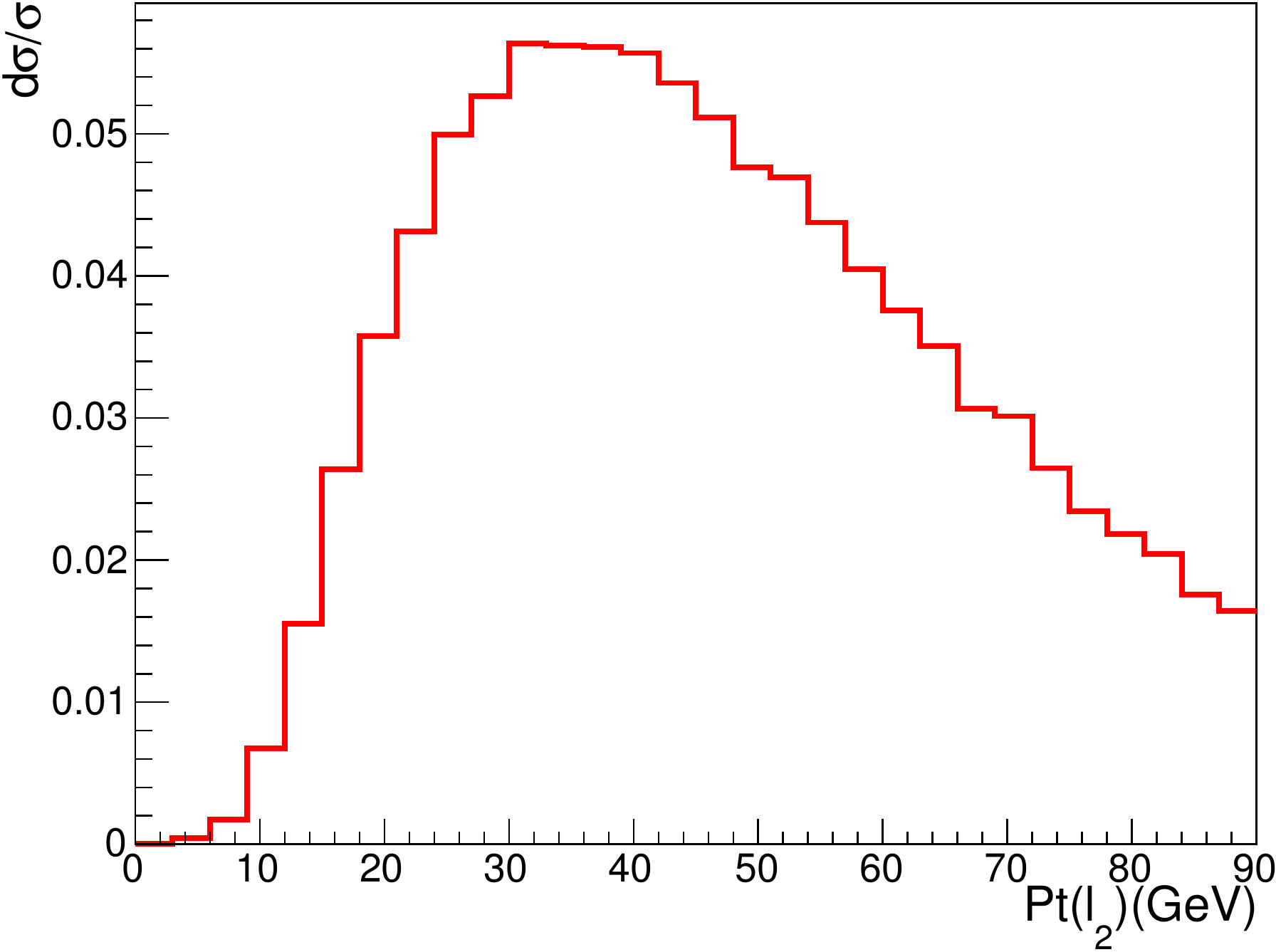}}
  \subfigure{
  \label{Fig1.sub.3}\thesubfigure
  \includegraphics[width=0.4\textwidth]{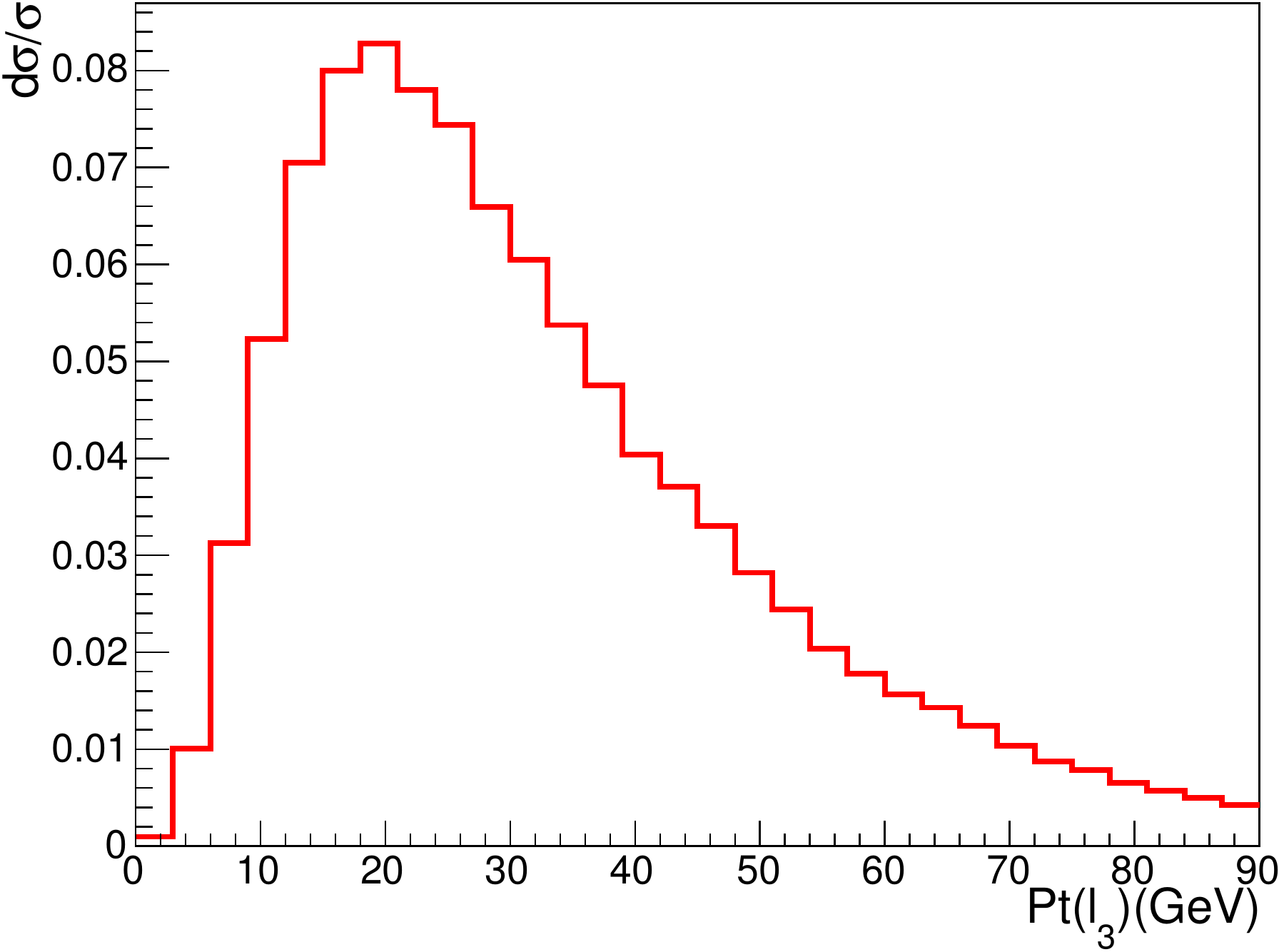}}
  \subfigure{
  \label{Fig1.sub.4}\thesubfigure
  \includegraphics[width=0.4\textwidth]{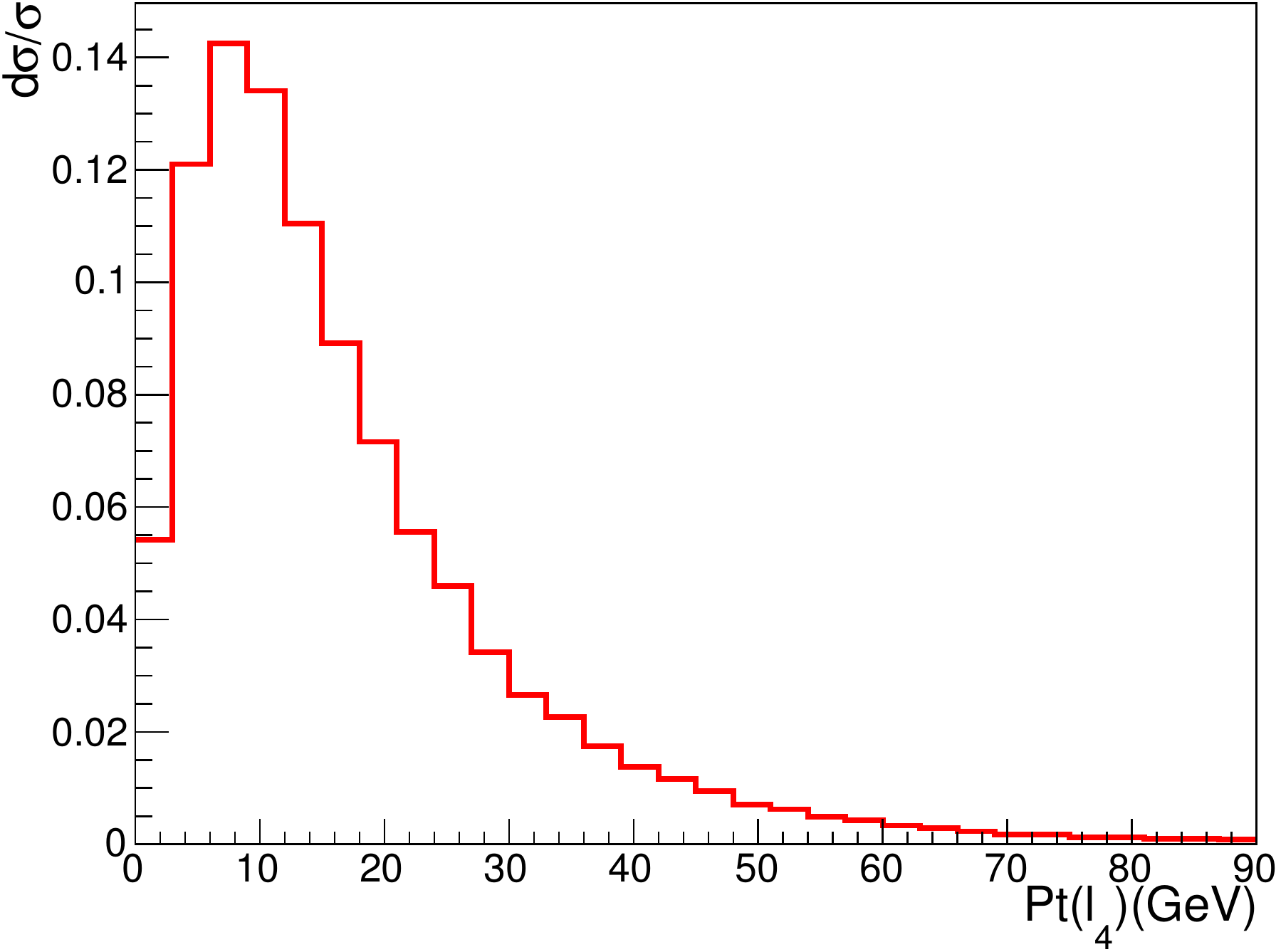}}
  \caption{The distributions of four leptons in the signal events are demonstrated before preselection cuts.}\label{fig1}
\end{figure}

\begin{figure}[htbp]
  \centering
  \subfigure{
  \label{Fig2.sub.1}\thesubfigure
  \includegraphics[width=0.4\textwidth]{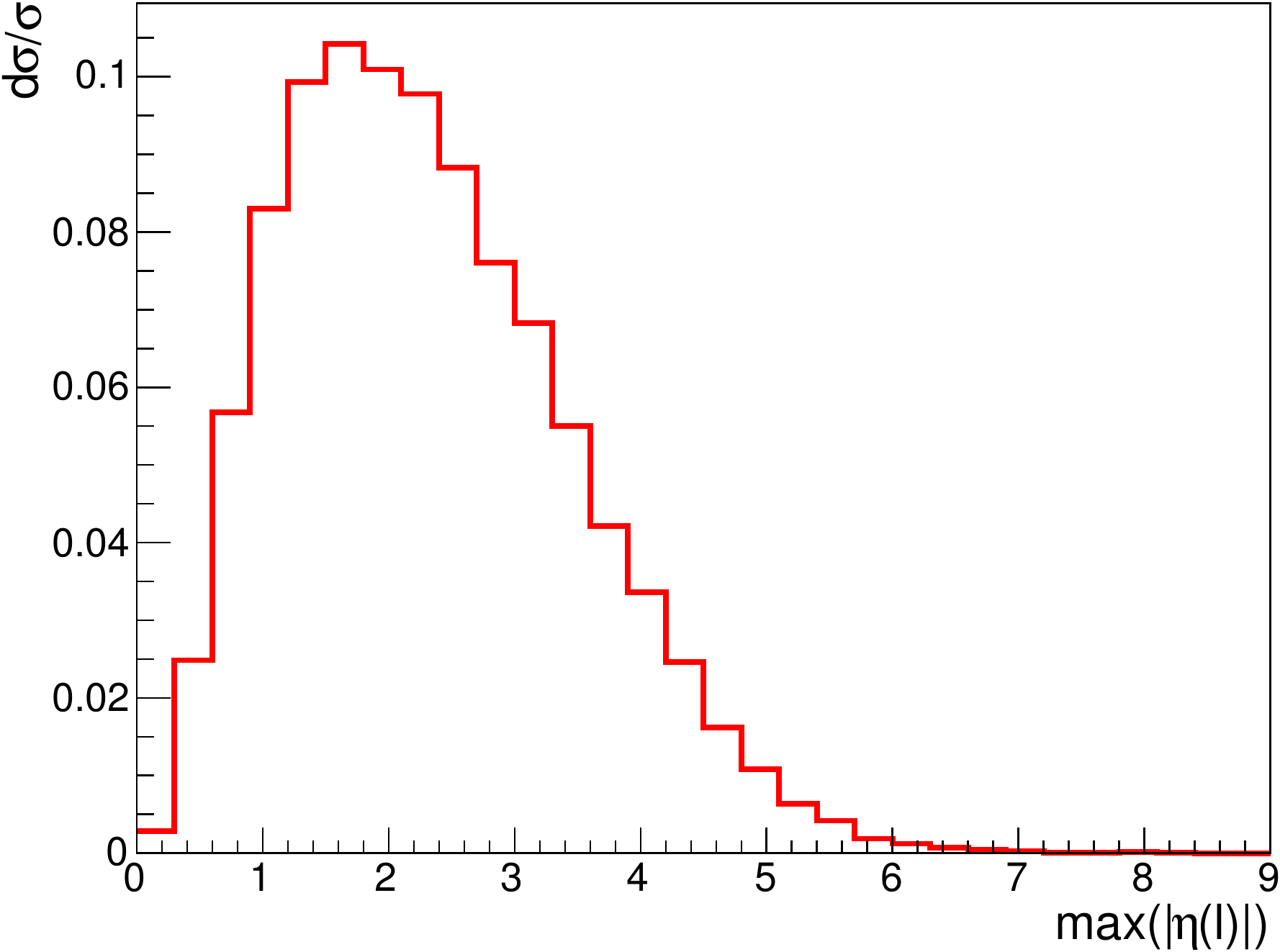}}
  \subfigure{
  \label{Fig2.sub.2}\thesubfigure
  \includegraphics[width=0.4\textwidth]{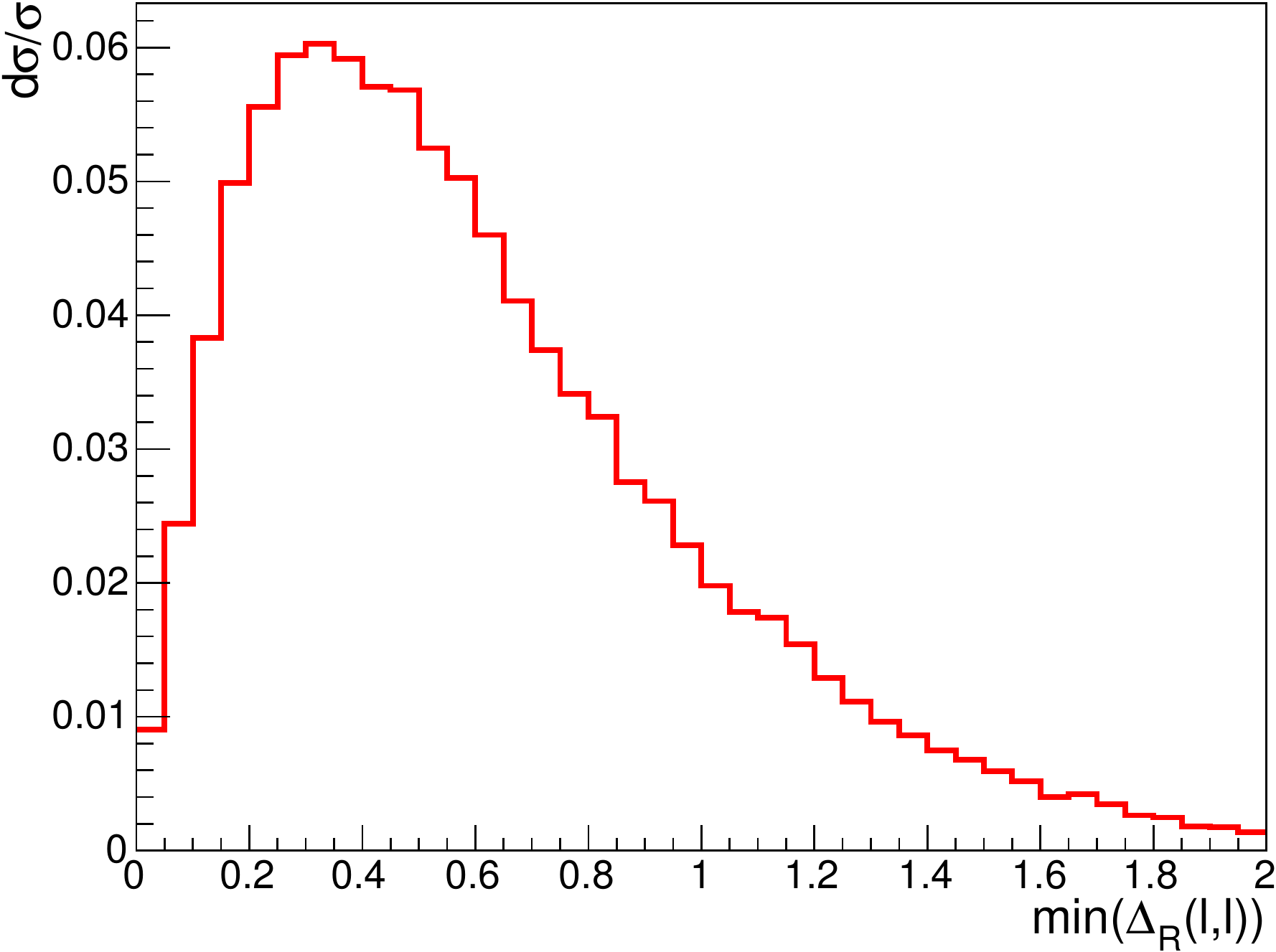}}
  \caption{The distributions of max$(\eta(\ell))$ and the minimal angular separation min$(R(\ell\,\ell))$ four leptons in each a signal event are demonstrated before preselection cuts.}\label{fig2}
\end{figure}

The backgrounds processes can be roughly categorized into the following three types:
\begin{itemize}
\item 1) The single Z processes include $p p \to Z h $, $p p \to Z W^+ W^-$, and $p p \to Z t \bar{t} $. The last process has a cross section around 35 times larger than the former two processes due to its QCD nature, while $Zh$ and $Z W^+ W^-$ have similar cross sections.
\item 2) The top pair processes include $p p \to t \bar{t} h$, $p p \to t \bar{t} t \bar{t}$, and $p p \to t \bar{t} W^+ W^-$. We notice that $h t \bar{t}$ is the dominant background in this category.
\item 3) The pure electroweak processes include $p p \to h W^+ W^-$ and $p p \to W^+ W^- W^+ W^-$. Each of their cross section is smaller than that of the signal, but the sum of them is comparable to that of the signal after taking into account the K factors.
\end{itemize}
The cross sections of these processes are listed in the Table \ref{lptn}. It is worthy of remarking that single-Higgs associated processes in the categories above are the main background events, and it turns out that $t\bar{t} h$ is the dominant background for the full leptonic mode, which can greatly affect the significance. 
\begin{center}
\begin{table}
  \begin{center}
\begin{tabular}{|c|c|c|c|c|c|c|}
\hline
     &Labels & Cross section &M1 & M2 & M3 & M4\\ 
   processes & in Figs.  &  in fb& $\ell^{+}\ell^{-}\ell^{+}\ell^{-}$ & $e^{+}e^{-}\mu^{+}\mu^{-}$ &
  $\ell^{+}\ell^{-}\ell^{\pm}\ell^{\prime\mp}$ & $\ell^{+}\ell^{\prime -}\ell^{+}\ell^{\prime -}$ \\
  \hline
  $h h$ & signal &0.29  &$\frac{1}{8}$ & $\frac{1}{4}$ & $\frac{1}{2}$ & $\frac{1}{8}$ \\ \hline
  $Z h$, $ZW^{+}W^{-}$, $Z t\bar{t}$ & $Z+$ &5.40,17.8,217 &$\frac{1}{4}$ & $\frac{1}{4}$ & $\frac{1}{2}$ & 0\\ \hline
  $ t\bar{t} h$, $t\bar{t}t\bar{t}$, $t\bar{t}W^{+}W^{-}$ & $t \bar{t}+$ & 16.9,7.12,2.3 &$\frac{1}{8}$ & $\frac{1}{4}$&$\frac{1}{2}$& $\frac{1}{8}$  \\ \hline 
  $h W^{+}W^{-}$, $W^{+}W^{-}W^{+}W^{-}$ &EW &8.4$\times 10^{-2}$, 7.7$\times 10^{-2}$&
  $\frac{1}{8}$ & $\frac{1}{4}$ & $\frac{1}{2}$ & $\frac{1}{8}$ \\ \hline
  $ZZ$ & & 485 &$\frac{1}{2}$ & $\frac{1}{2}$ & 0 & 0 \\ \hline
\end{tabular}
\end{center}
\caption{\label{lptn}The cross sections of four leptonic mode at a 100 TeV collider for different processes are tabulated, where $\ell=e,\,\mu$. Fraction of four modes in all the final states are shown.}
\end{table}
\end{center}
In order to select the most relevant events, we introduce the following preselection cuts:
\begin{itemize}
\item  For each event, there must be four leptons. The leading two leptons should have transverse momenta larger than 30 GeV and 20 GeV, respectively. While, the third and forth leptons should have transverse momenta larger than $15$ GeV and $8$ GeV. In Fig. \ref{fig1}, we show the transverse momenta of four leptons in the signal events.
\item We demand the future detector can have a better coverage of $\eta(\ell)$ as $|\eta(\ell)| < 4$ for identified leptons. Due to the good space resolution power and fine granularity in ECAL detector, we require that the minimal angular separation between two leptons is $\Delta R^{min}(\ell, \ell) \geq 0.15$. In Fig. \ref{fig2}, we demonstrate the distribution of the maximal $\eta^{max}(\ell)$ and the minimal angular separation of any a pair of two leptons $\Delta R^{min}(\ell\,\ell)$.  When these two cuts on leptons are applied, more than $90\%$ signal events can be accepted. We will discuss how the coverage of $\eta$ can affect our results in the discussion section.
\item Consider the fact that $t\bar{t}$ processes have quite large contributions to the background events, we introduce the "b-jet veto" to reject events with a tagged b-jet with $P_t(b) > 40$ GeV and $|\eta(b)|<5$. We assume that the b tagging efficiency is $60\%$ and simply time a factor $0.16$ to this type of background processes.
\item For the decay mode M3, in oder to suppress the background events from meson decays in the final states and Z boson decay, we demand the invariant mass of both lepton pairs should fall into two windows either 15 GeV $ \leq m_{\ell^+\, \ell^-}  \leq $ 80 GeV or $m_{\ell^+ \, \ell^-}  \geq $ 100 GeV. 
\end{itemize}
\begin{center}
\begin{table}
  \begin{center}
  \begin{tabular}{|c|c|}
  \hline
Preselection Cuts & Description  \\ \hline
1 &  $n_\ell = 4$ \\  
 & $P_t(\ell_1)>30$ GeV, $P_t(\ell_2)>15$ GeV \\ 
  & $P_t(\ell_3)>10$ GeV,  $P_t(\ell_4)>5$ GeV.  \\ 
 & $|\eta^{max}(\ell_i) | < 4$ \\ 
 &  $\Delta R^{min} (\ell,\,\ell)>0.15$ \\   \hline \hline 
2 & b jet veto   \\ \hline \hline
3 & low energy hadron veto and Z mass veto \\ 
  \hline
  \end{tabular}
  \end{center}
  \label{precuts}
  \caption{The preselection cuts in our analysis are tabulated.}
\end{table}
\end{center}
From the results given in Table \ref{lptn}, it is noticed that the background from $pp \to ZZ$ is huge which is about 2000 times larger in magnitude. Furthermore, due to the off shell Z and $\gamma^*$, the cut of invariant mass of two pairs of four leptons can only suppress the background down by 100 order at most. Our previous experiences in studying the $3\ell+2 j $ \cite{3l2j} reveals that background events from $Z,\gamma^*$-exchange processes are difficult to suppress. Therefore, the first two modes are challenging in the SM and will be neglected in this analysis. When there is a significant enhancement to the Higgs pair production by new physics, these two modes should also be considered. 

In this work, we will focus on the third and forth modes (labelled as "M3" and "M4", respectively) and we will perform a detailed analysis on these two modes. It is noticed that compared with the third mode although the forth mode has a smaller production rate, it enjoys less background contributions from the SM. 

\section{Kinematic features of signal and background events in M3 and M4 cases}
In this section, we explore the kinematic features of the signal and background events. It is impossible to fully reconstruct all the final particles due to 4 neutrinos in the final state.
\begin{figure}[htbp]
  \centering
  \subfigure{
  \label{Fig31.sub.1}\thesubfigure
  \includegraphics[width=0.4\textwidth]{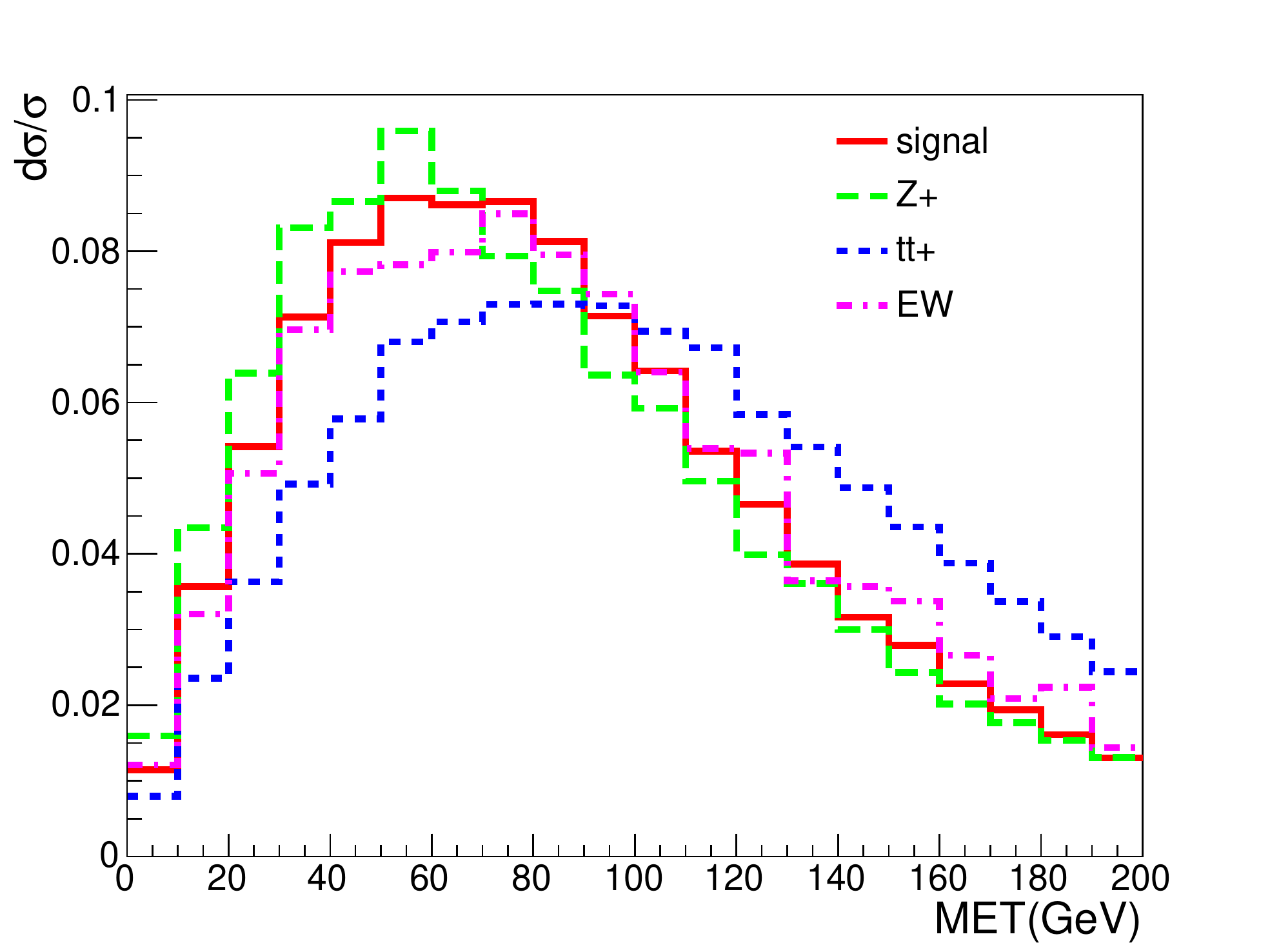}}
  \subfigure{
  \label{Fig31.sub.2}\thesubfigure
  \includegraphics[width=0.4\textwidth]{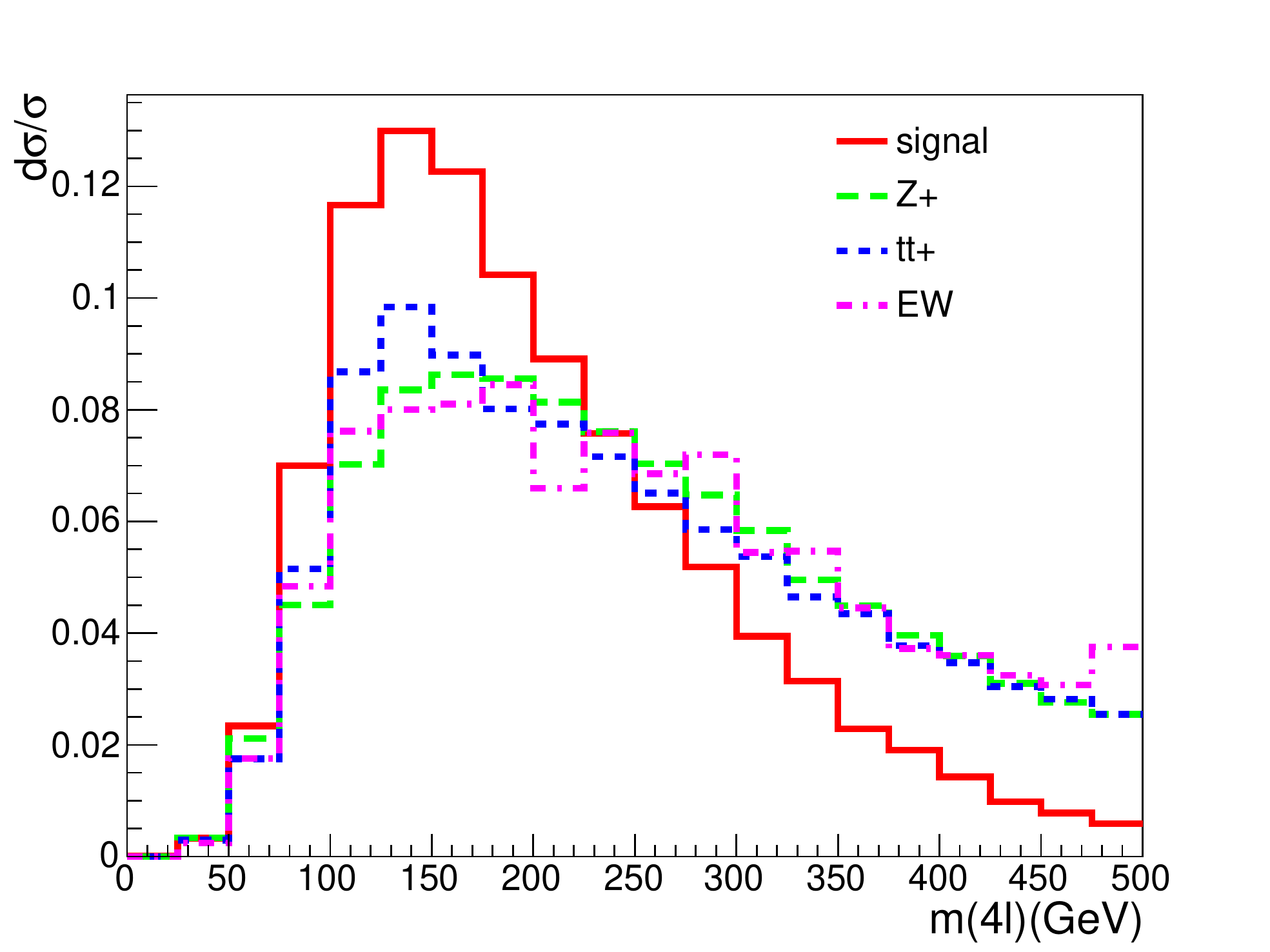}}
  \subfigure{
  \label{Fig31.sub.3}\thesubfigure
  \includegraphics[width=0.4\textwidth]{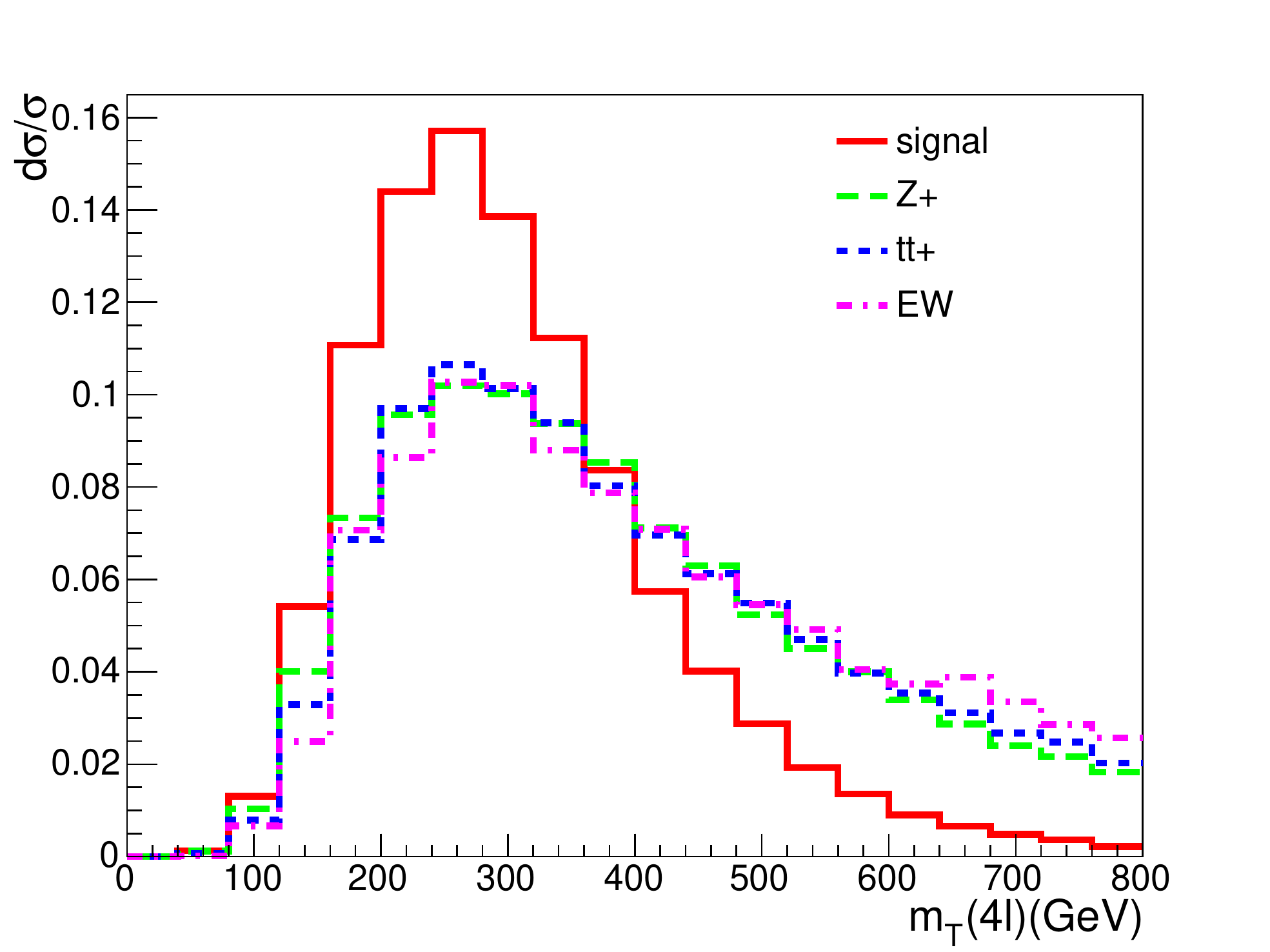}}
  \subfigure{
  \label{Fig31.sub.4}\thesubfigure
  \includegraphics[width=0.4\textwidth]{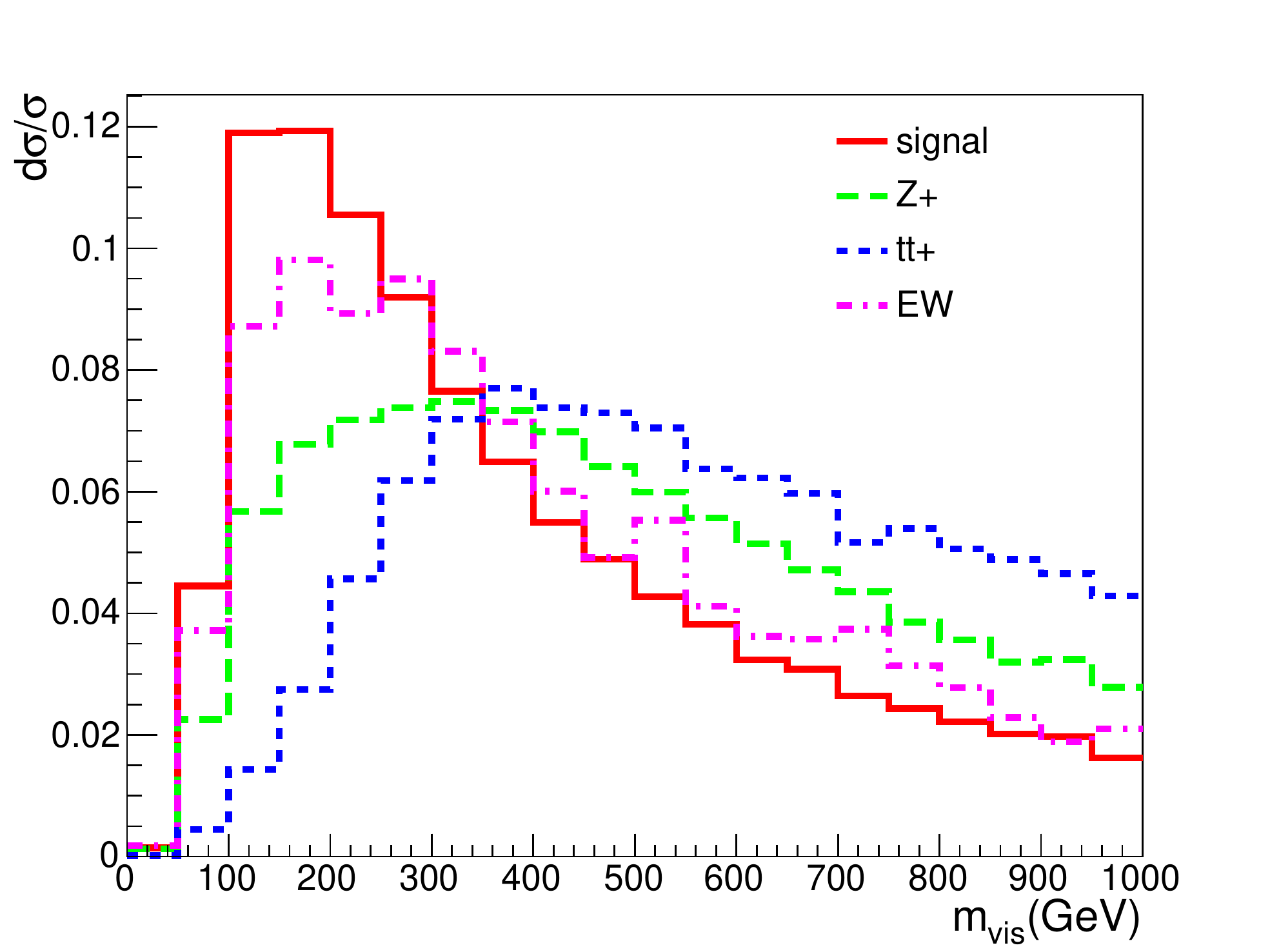}}
  \subfigure{
  \label{Fig31.sub.5}\thesubfigure
  \includegraphics[width=0.4\textwidth]{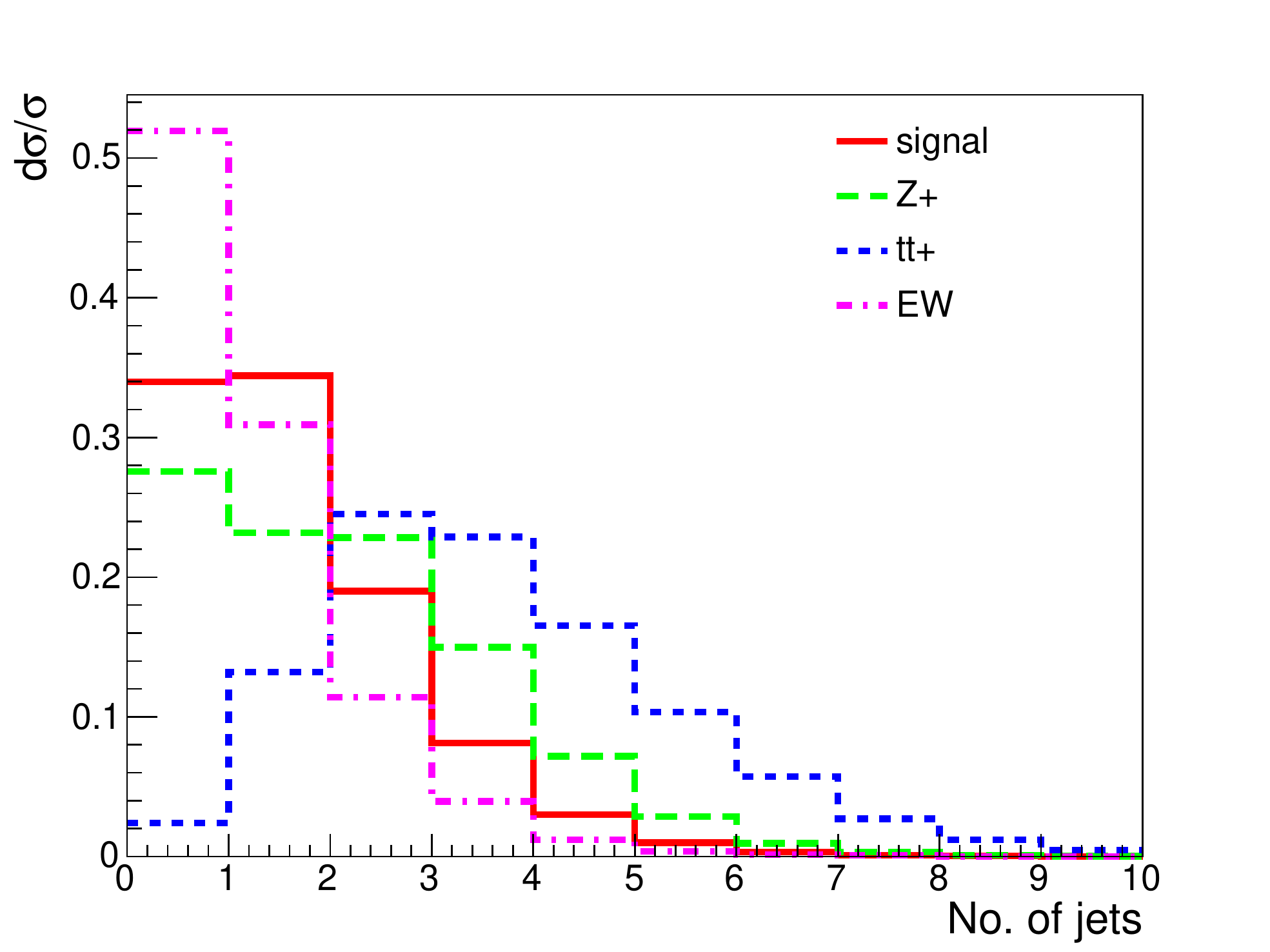}}
  \subfigure{
  \label{Fig31.sub.6}\thesubfigure
  \includegraphics[width=0.4\textwidth]{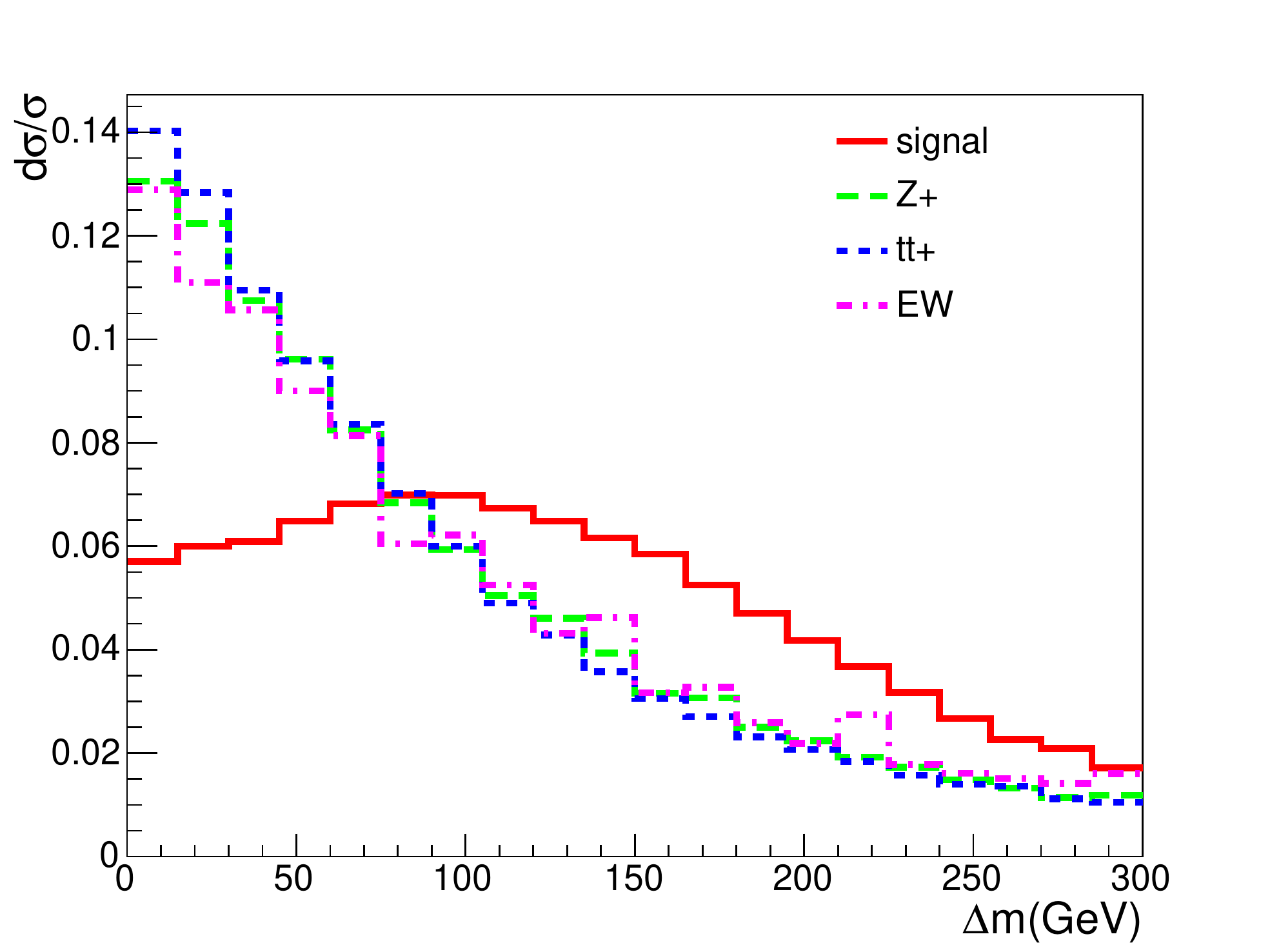}}
    \subfigure{
  \label{Fig31.sub.7}\thesubfigure
  \includegraphics[width=0.4\textwidth]{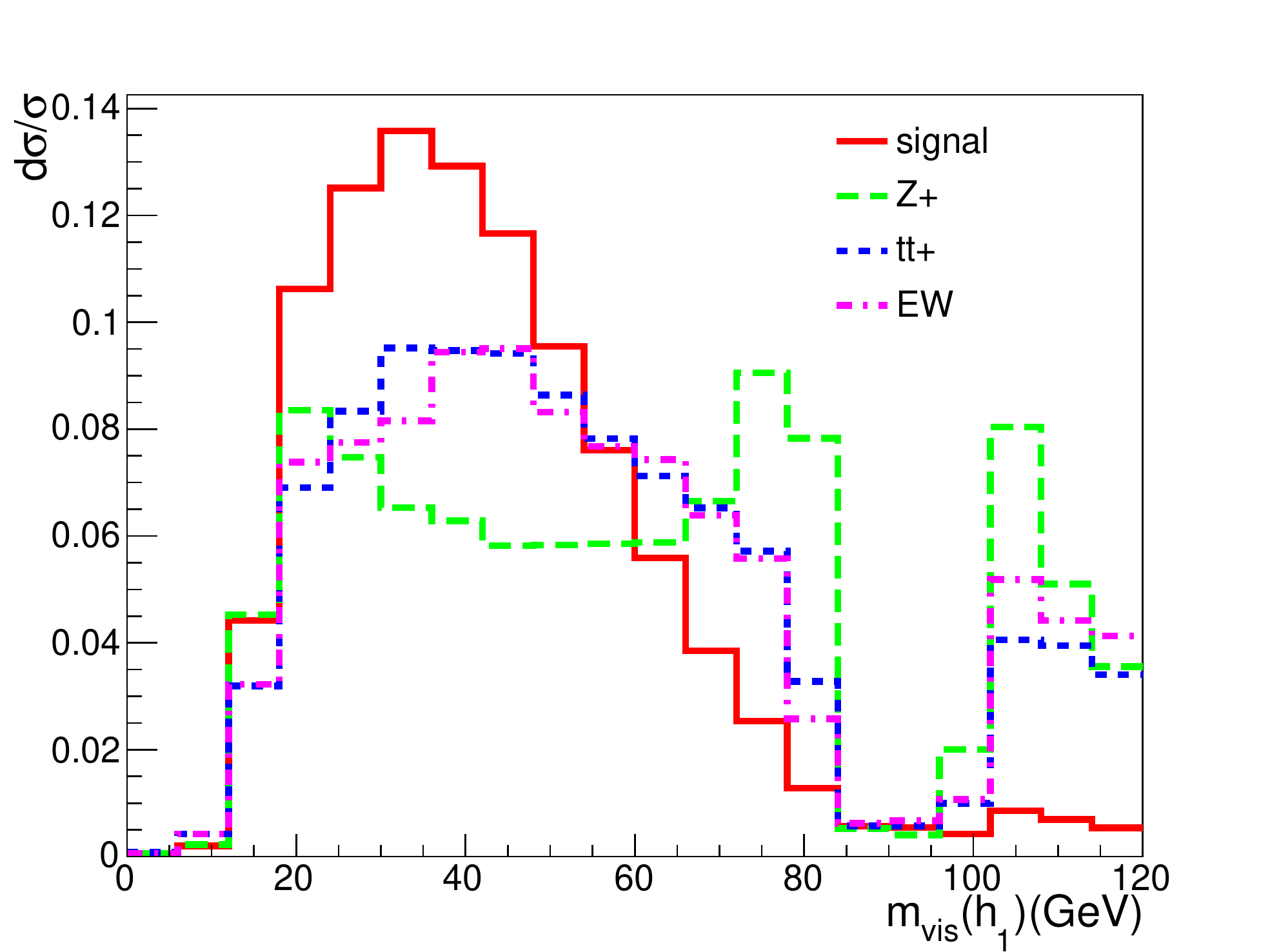}}
  \subfigure{
  \label{Fig31.sub.8}\thesubfigure
  \includegraphics[width=0.4\textwidth]{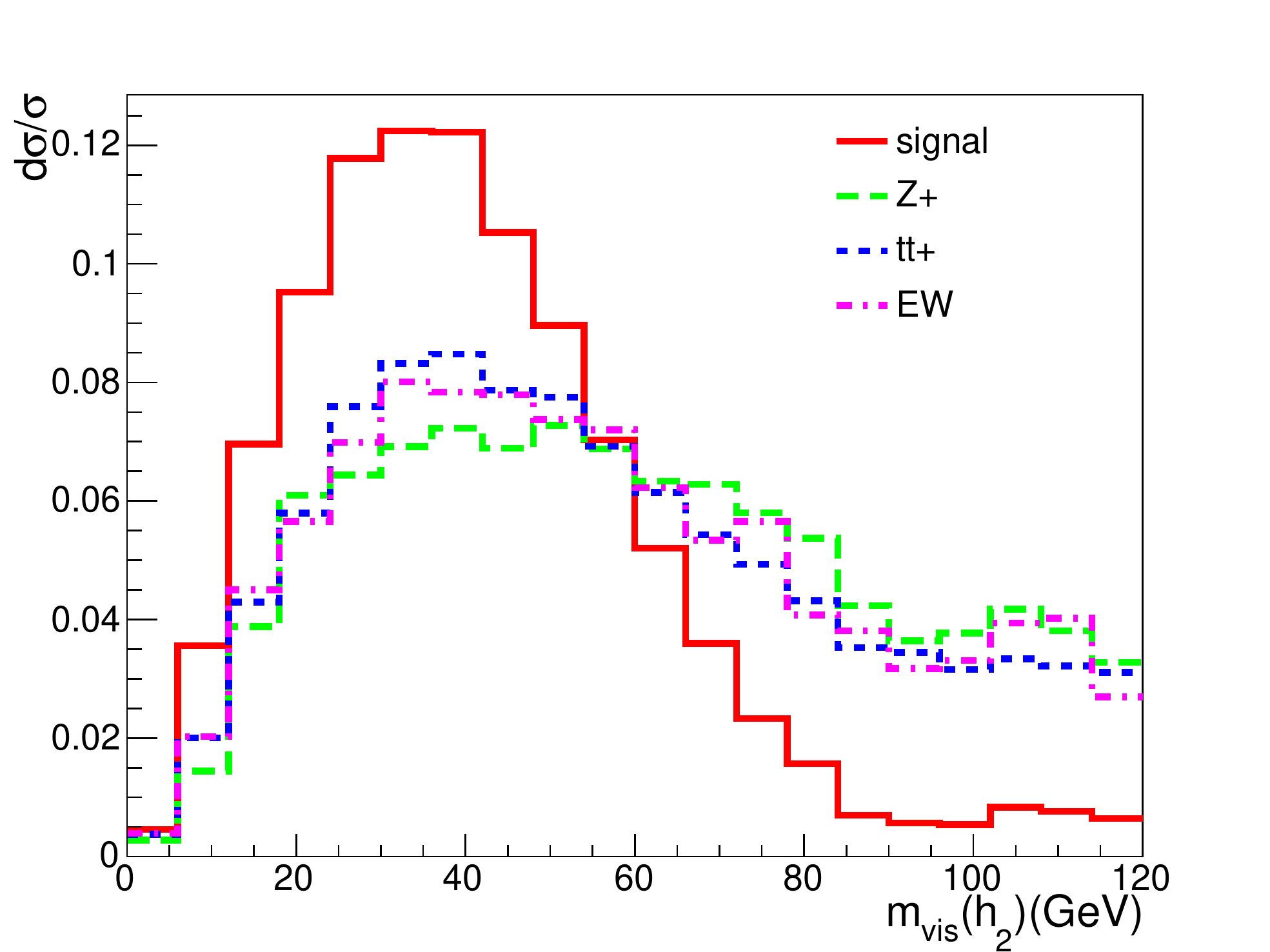}}

  \caption{Some useful kinematic observables for the M3 case which can separate signal and background events are shown. }\label{fig31}
\end{figure}

The physical observables can be divided into two types. {\bf The first type} is defined as the global and topological event shape observables for each event, and {\bf the second type} is defined from the partial reconstruction method introduced later. {\bf The first type} includes the following observables listed below.
\begin{itemize}
\item o1) The missing transverse momentum spectrum. Since there are 4 neutrinos in the final states, we expect that there should be a large transverse momentum. For the signal, as shown in Fig. \ref{Fig31.sub.1}, the distribution peaks near 60-80 GeV.
\item o2)) The invariant mass of four leptons. This quantity is expected to capture the mass of mother particles. For the signal, as shown in Fig. \ref{Fig31.sub.2}, the distribution peaks near 100-150 GeV. Since four neutrinos can take away half of the energy of Higgs pair, so this quantity is expected to be close the mass of one Higgs boson.
\item o3) The transverse mass of each event is constructed from the sum of 4-momentum of leptons (denoted as $P_{4 \ell}$, which has components $(E_{4\ell}, P_{4 \ell}^x,P_{4 \ell}^y,P_{4 \ell}^z)$ ) and the missing transverse momentum ($\missP_T$ which has components $(\missE_T, \missP_x, \missP_y, 0)$) where $\missE_T = \sqrt{\missP_x^2 + \missP_y^2}$. To construct this observable,  we boost the 4-momentum of leptons such that the $P^z_{4 \ell}=0$, i.e. $(\tilde{E}_{4\ell}, P_{4 \ell}^x,P_{4 \ell}^y,0)$. Then we construct the observable as $\sqrt{\tilde{E}_{4\ell} \missE_T - 2 P_{4\ell}^T \missE_T cos[\phi]}$, where $\phi$ denotes the athemuthal angle between the $\tilde{P}_{4 \ell}$ and $\missP_T$. For the signal, as shown in Fig. \ref{Fig31.sub.3}, the distribution peaks near 150-250 GeV.
\item o4) The invariant mass of visible objects is defined as the momentum sum of four leptons and jets. For the signal, as demonstrated in Fig. \ref{Fig31.sub.4}, the distribution peaks near the region 100-300 GeV. 
\item o5) The number of jets in each event with $P_t(j) >  40 $ GeV. As demonstrated in Fig. \ref{Fig31.sub.5}, we noticed that when demand $n_j \leq 2$, around $90\%$ signal events can be selected out, though with a considerable background events from $t\bar{t}$ processes.
\end{itemize}

Below we explain how to construct {\bf the second type} of observables.

Obviously, the most important information about the the signal events is the mass of Higgs boson. So it is crucial to extract this useful observable. Due to the fact that four neutrinos can not be fully reconstructed, instead we can only determine the visible masses of each Higgs boson from the identified leptons. 

To determine the visible mass of each Higgs boson mass, we encounter a minor combinatorics issue: there are two possible combinations in each event. To determine which one is correct, we follow the minimal mass method introduced in \cite{3l2j} to determine the right combination, which can yield a correctness up to $94\%$ here as we have checked this by using the parton level data sample. The method evaluates the sum of two visible masses of Higgs bosons for each combination, and picks out the smaller one as the correct combination. In contrast, we examine  a method by using the angular separation of leptons, which can only yield a correctness up to $85\%$ at most.

After having found the visible mass of Higgs boson in the signal, by using the standard M$_{T2}$ method we can split the missing transverse momentum into two parts and exploit the kinematic feature of pair production to reconstruct the transverse mass of Higgs boson. Then we can construct {\bf the second type} of observables as listed below.
\begin{itemize}
\item o6) The observable $\Delta m$, which is defined as the mass difference between two mass sums of reconstructed Higgs bosons in  two possible combinations. The distribution is shown in Fig. \ref{Fig31.sub.6}, which shows a large shape difference between the signal and background processes in shape.
\item o7)) The first reconstructed partial mass of Higgs boson with two leptons of the same flavour in M3 case, which is labelled as $h_1$. In the M4 case, the one with the hardest lepton is labelled as $h_1$. The distribution of this quantity is shown in Fig. \ref{Fig31.sub.7} and Fig. \ref{Fig33.sub.7}. In Fig. \ref{Fig31.sub.7}, the Z mass window cut is clearly shown for the processes with a single Z.
\item o8)) The second reconstructed partial mass of Higgs boson with two leptons of different flavour in M3 case, which is labelled as $h_2$. In the M4 case, the one reconstructed not with the hardest lepton is labelled as $h_2$. The distribution of this quantity is shown in Fig. \ref{Fig31.sub.8} and Fig. \ref{Fig33.sub.8}.
\item o9) The transverse mass of Higgs bosons reconstructed by using the $M_{T2}$ method, which is shown in Fig. \ref{Fig32.sub.1} and Fig. \ref{Fig34.sub.1}.
\end{itemize}

It is noticed that all of these reconstructed observables are crucial and important for both M3 and M4 cases. There exist strong correlations among these observables for signal events and they are related to the mass of Higgs boson while for the background they are not necessarily related to the mass of Higgs boson. Such a fact can be utilised to separate signal and background, which is the spirit of multivariate analysis methods.

\subsection{The M3 case}

In the M3 case, the dominant background processes is the single Z associated processes, as clearly demonstrated in the Table \ref{m3cuts} in the cut-based analysis. From Fig. \ref{Fig31.sub.7}, it is noticed that the mass cut can clearly affect the single Z processes.

In the cut-based analysis, we choose the sequential 4 cuts: 1) a cut on the number of jets $n_j \leq 2$, which is supposed to suppress background processes associated with a top pair; 2) a cut on the variable $m_{T2}$, as we demand $m_{T_2}< 110 $ GeV, which can greatly suppress the background processes like three body and four body productions; 3) a cut on the reconstructed visible mass of Higgs boson is imposed as $m_{h_{1,2}} \leq 60 $ GeV, which can suppress background events from the $Z h$ and $h t \bar{t}$ processes; 4) a cut on the mass difference $\Delta m>50$ GeV. By using these cuts, we can achieve the S/B and significance 0.25 and 4.2, respectively.

\begin{figure}[htbp]
  \centering
  \subfigure{
  \label{Fig32.sub.1}\thesubfigure
  \includegraphics[width=0.4\textwidth]{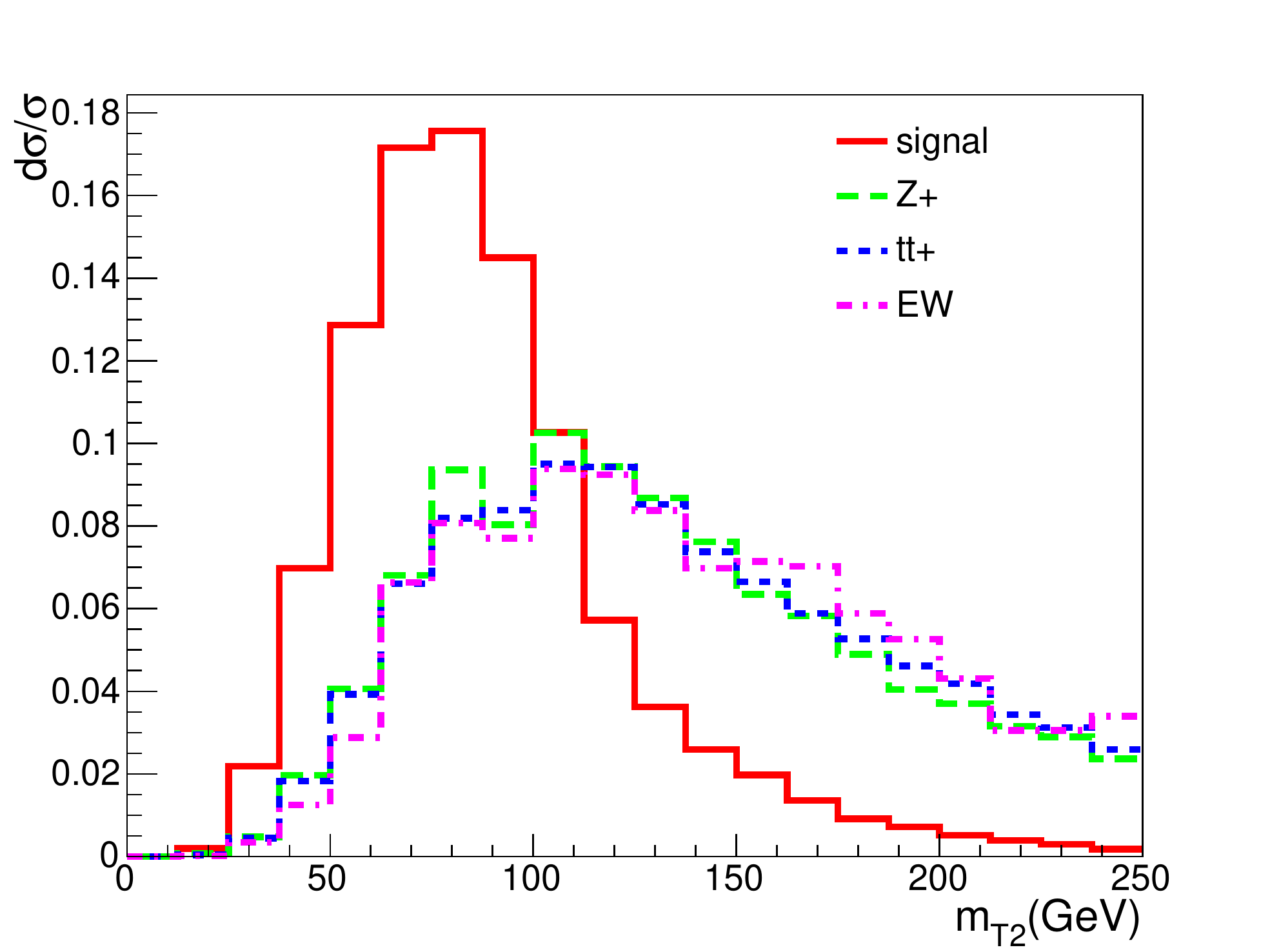}}
  \subfigure{
  \label{Fig32.sub.2}\thesubfigure
  \includegraphics[width=0.4\textwidth]{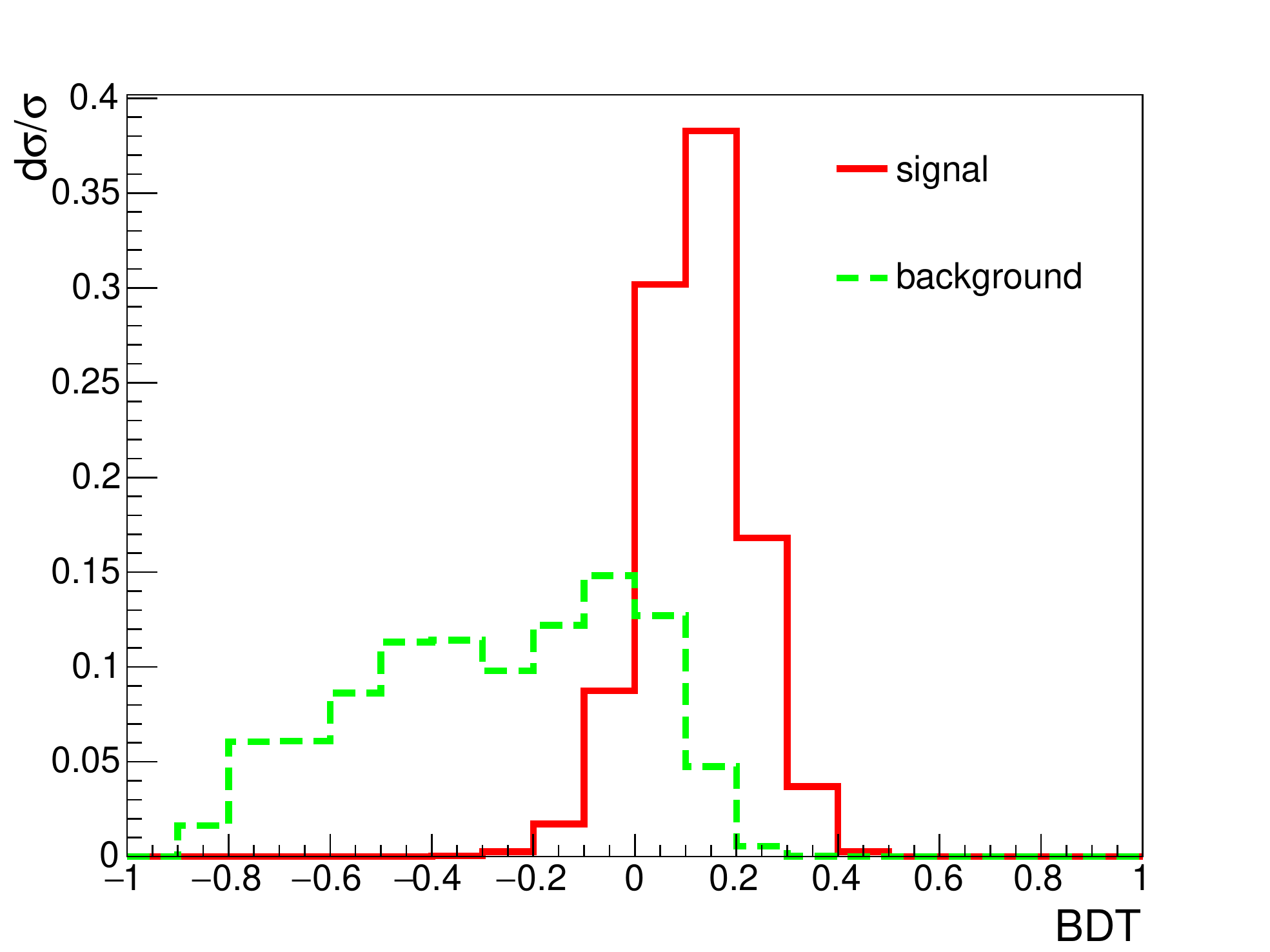}}
  \caption{The reconstructed visible masses of two Higgs bosons and $m_{T2}$ for the M3 case and the BDT discriminant are shown. }\label{fig32}
\end{figure}

According to the results based on the cut-based method, it is noticed that the processes associated with top pair are the main background for the "M3" mode. In order to suppress this type of background, we introduce a few top veto observables in our BDT analysis. For example, the visible invariant mass of the whole event, as demonstrated in Fig. \ref{Fig31.sub.4}, can help to separate background events with and without top quark pair. Furthermore, the number of jets in each events can help to suppress the background events with top quark pair, as demonstrated in Fig. \ref{Fig31.sub.5}. It is noticed that these two observables are correlated with each other for the background processes.

In Table \ref{m3cuts}, both a cut-based analysis and a BDT analysis are presented to compare. As demonstrated in Fig. \ref{Fig32.sub.2}, after taking into account these observables to reject top quarks final states, we can further suppress background and gain in the S/B and significance up to 0.38 and 6.1, respectively, as represented in the Table \ref{m3cuts}. As discussed in \cite{Hinchliffe:2015qma}, if an integrated luminosity 20-30 fb$^{-1}$ for a 100 TeV collider can be achieved, then a significance 20.0 is expected.

\begin{center}
\begin{table}
  \begin{center}
\begin{tabular}{|c||c|c|c|c|c||c|}
\hline
        Processes &  Pre-Sel. Cuts & $n_j \leq 2$ &  $M_{T2}<\SI{115}{GeV}$ & $m({h}^{vis})<\SI{60}{GeV}$ & $\Delta m >\SI{50}{GeV}$  & BDT\\
        \hline \hline
       hh & 172 & 150 & 124 & 91.2 & 68.8 & 99.8 \\
        \hline \hline
        Z h   &   243 & 238 & 197 & 66.6 & 23.4 & 27.4 \\
        $Z W^+ W^-$ &  $1.60\times 10^3$  &  $1.54\times 10^3$ & 444 & 173 & 91.4 & 111.7 \\
        $Zt\bar{t}$ &  $2.55\times 10^3$ & $1.45\times 10^3$& 542 & 222 & 117 & 89.7 \\
        \hline \hline
        $t\bar{t} h$ & 446 & 245 & 128 & 68.0 & 31.1 & 29.5 \\
        $ t\bar{t}t\bar{t}$ & 254 & 24.4 & 3.96 & 1.44 &1.04 & 3.63 \\
        $ t\bar{t} W^+ W^-$  & 151 & 71.2 & 12.5 & 4.16 & 2.48 & 4.30 \\
        \hline \hline
        $  h W^+ W^-$ & 44.5 & 42.7 & 20.3 & 10.3 & 4.94 & 6.11\\
        $W^+ W^- W^+ W^-$  & 50.9 & 47.3 & 5.67 & 1.98 & 1.20 & 1.98\\
        \hline \hline
        S/B  & $3.2 \times 10^{-2}$    &  $4.1 \times 10^{-2}$ & 0.10 & 0.17 & 0.25 & 0.38\\
        \hline
        $S/\sqrt{B}$  & $2.35$  &  $2.48$   &   $3.37$    &    $3.90$ & $4.20$ & $6.1$ \\ \hline
   \end{tabular}
  \end{center}
  \label{precuts}
  \caption{\label{m3cuts}A cut-based analysis and a BDT analysis for the M3 case are presented. Efficiencies of each cut in the cut-based analysis are demonstrated. We assume the integrated luminosity as 3 ab$^{-1}$.}
\end{table}
\end{center}

\subsection{The M4 case}
In the M4 case, similar to the M3 case, we can reconstruct the visible mass of Higgs boson, $h_1(\ell \ell^{\prime})$ and $h_2(\ell\ell^{\prime})$, respectively, where subscript 1 and 2 is assigned according to the simple rule: the one which includes the most energetic lepton is assigned to be $h_1$ and the other is assigned to be $h_2$. It is noticed that these two masses peak near 30-50 GeV and 20-40 GeV, respectively, and both have a edge is near 60 GeV.

\begin{figure}[htbp]
  \centering
  \subfigure{
  \label{Fig33.sub.1}\thesubfigure
  \includegraphics[width=0.4\textwidth]{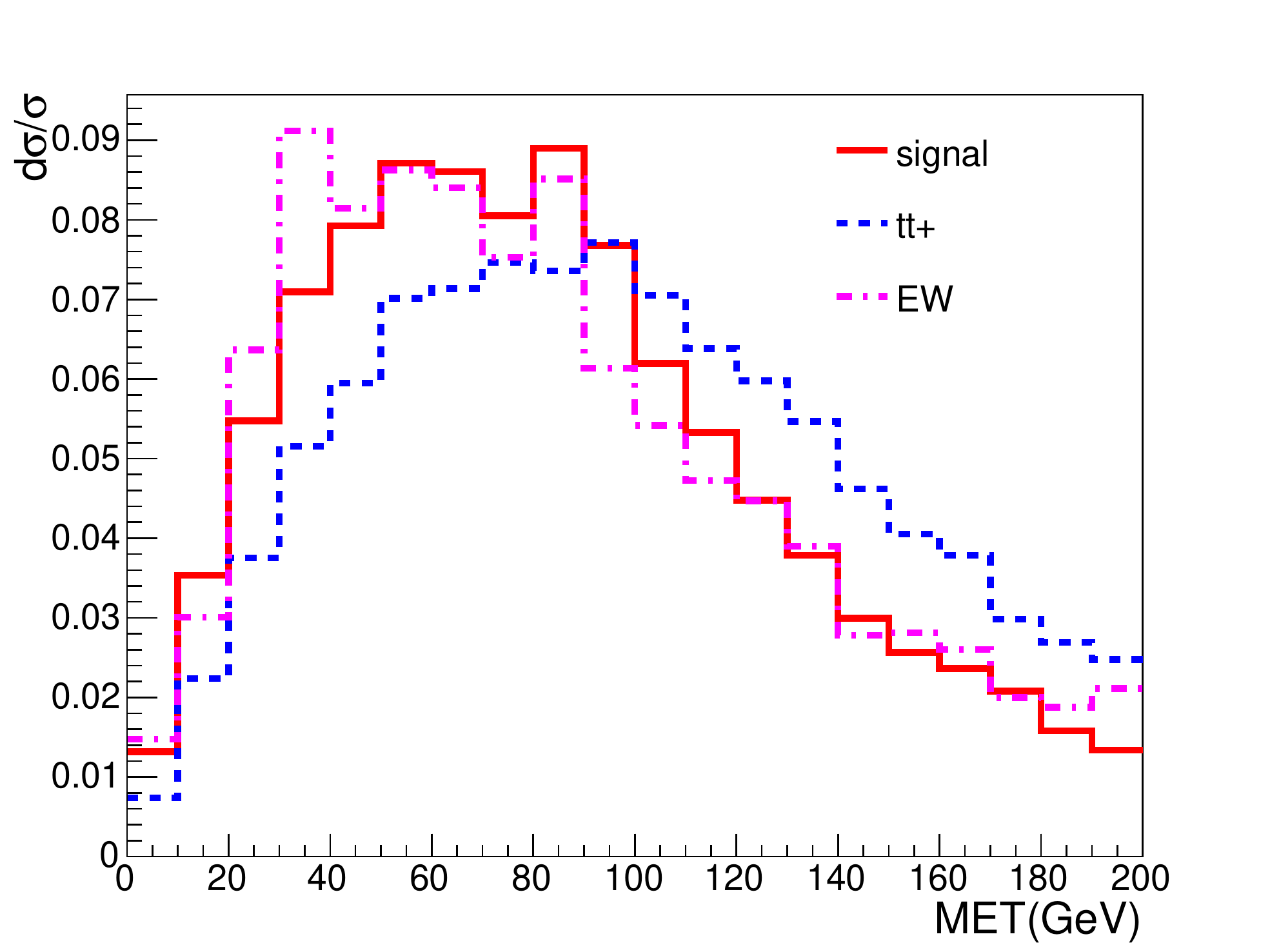}}
  \subfigure{
  \label{Fig33.sub.2}\thesubfigure
  \includegraphics[width=0.4\textwidth]{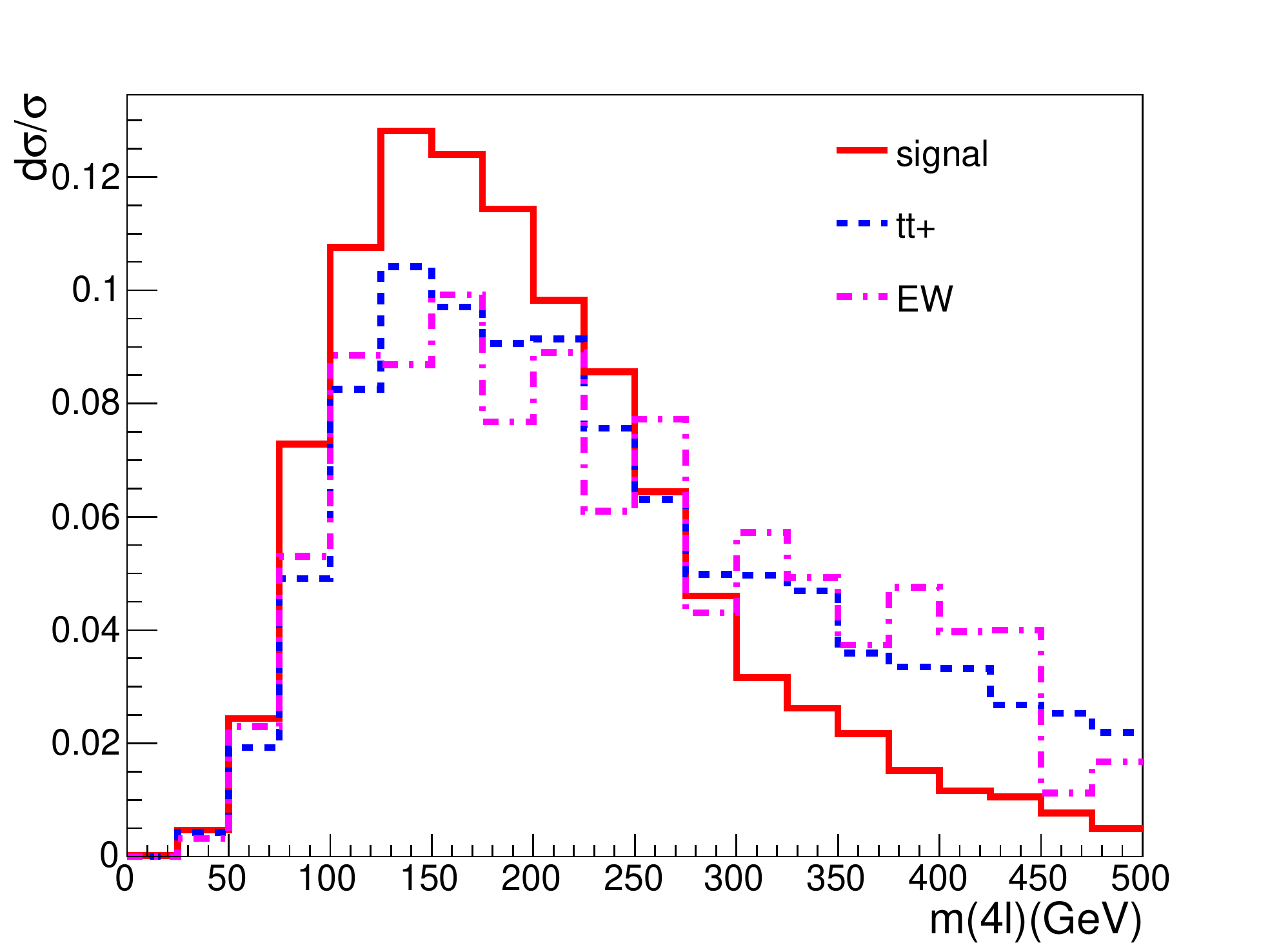}}
  \subfigure{
  \label{Fig33.sub.3}\thesubfigure
  \includegraphics[width=0.4\textwidth]{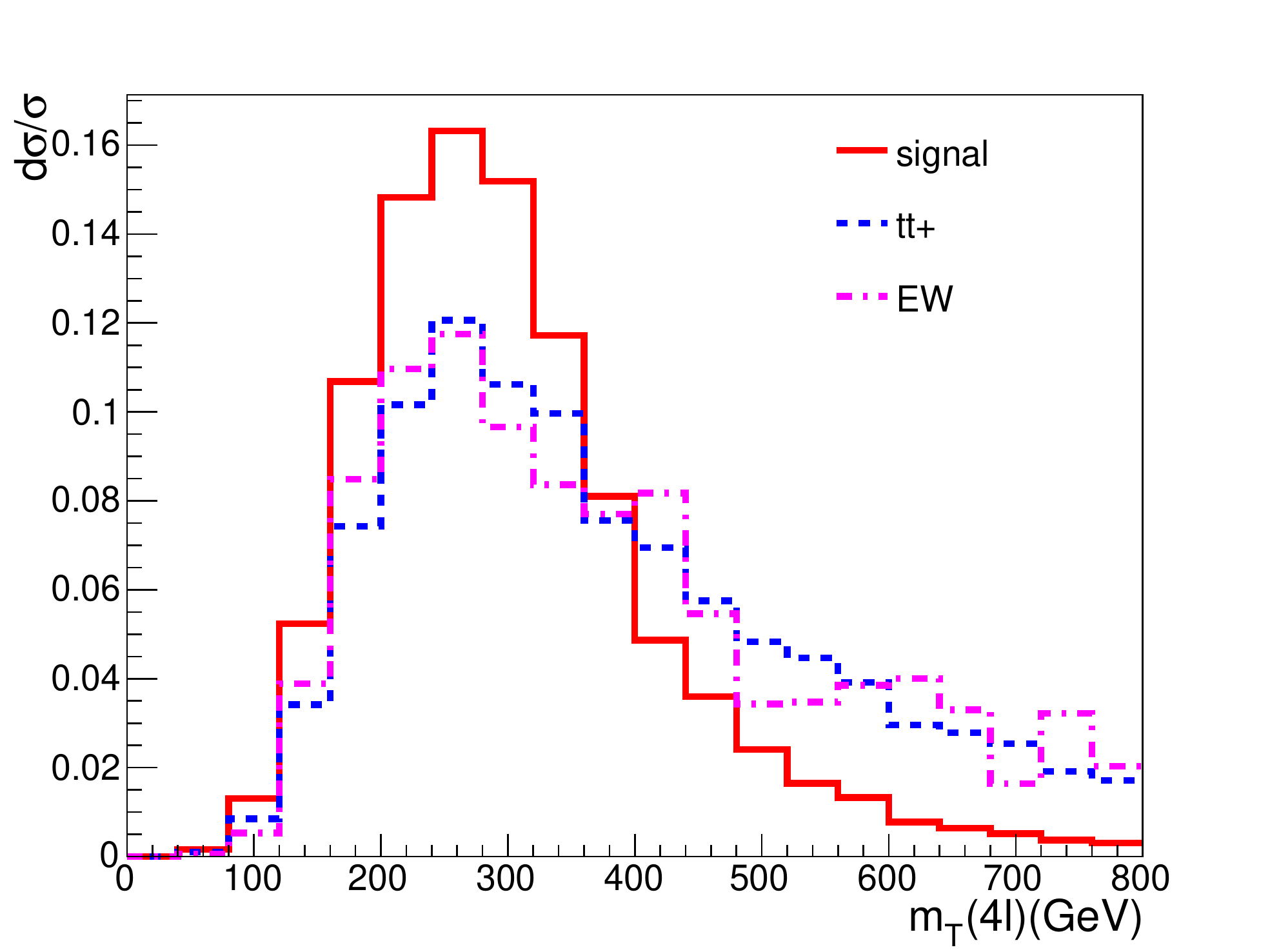}}
  \subfigure{
  \label{Fig33.sub.4}\thesubfigure
  \includegraphics[width=0.4\textwidth]{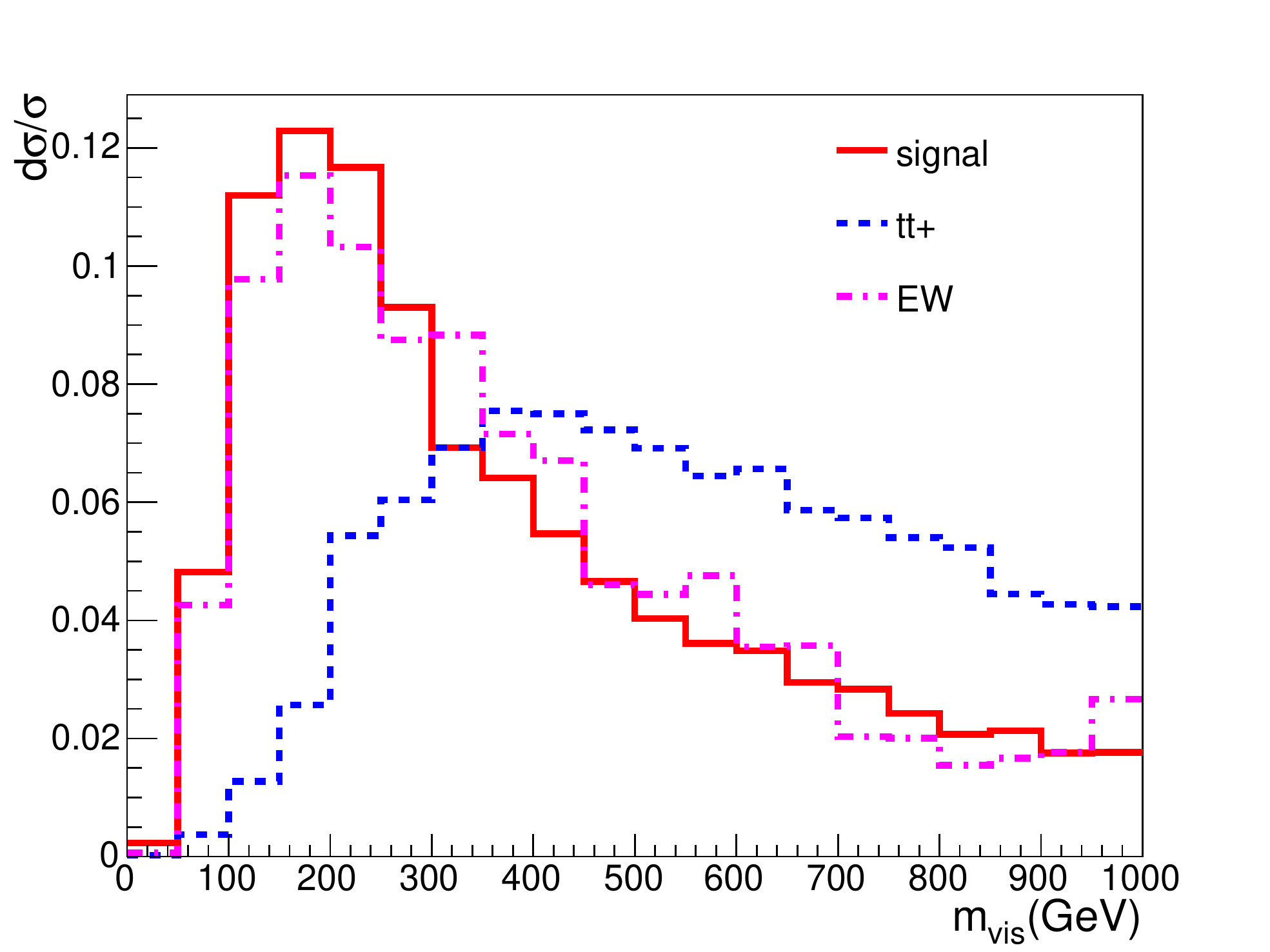}}
  \subfigure{
  \label{Fig33.sub.5}\thesubfigure
  \includegraphics[width=0.4\textwidth]{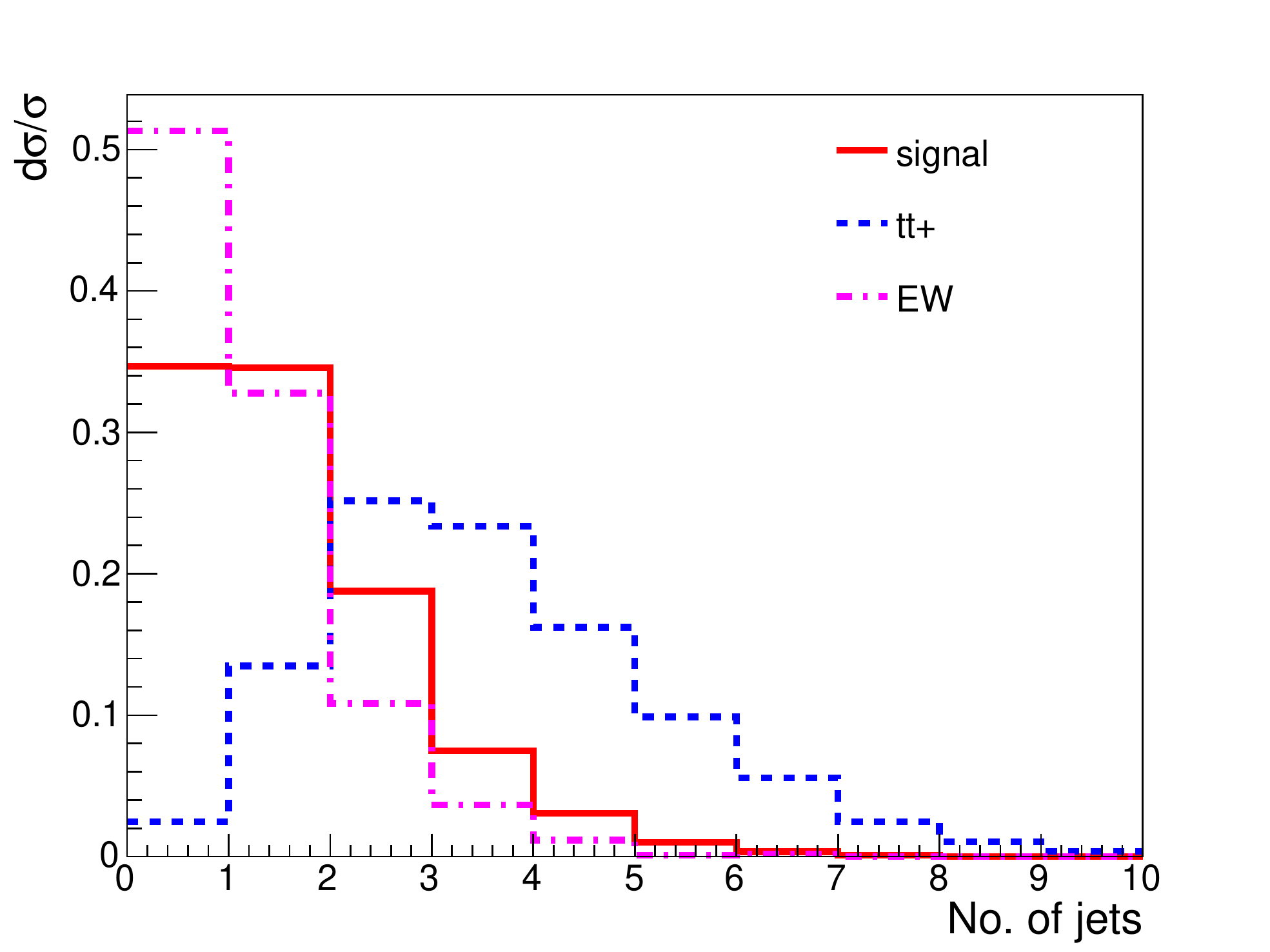}}
  \subfigure{
  \label{Fig33.sub.6}\thesubfigure
  \includegraphics[width=0.4\textwidth]{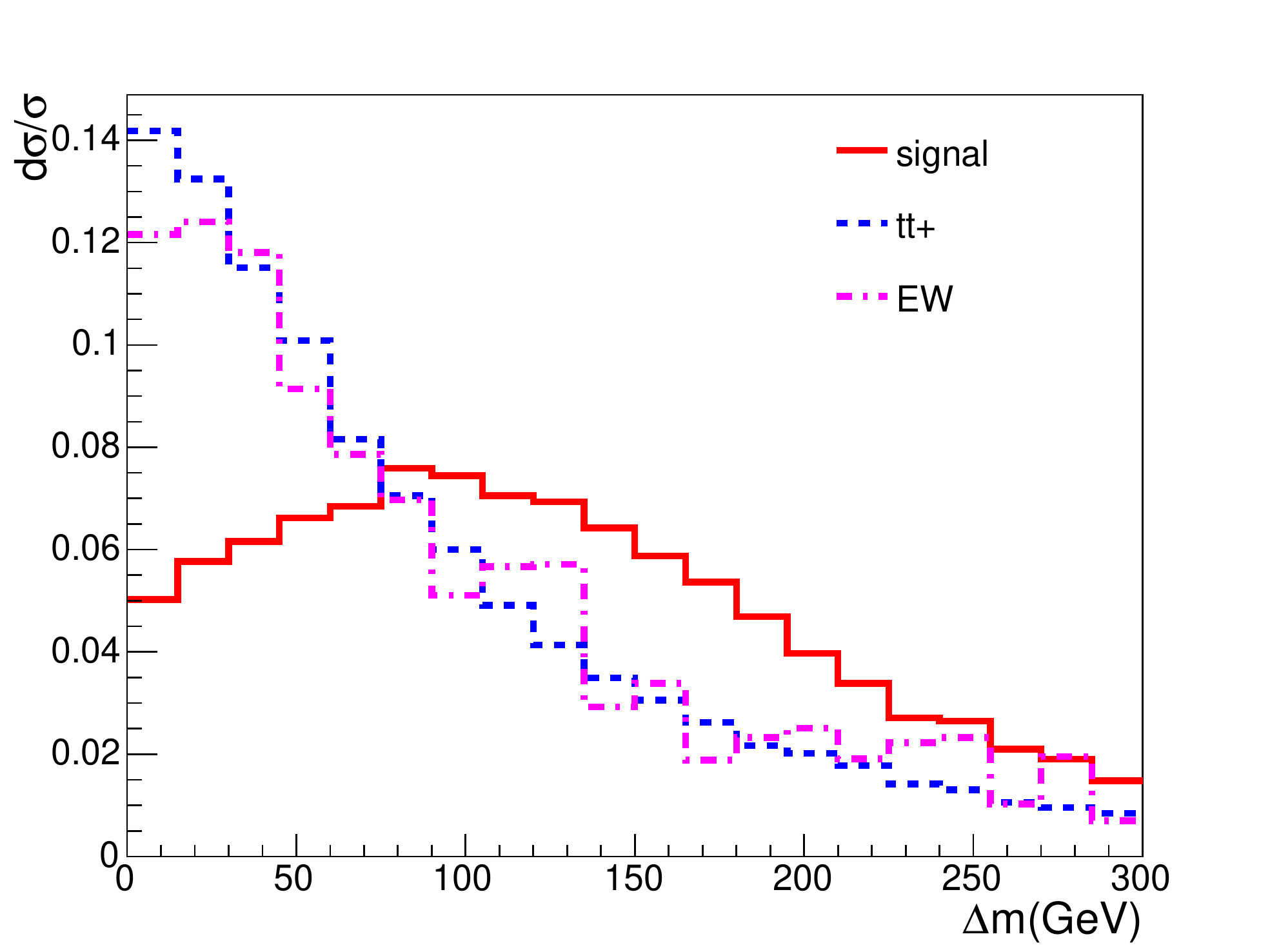}}
  \subfigure{
  \label{Fig33.sub.7}\thesubfigure
  \includegraphics[width=0.4\textwidth]{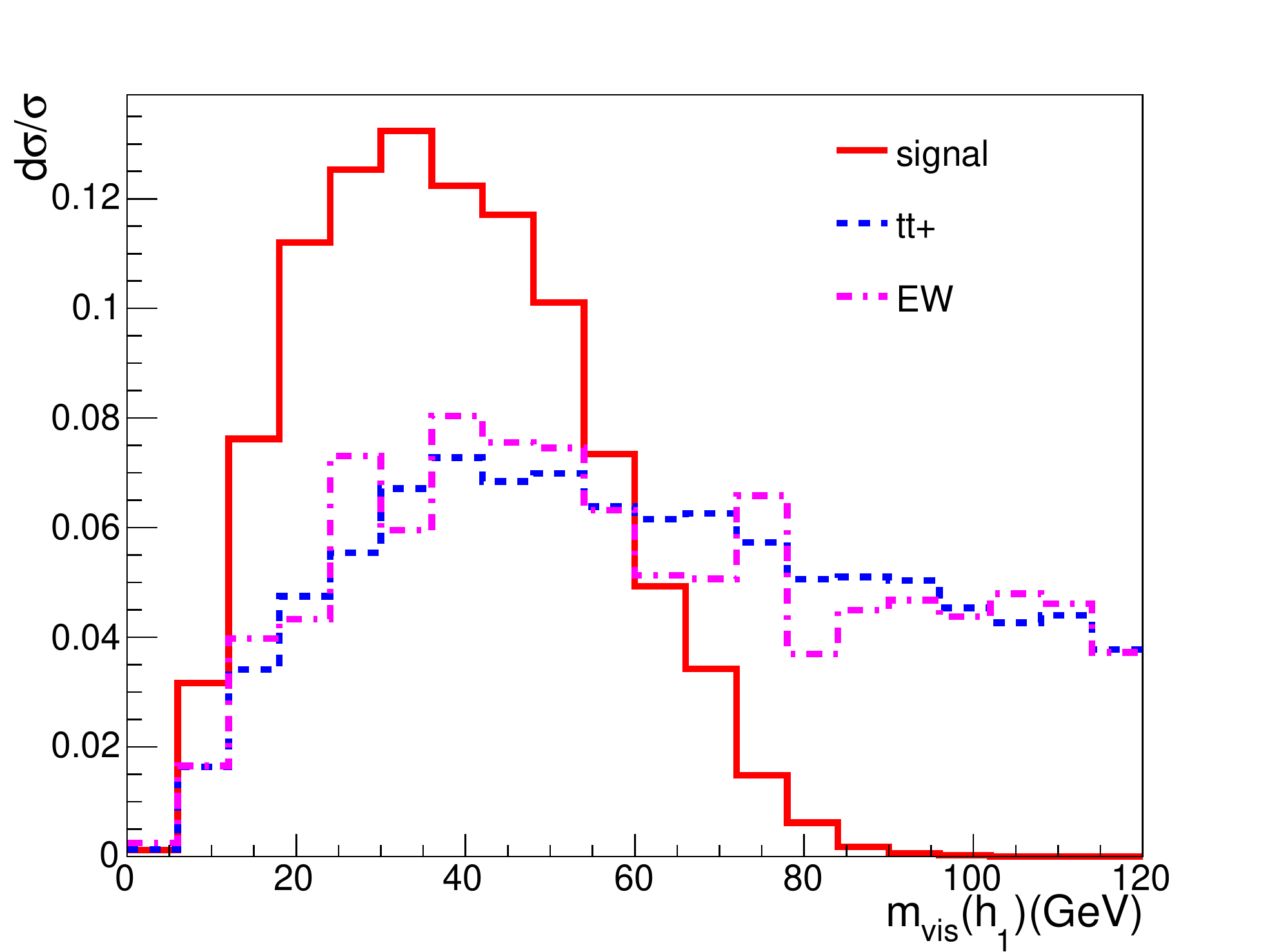}}
  \subfigure{
  \label{Fig33.sub.8}\thesubfigure
  \includegraphics[width=0.4\textwidth]{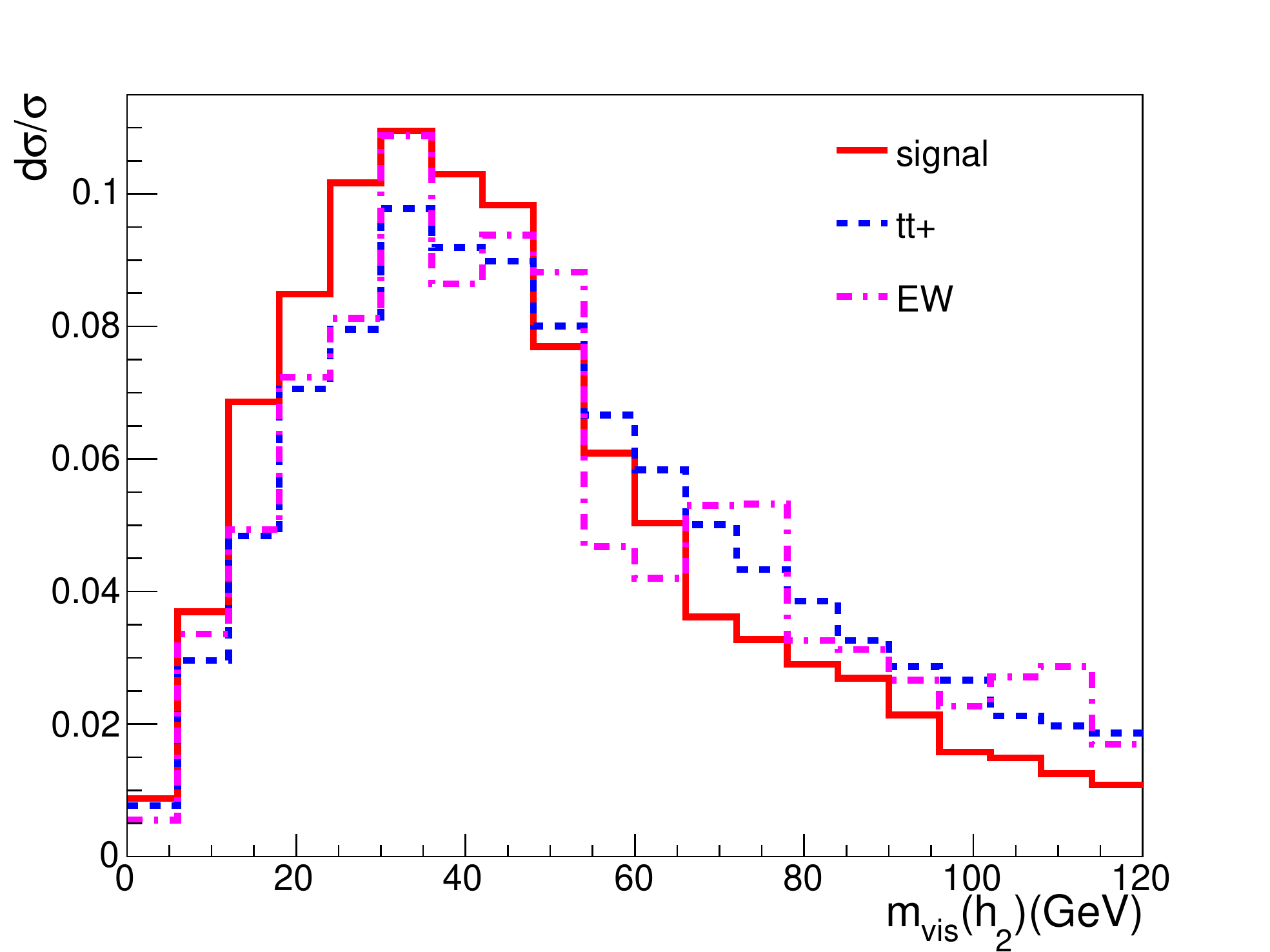}}
  \caption{Some useful kinematic observables for the M4 case which can separate signal and background events are shown. }\label{fig33}
\end{figure}

\begin{figure}[htbp]
  \centering
  \subfigure{
  \label{Fig34.sub.1}\thesubfigure
  \includegraphics[width=0.4\textwidth]{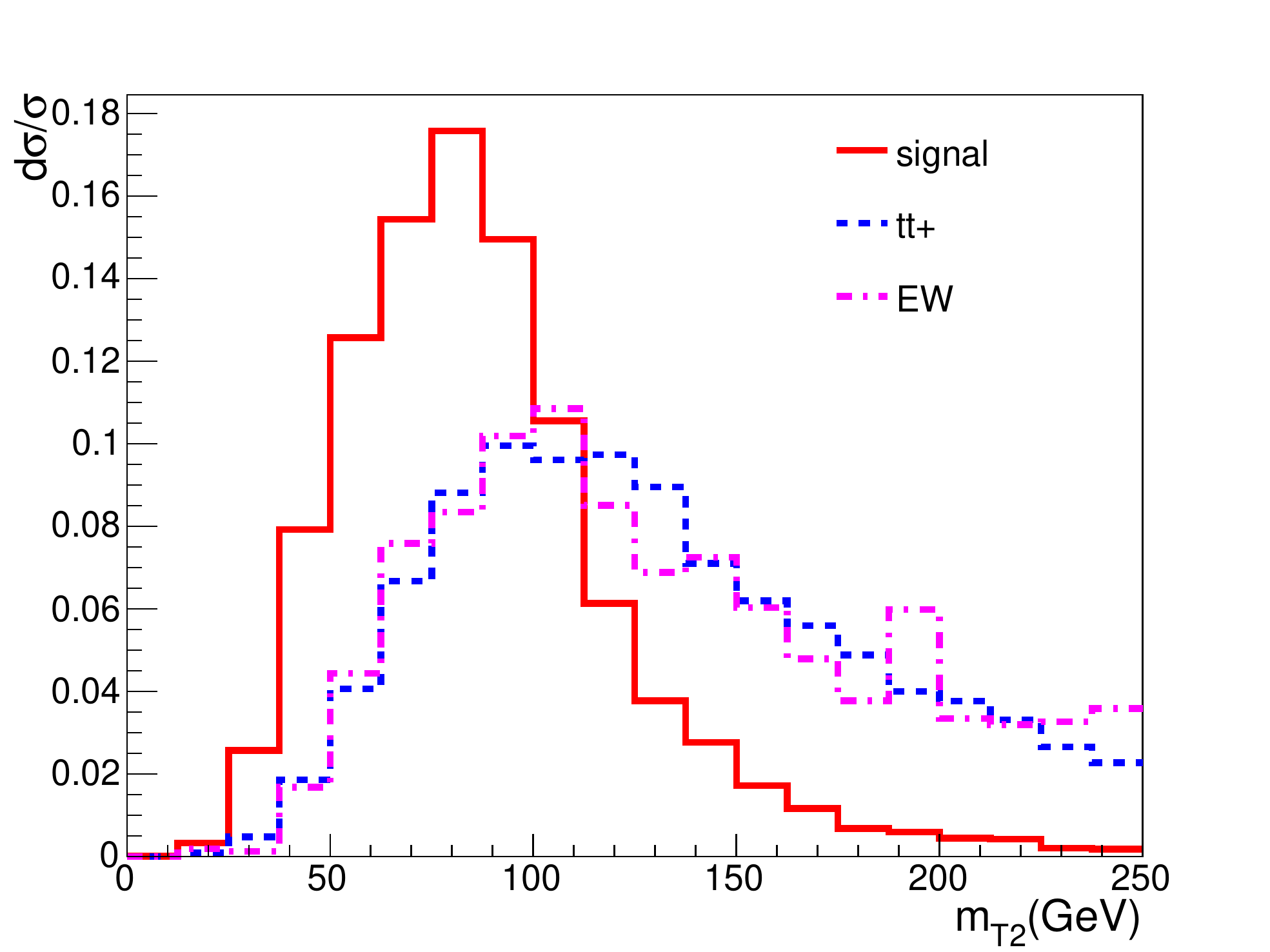}} 
  \subfigure{
  \label{Fig34.sub.2}\thesubfigure
  \includegraphics[width=0.4\textwidth]{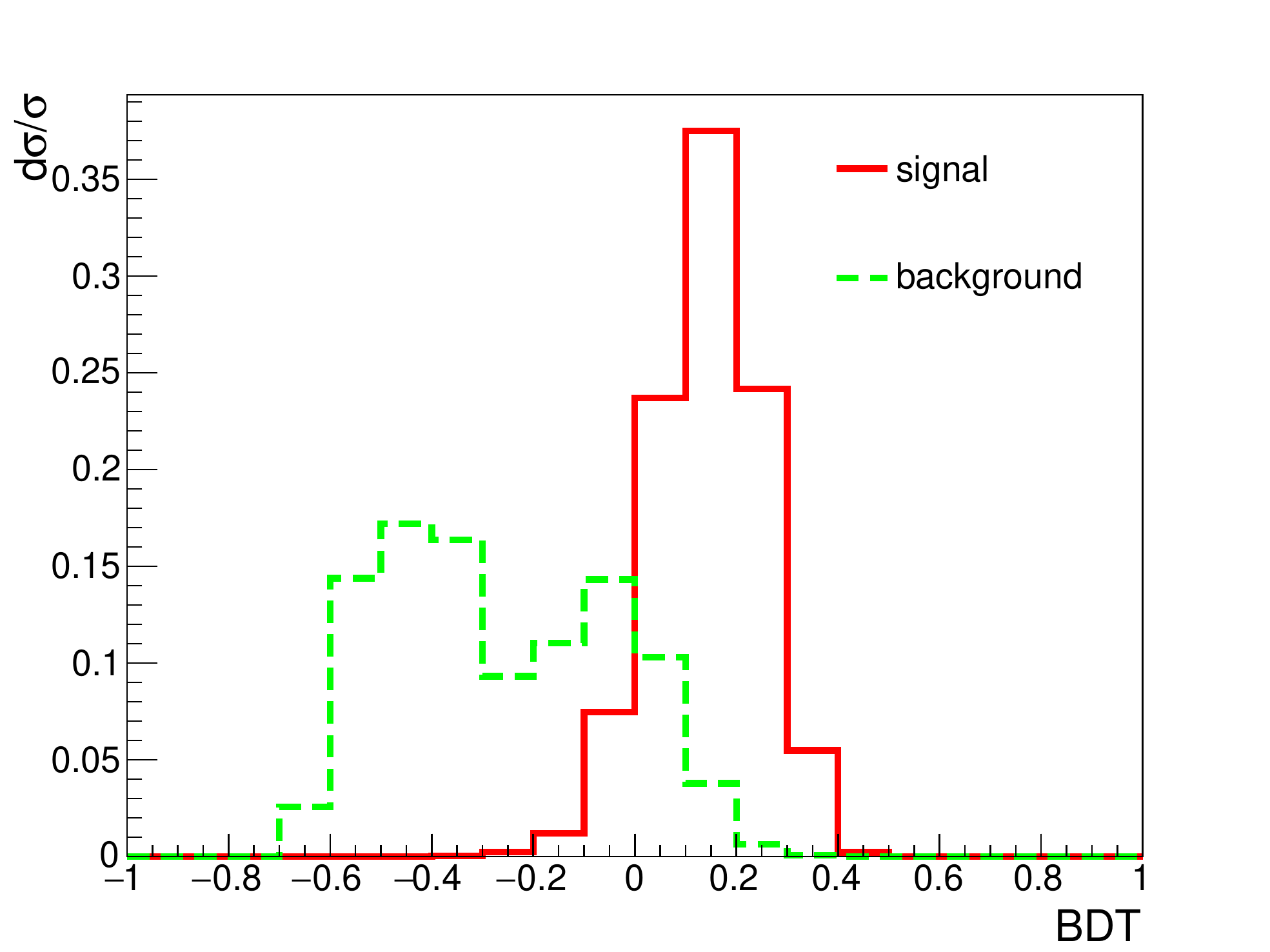}}
  \caption{The $m_{T2}$ observable and the BDT discriminant for the M4 case are shown. }\label{fig34}
\end{figure}

Different from the M3 case, the main task in the M4 mode is to suppress the background processes from $h t\bar{t}$ final states, while the singlet Z processes can be negligible and are omitted in the Table \ref{m4cuts}. The final state $h t \bar{t}$ is the dominant background. We noticed that our reconstruction method can successfully reconstruct the Higgs boson in $h t \bar{t}$ processes, as demonstrated in Fig. \ref{Fig33.sub.8}, where the shapes of background and signal look similar. Therefore, we impose a cut on $m(h_1^{vis})$, instead of both as in the M3 mode case.

A cut-based analysis and a BDT analysis are presented in Table \ref{m4cuts}. For the third cut, in this case, as we emphases that we only impose a cut on the visible mass $m(h_1^{vis}) < 60 $ GeV. As demonstrated in Fig. \ref{Fig34.sub.2}, after taking into account these observables to reject top quarks final states, we can further suppress the main background and gain in the S/B and significance up to 1.6 and 6.8, respectively, as represented in the Table \ref{m4cuts}. 

After using the same methods to veto top pair associated background in the MVA method, a better result is yield, 
the S/B and significance can reach 1.9 and 9.2, respectively. If an integrated luminosity 20-30 fb$^{-1}$ is assumed, then a significance 30.0 is expected.

When comparing the results given in Table \ref{m3cuts} and Table \ref{m4cuts}, we notice that the S/B of the M4 case is much better than that of the M3 case, due to the lack of huge  background processes as in the M3 case. Furthermore, the analysis for the M4 case is relatively simpler than the M3 case due to the background from single Z processes can be efficiently suppressed.

\begin{center}
\begin{table}
  \begin{center}
\begin{tabular}{|c||c|c|c|c|c||c|}
\hline
         Processes &  Pre-Sel. Cuts & $n_j \leq 2$ &  $M_{T2}<\SI{110}{GeV}$ & $m({h_1}^{vis})<\SI{60}{GeV}$ & $\Delta m >\SI{50}{GeV}$  & BDT\\
       \hline
        hh  &     55.2   & 48.5  & 40.5 & 36.3 & 28.8 & 43.9\\
        \hline\hline
        $t\bar{t} h$  &    147.1    &  80.5 & 44.4 & 27.1 & 13.7 & 14.3\\
        $t\bar{t}t\bar{t}$&   77.4 & 7.93 & 1.27 & $5.13\times10^{-1}$ & $3.59\times10^{-1}$ & 1.24\\
        $ t\bar{t}W^+ W^-$& 47.1 & 22.6 & 4.21 & 1.94 & 1.21 & 1.94\\
        \hline\hline
        $hW^+ W^-$  &  15.1 & 14.5 & 7.41 & 4.36 & 2.11 & 4.46\\
        $W^+ W^- W^+ W^-$  & 15.9 & 15.0 & 2.37 & $1.08$ & $6.22\times10^{-1}$ & 1.86\\
        \hline\hline
        S/B  &    0.18   &  0.35  &   0.68    &  1.04 & 1.6 & 1.9\\
        $S/\sqrt{B}$  &  3.17    & 4.09   & 5.24  & 6.13  & 6.79  & 9.2 \\
        \hline\hline
   \end{tabular}
  \end{center}
  \label{precuts}
  \caption{\label{m4cuts} A cut-based analysis and a BDT analysis for the M4 case are presented. Efficiencies of each cut in the cut-based analysis are demonstrated. We assume the integrated luminosity as 3 ab$^{-1}$.}
\end{table}
\end{center}

\section{Interplay between $gg \to hh$ and $pp \to t\bar{t} h$}
It is noticed that events from the $t \bar{t} h$ final state are the dominant background for the M4 case, which is also true for the $ p p \to h h \to 3\ell + 2 j + \missE_T$ mode explored in \cite{3l2j} at the HCs. 
Therefore, the measurement on the top Yukawa coupling can significantly affect the detection of Higgs pair production. Below we examine how the measurement of $t \bar{t} h$ couplings can affect the determination of $\lambda_3$. The correlation between the measurement of $t \bar{t} h$ and the measurement of Higgs pair production can be investigated by the cross section of $gg \to hh$ given in Eq. (\ref{xsgghh}). This issue has not been addressed in literatures.

\begin{figure}[htbp]
  \center
  \subfigure{
  \label{Fig4.sub.1}\thesubfigure
  \includegraphics[width=0.4\textwidth]{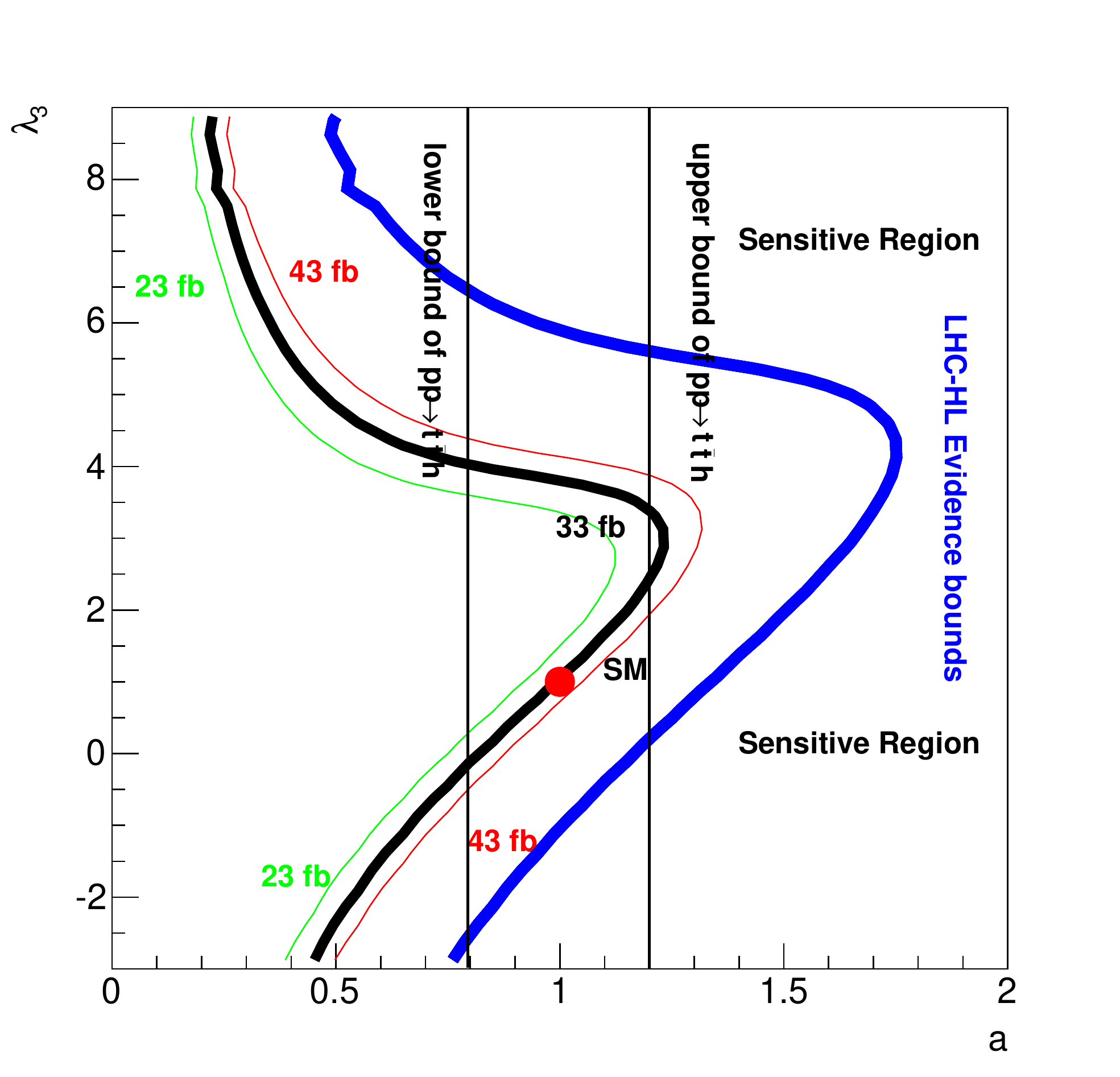}}
  \subfigure{
  \label{Fig4.sub.2}\thesubfigure
  \includegraphics[width=0.4\textwidth]{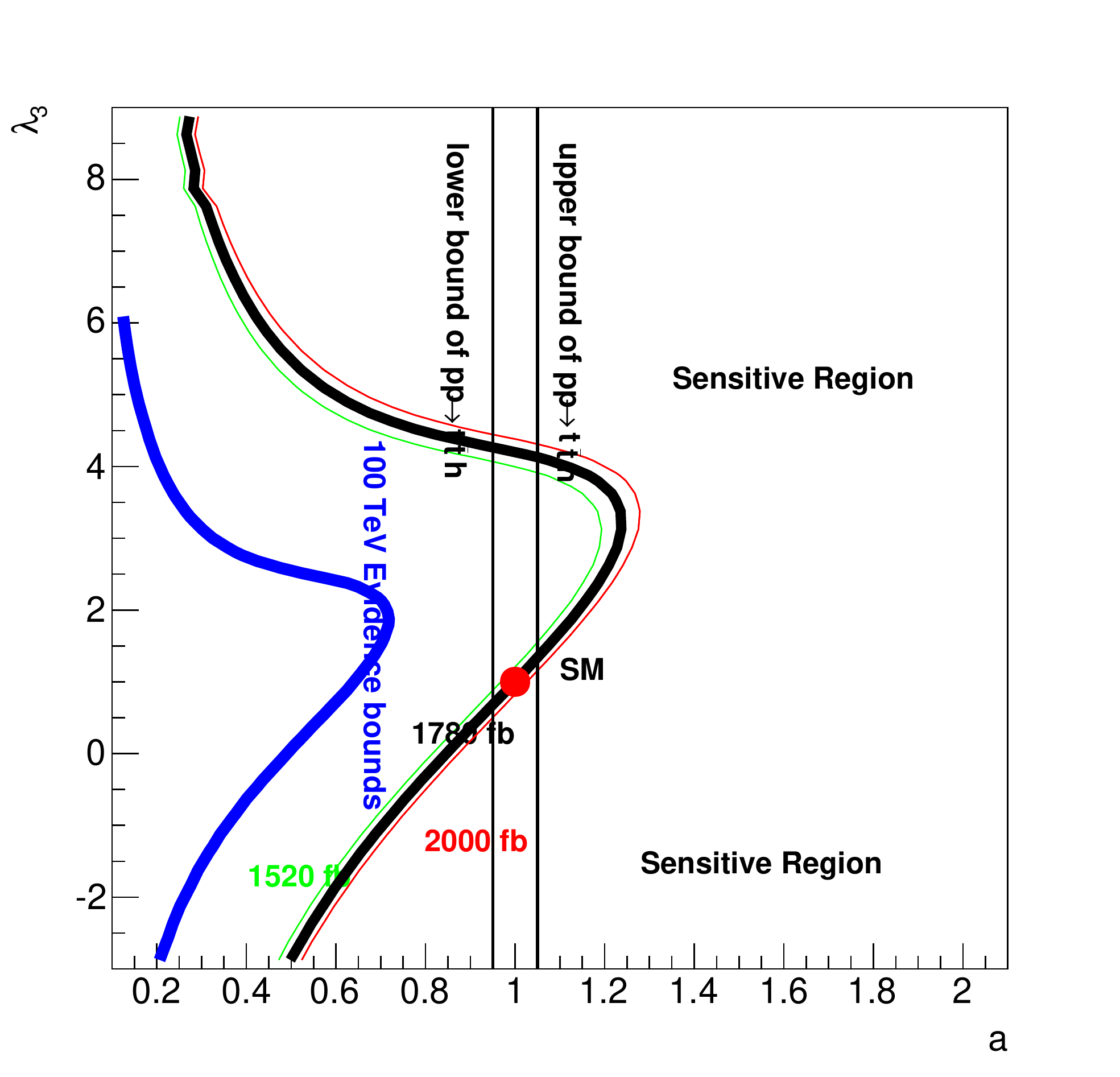}}
  \caption{\label{fig4}The sensitivity in the $a-\lambda_3$ plane between the LHC high-luminosity run and a 100 TeV collider. We fix $b=0.0$.}
\end{figure}

In Fig. \ref{fig4}, we demonstrate the correlation between the determination of $t\bar{t}h$ at the LHC and a future 100 TeV collider, where the bounds are estimated from the measurement of $3 \ell 2 j + \missE$. These bounds have not optimised for each value of $a$ and $\lambda_3$, we simply use the information of cross sections in this projection. As demonstrated in \cite{3l2j}, when the discrimination of signal and background is optimised, we can expect better bounds.

For the LHC with a luminosity 3 fb$^{-1}$, we assume that $t\bar{t} h$ couplings can be determined up to $20\%$ ($10\%$ is estimated by using the boosted techniques for $t \bar{t} h$ and $h\to b \bar{b}$ \cite{Plehn:2009rk,Klute:2013cx}).  We deliberately take a larger value for this since there only statistics are taken into account, which is denoted by two solid lines in Fig. \ref{Fig4.sub.1} as upper and lower bounds from $t \bar{t} h$ measurements. 

For a 100 TeV collider with a luminosity 3 fb$^{-1}$, we assume that a $5\%$ precision can be achieved, which is denoted by two solid lines in Fig. \ref{Fig4.sub.2} as upper and lower bounds from $t \bar{t} h$ measurements.  According to the study of \cite{Plehn:2015cta}, where by using the production ratio $t\bar{t}h$/$t\bar{t}Z$ measurement and boosted Higgs and top quark, it is argued that this coupling can be measured up to a precision $1\%$ or so when only statistics is considered. In reality, more background processes and detector effects must be considered for each $t\bar{t}$ decay final states, so we assume a precision up to $5\%$ as a relatively conservative and loose estimation.

Comparing Fig. \ref{Fig4.sub.1} and Fig. \ref{Fig4.sub.2}, it is noticed that a 100 TeV collider can greatly shrink the uncertainty in determining the $\lambda_3$ and $a$ parameters. Due to $a^4$ dependence of the cross section $\sigma(p p \to h h)$, a $5\%$ uncertainty of $\delta a$ can induce an uncertainty of $\lambda_3$ up to $20 \%$ or so. If the coupling $a$ can be determined to a precision $1\%$, that will be undoubtedly crucial to pinpoint the value of $\lambda_3$ down to $5\%$. It is worthy of mentioning that the two-fold ambiguity in $a$ with the same cross section can be removed by using the method to check the differential distribution of leptons in the final state, as demonstrated in \cite{3l2j}, which is robust against the contamination of underlying events and pileup effects.

\begin{figure}[htbp]
  \centering
  \subfigure{
  \label{Fig5.sub.1}\thesubfigure
  \includegraphics[width=0.4\textwidth]{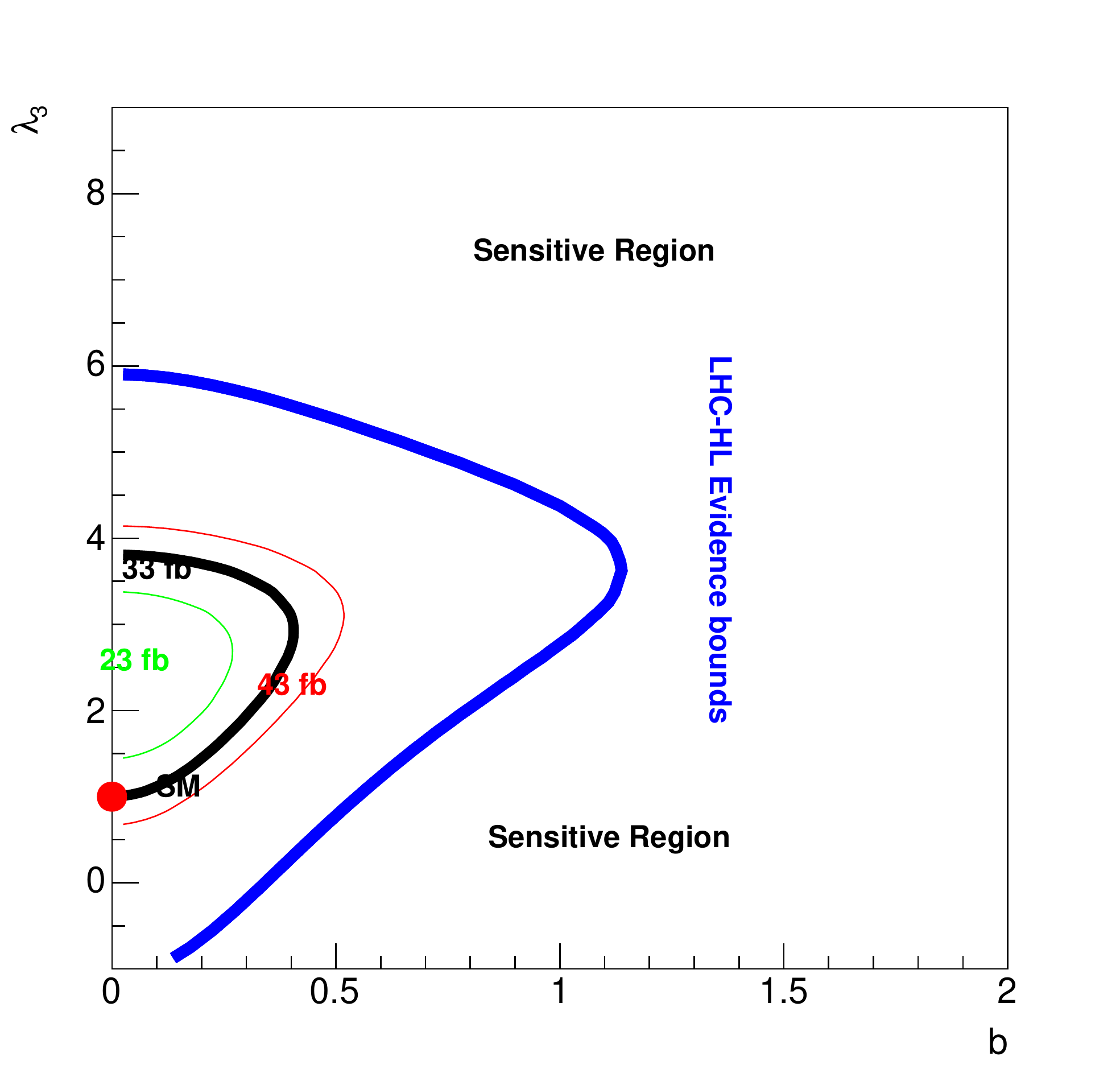}}
  \subfigure{
  \label{Fig5.sub.2}\thesubfigure
  \includegraphics[width=0.4\textwidth]{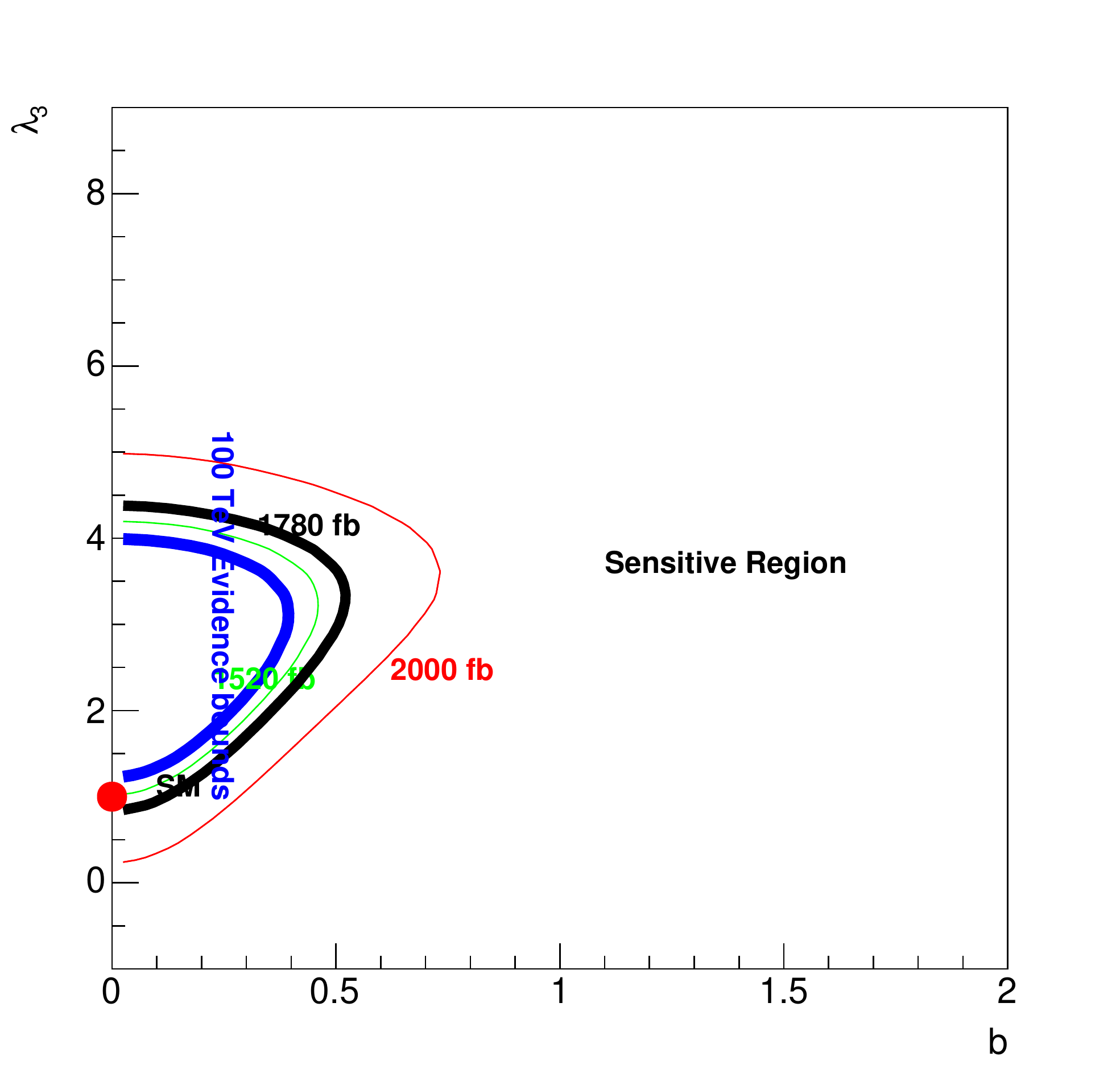}}
  \caption{\label{fig5}The sensitivity in the $b-\lambda_3$ plane between the LHC high-luminosity run and a 100 TeV collider. To project the feasibility, we fix $a=1.0$.}
\end{figure}

In the effective Lagrangian given in Eq. (\ref{efl}), the parameter $b$ is related with the strength of CP violation. We explore the determination of $b$ parameter from Higgs pair production, which is given in Fig. \ref{Fig5.sub.1}.  From Fig. \ref{Fig5.sub.2}, it is easy to read out that the potential of a 100 TeV collider to probe the parameter space expanded by $b$ and $\lambda_3$ is obviously better than that of the LHC. 

Higgs pair production at a 100 TeV collider can bound $b$ down to $0.4$. In contrast,the LHC runs can only constrain this parameter to $1.2$ or so at most. When we quote these number, we have not taken into account the uncertainty in $a$. Obviously, the direct measurement from $t\bar{t}h$ could impose a better constraint to the value of $b$, either at the LHC or at a 100 TeV collider. 

\begin{figure}[htbp]
  \centering
  \subfigure{
  \label{Fig6.sub.1}\thesubfigure
  \includegraphics[width=0.4\textwidth]{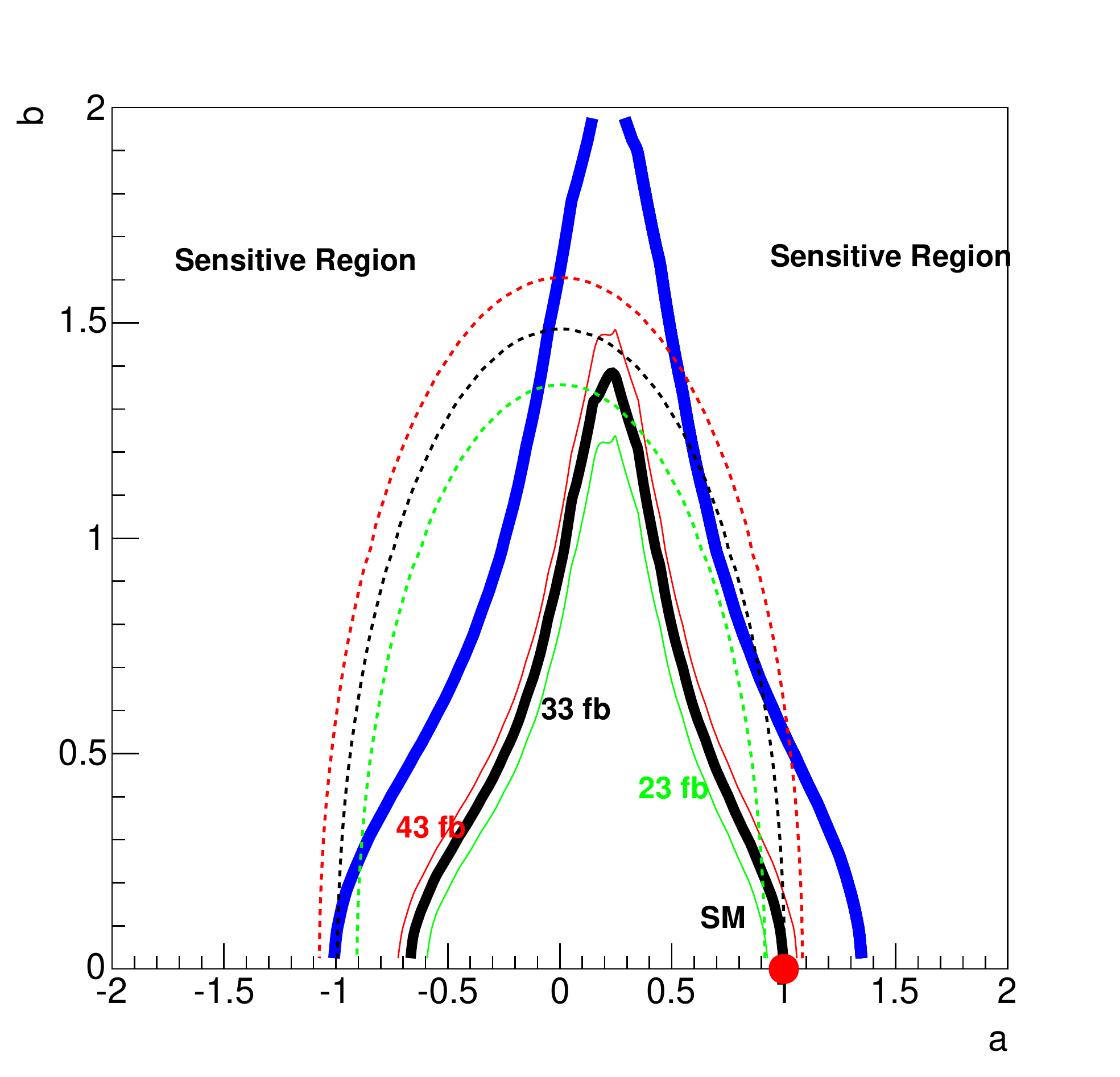}}
  \subfigure{
  \label{Fig6.sub.2}\thesubfigure
  \includegraphics[width=0.4\textwidth]{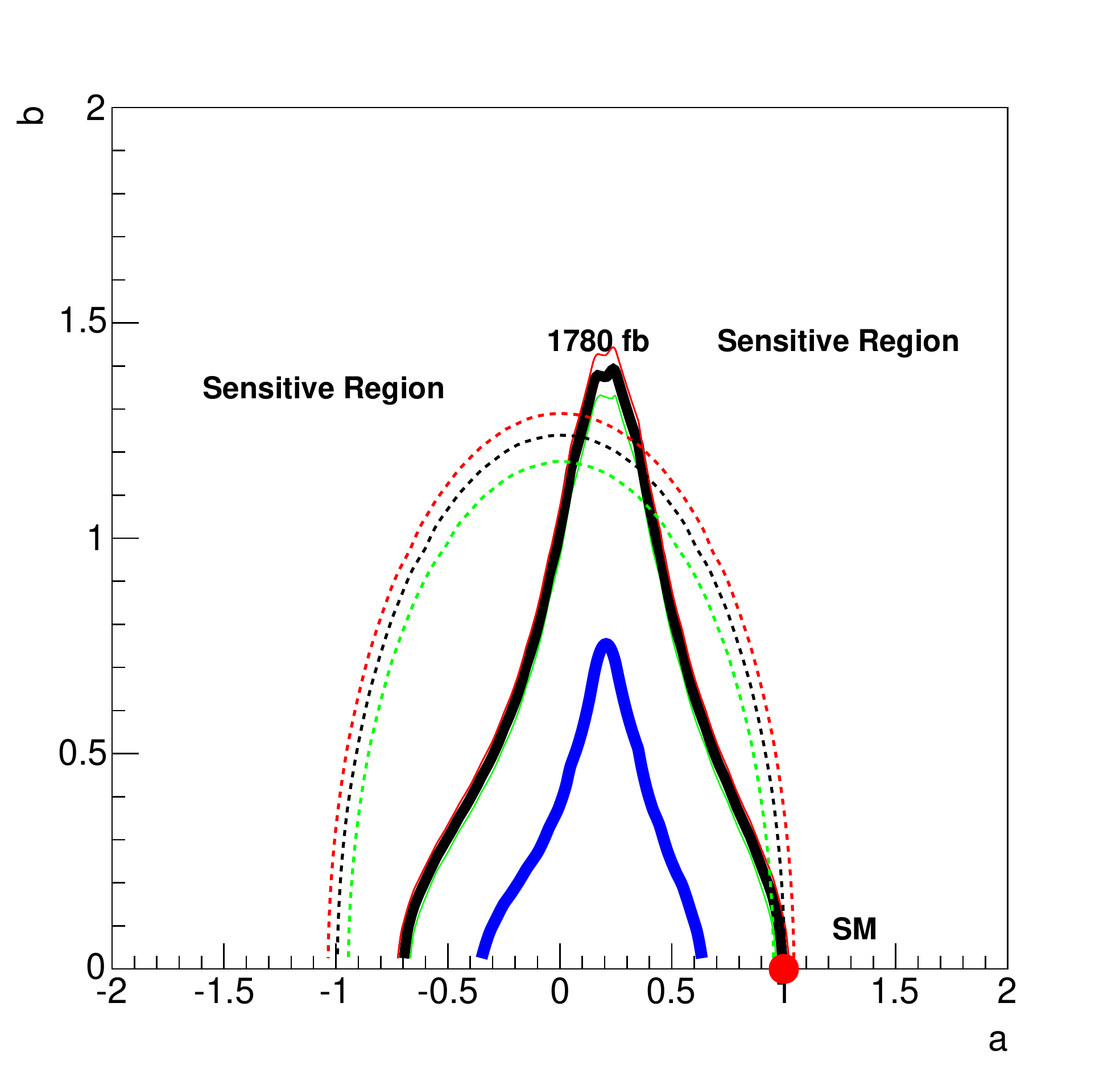}}
  \caption{\label{fig6} The sensitivity in the $a-b$ plane between the LHC high-luminosity run and a 100 TeV collider, where the constraints from $t\bar{t}h$ are depicted as dotted lines. For the LHC 14 TeV, the bands of  $t\bar{t}h$ are determined by assuming an error bar $20\%$ on the cross section. For a100 TeV collider, the bands are determined by assuming an error bar $10\%$ on the cross section for the purpose of visional effects. To demonstrate the bounds, we fix $\lambda_3=1.0$.}
\end{figure}

The correlation of $a$ and $b$ in Higgs pair is provided in Fig. \ref{Fig6.sub.1}. We plot the dependence of cross section of $t\bar{t}h$ upon $a$ and $b$ in the same plots. From Fig. \ref{Fig6.sub.2}, it is noticed that the LHC can be sensitive to the region where $a<0$ and b is sizeable, which corresponds a large production rate of $g g \to h h$ due to a constructive interference.

To consider the bounds from the $t \bar{t} h$ measurement in Fig. \ref{Fig6.sub.1} and Fig. \ref{Fig6.sub.2}, we have parametrised the cross section of $t \bar{t} h$ at the LHC 14 TeV and at a 100 TeV collider as
\bea
\sigma(p p \to t {\bar t} h) = t_1 a^2 + t_2 b^2\,,
\label{pp2tth}
\eea
where the values of $t_1$ and $t_2$ for 14 TeV and 100 TeV are computed by using fit and are provided in Table (\ref{cetth}),
\begin{table}[th]
\begin{center}
\begin{tabular}{|c|c|c|}
\hline
&$ t_1$ (pb) & $t_2$ (pb) \\ \hline
14 TeV &$0.58$  &  $0.26$  \\ \hline
100 TeV &$33.2 $  &  $21.6$ \\ \hline \hline
R & $57.71$ & $82.95$ \\\hline
\end{tabular}
\end{center}
\caption{The fit coefficients of $t\bar{t} h$ cross section for the LHC 14 TeV and a 100 TeV collider are tabulated. $R$ is defined as $K^{100}$/$K^{14}$, where K denotes $t_1$ and $t_2$. \label{cetth}}%
\end{table}
Here the NLO correction has been taken into account.

For a 100 TeV collider, it is noticed that there is a 3-fold ambiguity when we combine the measurements of $t\bar{t}h$ and $h h$. To remove this 3-fold ambiguity, the differential distribution of $t \bar{t} h$ final state should be carefully analysed, as demonstrated in \cite{He:2014xla} where quite a few differential observables are proposed.

It is interesting to observe that both single Higgs production and Higgs pair production can indirectly help to determine the interaction between top quark and Higgs boson at a 100 TeV collider, and it is pointed out that the Higgs pair production can also help to disentangle and resolve the nature of ultraviolet contributions to Higgs couplings to two gluons \cite{Dawson:2015oha}. 

\begin{figure}[htbp]
  \centering
  \subfigure{
  \includegraphics[width=0.4\textwidth]{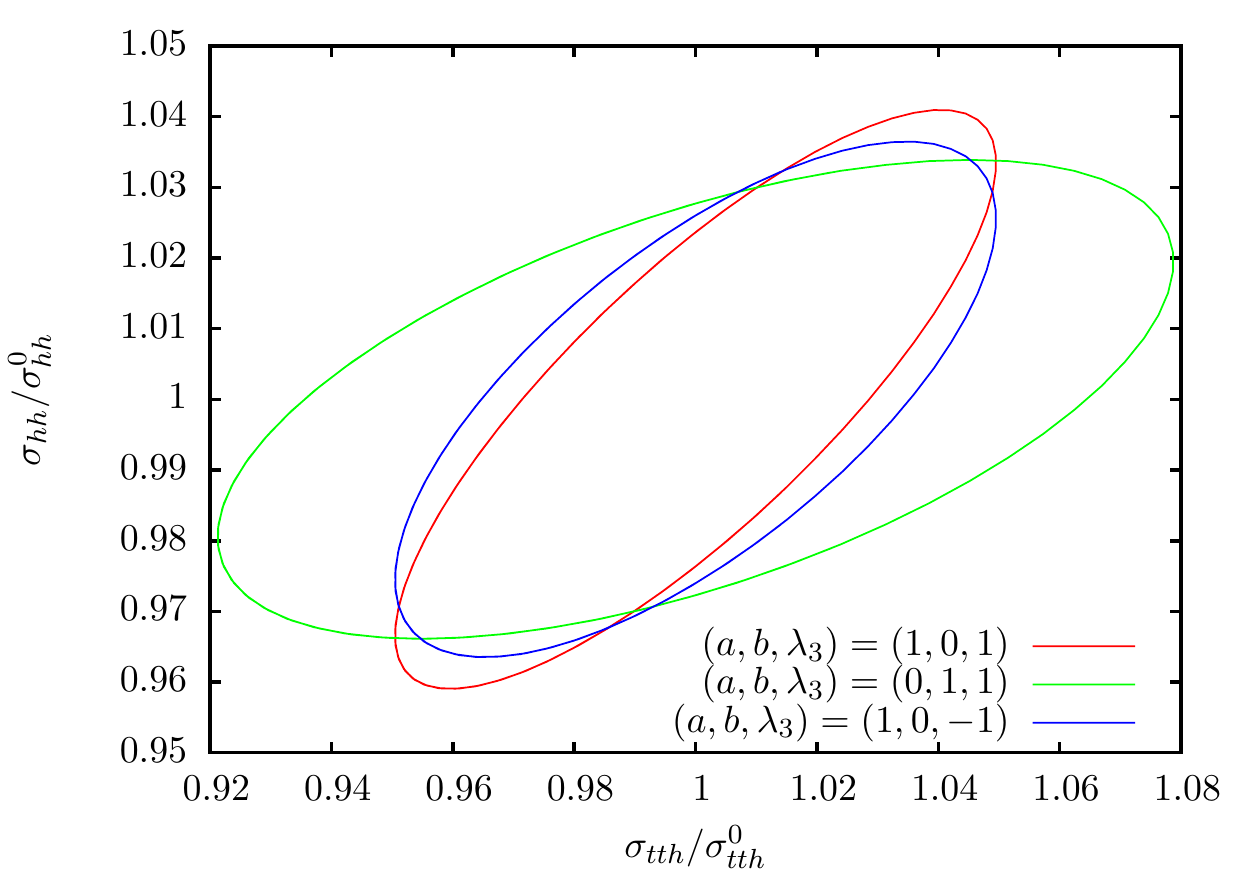}}
  \subfigure{
  \includegraphics[width=0.4\textwidth]{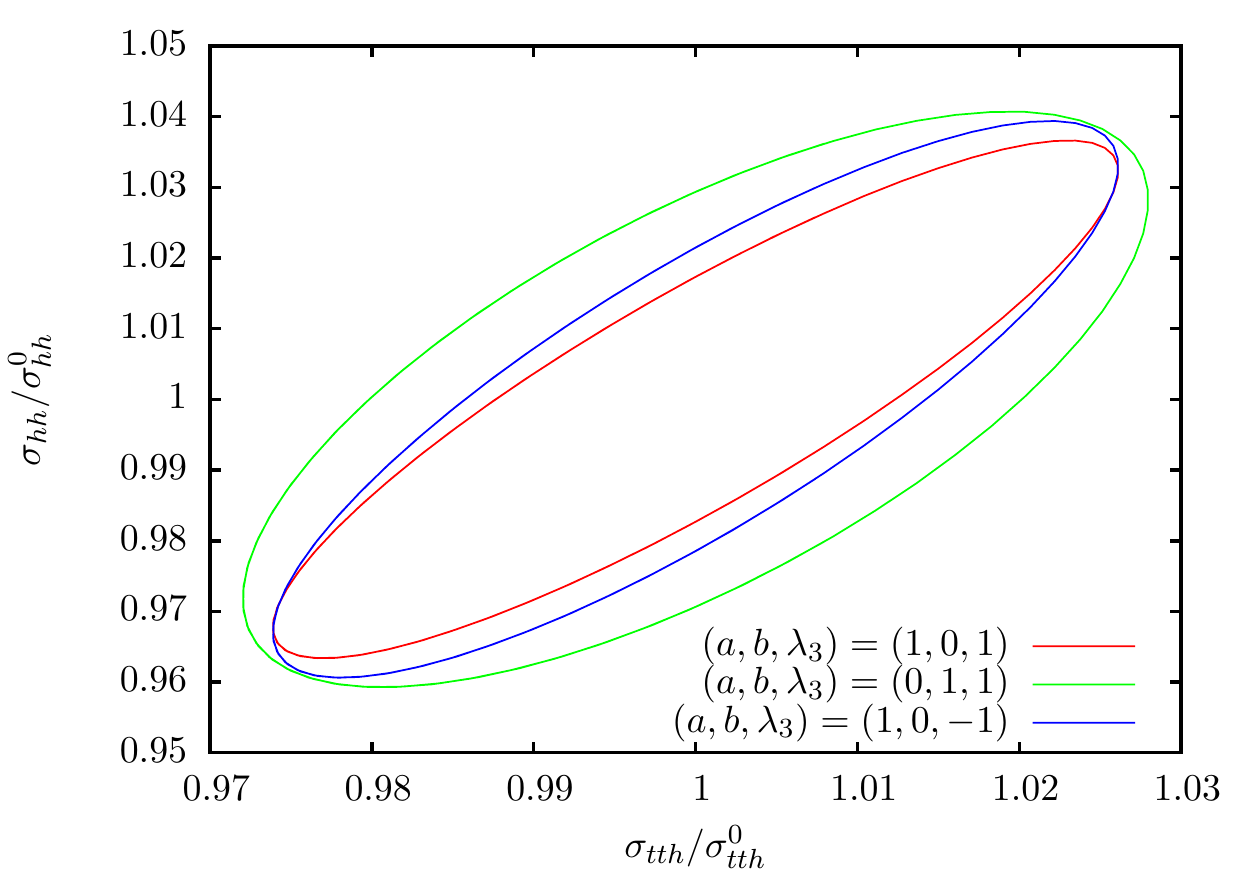}}
  \caption{\label{xscor}The PDF correlations between $\sigma(pp\to t\bar{t}h)$ and $\sigma(pp\to hh)$ at 14TeV LHC  and 100 TeV hadron collider. PDF CT10 is adopted, and all cross sections are normalized by the central values of each case.}
\end{figure}
It is also interesting to study the correlations of the cross section $\sigma(pp\to t {\bar t} h)$
and $\sigma(pp\to hh)$ when the uncertainty from parton distribution function will be considered. We use the dataset of CT10 \cite{Gao:2013xoa} to examine the uncertainty of PDF. A recent analysis on CT14 NNLO PDF can be found in \cite{Dulat:2015mca}. 

We consider the correlations between the processes $gg \to h h $ and $ p p \to t \bar{t} h$ in three cases. The first case is $(a,b,\lambda_3)=(1,0,1)$, the second case is $(a,b,\lambda_3)=(1,0,-1)$, and the third case $(a,b,\lambda_3)=(0,1,1)$. For  these three cases, the central values of cross sections of $gg \to h h $ and $ p p \to t \bar{t} h$ are so different that the uncertainty from PDF can not lead to an overlap among them. In order to examine the correlation caused by the uncertainty of PDF, we normalise the cross sections with the central values in each case and plot the $90\%$ confidence level contours in Fig. \ref{xscor}. 

In the first case, it is noticed that the uncertainty of PDF at the LHC 14 TeV can at most lead to an uncertainty $5\%$ in the cross  section of $pp \to t\bar{t}H$ and $4\%$ or so in that of $gg \to h h$.  The uncertainty of $gg \to h h$ is almost the same at a 100 TeV, while that of $p p \to t \bar{t} h$ is shrunk to $3\%$. 

At the LHC 14 TeV, when comparing the second case with the first case, we notice that the uncertainty of $pp
\to t\bar{t}H$ is enhanced due to the cross section becomes smaller. The correlation angle of the second case
is different from that of the first case due to form factors proportional to $b^4$ and $b^2$ are different from
those proportional to $a^4$, $a^3$, and $a^2$.  When the collision energy is 100 TeV, the correlation starts to
become stronger.

In the third case, the uncertainty of $pp \to t\bar{t}H$ is smaller due to normalisation when compared with the first case. Meanwhile, the correlation becomes weaker due to the interference and form factors changing, the area of the third case is fatter than that of the first case, in both the 14 TeV and 100 TeV collisions.

\section{Complementarity between hadron and electron-positron colliders}
Below we briefly comments on the sensitivity to probe the Higgs trilinear coupling $\lambda_3$ from other production processes at hadronic colliders simply based on the cross sections. A recent study on the interferences between the contribution of trilinear coupling and those independent of trilinear coupling can be found in \cite{Dicus:2015yva}, where it was found that the interference is almost insensitive to the change of collision energy of hadronic colliders. 

We first examine the vector boson associated production processes. To examine how the couplings between Higgs boson and the weak bosons can affect the determination of $\lambda_3$, we parameterise the couplings of the interactions $VVh$ and $VVhh$ as $d_1$ and $d_2$, respectively. 
\bea
{\cal  L}_{2} = \frac{d_1}{2} \left ( \, g_w^2\, v \, h \,W \cdot W + \frac{g_z^2}{2}  \,  v \, h \,Z \cdot Z \right ) + \frac{d_2}{4} \, \left ( g^2_w \,h \,h \,W \cdot W + \frac{g^2_z}{2} \,h \,h \,Z \cdot Z \right )\,.
 \label{hvv}
\eea
In the standard model, both $d_1$ and $d_2$ are equal to one. According to the current analysis on the single Higgs boson production and decay final states, it is known that the $d_1$ is close to 1. But $d_2$ has not yet measured.

With this parameterisation given in Eq. (\ref{hvv}) and Eq. $(\ref{efl})$, the cross section of $W/Z hh$ at  hadron colliders can be parametrised as
\bea
\sigma(p p  \to W/Z \,\, h h) & =  &
V_1 \, d_1^4 + V_2  \,d_1^2 d_2 + V_3  \, d_2^2  \nonumber \\
& + &  V_4  \, d_1^3 \lambda_3  + V_5  \, d_2 d_1 \lambda_3 +  V_6 \, d_1^2 \lambda_3^2  \,,
\label{vahh}
\eea
where form factors $V_i$ are evaluated numericaly and are given in Table \ref{wcoetev}.
\begin{table}[th]
\begin{center}
\begin{tabular}{|c|c|c|c|c|c|c|}
\hline
&$ V_1(fb)$ & $V_2(fb) $& $V_3(fb)$ & $V_4(fb) $& $V_5(fb) $ & $V_6(fb) $ \\ \hline
14 TeV &$0.093/0.077$  & $-0.094/-0.047$ &$0.153/0.122$& $-0.027/-0.011$ & $0.114/0.094$& $0.030/0.025$ \\ \hline
\hline
33 TeV& $0.447/0.367$ & $-0.518/-0.326$& $0.631/0.541$ & $-0.122/-0.065$ & $0.407/0.362$ &  $0.103/0.092$  \\
$R^{33}$ & $4.51/4.76$ & $5.52/6.96$ & $4.12/4.44$& $4.53/5.72$ & $3.57/3.87$ & $3.41/3.66$ \\\hline \hline
100 TeV &$2.227/2.012$  & $-2.847/-2.150$ & $2.943/2.719$&$-0.581/-0.386$& $1.677/1.607$ &$0.414/0.395$ \\ 
$R^{100}$ & $22.4/26.1$ & $30.3/45.9$ & $19.2/22.3$ & $21.6/34.2$  & $14.67/17.15$ & $13.70/15.77$  \\ \hline 
\end{tabular}
\end{center}
\caption{The fitting coefficients of the cross section of $p p \to W/Z hh$ for the LHC 14 TeV, 33 TeV and 100 TeV collider are tabulated. $R^{33}$($R^{100}$) is defined as $K^{33}$/$K^{14}$($K^{100}$/$K^{14}$), where K denotes $V_1-V_6$. \label{wcoetev}}%
\end{table}
We notice that signs of $V_2$ and $V_5$ are different from those of others and there exist interference cancellations between $V_1$($V_4$) and $V_2$($V_5$). This process should be sensitive to $\lambda_3$ due to the large fraction of contribution from terms proportional to $\lambda_3$, but the production rate is 3-order smaller than the process $gg \to hh$. This process has recently been studied in \cite{Cao:2015oxx}. It is sensitive to a specific window of $\lambda_3$ and is complimentary to other channels.

Here we examine the cross section of the process $t\bar{t}hh$, which can be parametrised similar to Eq. (\ref{xsgghh}), but with a different set of free parameters $T_1-T_7$. 
\bea
\sigma( p p \to t \bar{t} h h ) = T_1 \, a^4 + T_2 \, b^4 + T_3 \, a^2 \,  b^2 
+ (T_4 \,  a^3  + T_5 \, a \, b^2  ) \,  \lambda_3  + + ( T_6 \, a^2 + T_7 \,  b^2 ) \,  \lambda_3^2 \,.
\label{xspptthh}
\eea
The values of these free parameters are evaluated numerically and are tabulated in Table \ref{ce14tev}.  It is noticed that all these parameters are positive and the interference can occur when $a$ and $\lambda_3$ opposite signs. The interference terms proportional to $T_4$ and $T_5$ have largest enhancement factors than that of other terms. The feasibility for hadron colliders of this process has been studied in \cite{Englert:2014uqa,Liu:2014rva}.

\begin{table}[th]
\begin{center}
\begin{tabular}{|c|c|c|c|c|c|c|c|}
\hline
&$ T_1$  (fb)& $T_2$ (fb)& $T_3$ (fb)& $T_4$ (fb)& $T_5$ (fb)& $T_6$ (fb)& $T_7$ (fb)\\ \hline
14 TeV &$0.74$  & $0.12$ &$0.46$& $0.04$ & $0.06$& $0.13$ &$0.12$ \\ \hline
\hline
33 TeV& $5.94$ & $1.41$& $4.09$ & $0.40$ & $0.65$ &  $0.80$ & $0.96$  \\
$R^{33}$ & $8.02$ & $11.75$ & $8.89$& $10.00$ & $10.83$ & $6.15$ & $8.00$ \\\hline \hline
100 TeV &$56.29$  & $18.11$ & $44.09$&$4.64$& $7.43$& $6.26$ &$9.09$ \\ 
$R^{100}$ & $76.07$ & $92.58$ & $95.85$ & $116.00$ & $123.83$ & $48.15$ & $75.75$  \\ \hline 
\end{tabular}
\end{center}
\caption{The fitting coefficients of $p p \to t \bar{t} h h$ for the LHC 14 TeV and a 100 TeV collider are tabulated. $R^{33}$($R^{100}$) is defined as $K^{33}$/$K^{14}$($K^{100}$/$K^{14}$), where K denotes $T_1-T_7$. \label{ce14tev}}%
\end{table}

The cross section of vector boson fusion processes at hadron colliders can be 
parametrised as follows
\bea
\sigma(p p  \to WW/ZZ jj \to  h h j j ) & =  &
f_1 \, d_1^4 + f_2  \,d_1^2 d_2 + f_3  \, d_2^2  \nonumber \\
& + &  f_4  \, d_1^3 \lambda_3  + f_5  \, d_2 d_1 \lambda_3 +  f_6 \, d_1^2 \lambda_3^2  \,,
\label{vvfhh}
\eea
where $f_1-f_6$ are given in Table \ref{fcoetev}.
\begin{table}[th]
\begin{center}
\begin{tabular}{|c|c|c|c|c|c|c|}
\hline
&$ f_1$ (fb)& $f_2$ (fb)& $f_3$ (fb)& $f_4$ (fb)& $f_5$ (fb)& $f_6$ (fb)\\ \hline
14 TeV &$19.6/7.5$  & $-27.9/-10.5$ &$10.6/3.9$& $-5.6/-2.0$ & $3.5/1.3$& $0.6/0.23$ \\ \hline
\hline
33 TeV& $120.5/47.6$ & $-188.5/-73.0$& $77.9/29.7$ & $-24.5/-9.3$ & $16.2/6.0$ &  $2.64/0.96$  \\
$R^{33}$ & $6.15/6.29$ & $6.75/6.93$ & $7.33/7.61$& $4.41/4.52$ & $4.63/4.72$ & $4.09/4.20$ \\\hline \hline
100 TeV &$634.1/263.6$  & $-1062/-436.4$ & $466.2/189.9$&$-93.3/-36.3$& $62.8/24.1$ &$9.75/3.65$ \\ 
$R^{100}$ & $32.3/34.8$ & $38.0/41.5$ & $43.9/48.6$ & $16.8/17.7$  & $17.9/19.1$ & $15.1/15.9$  \\ \hline 
\end{tabular}
\end{center}
\caption{The fitting coefficients of VBF processes $p p \to  WW/ZZ  j j \to hh j j$ for the LHC 14 TeV, 33 TeV and 100 TeV collider are tabulated. $R^{33}$($R^{100}$) is defined as $K^{33}$/$K^{14}$($K^{100}$/$K^{14}$), where K denotes $f_1-f_6$. \label{fcoetev}}%
\end{table}

One interesting observation is that the interference among $f_1-f_3$ and $f_4-f_5$ is severe. Although each term is increasing with the increase of collision energy and is large, the total cross section is smaller than that of the process $pp \to t \bar{t} h h$ at a 100 TeV collider due to the strong cancellation induced by the interference. There is one tradeoff from this interference: the term proportional to $\lambda_3^2$ determines the total cross section. In other words, the irreducible background is much smaller than the signal and the signal is very dependent upon the trilinear coupling $\lambda_3$. In order to determine the trilinear coupling $\lambda_3$,  it is crucial to suppress the reducible backgrounds. It is also noticed that the interference between $WWjj$ and $ZZjj$ is suppressed by colour and in-flow quark flux and can be safely neglected. Our results are consistent with the results presented in \cite{Dicus:2015yva}.

As pointed in \cite{Dolan:2015zja}, this process can be used to probe the quartic coupling $g_{VVhh}$ vector boson and Higgs boson and the LHC can constrain this quartic coupling to a precision $50\%$. A 100 TeV collider can help to fix the quartic coupling $g_{VVhh}$.

It is also interesting to mention the process of single Higgs production associated with a single top. As demonstrated in \cite{Maltoni:2001hu,Barger:2009ky,Biswas:2012bd,Chang:2014rfa,Englert:2014pja,Yue:2014tya,Demartin:2015uha}, single top and Higgs processes can be useful to probe the relative sign between $a$ and $d_1$. Here we can parameterise the cross section as
\bea
\sigma(p p \to h t j ) = s_1 \, a^2 + s_2 \, b^2 + s_3 \, a \, d_1 + s_4 \,  d_1^2 \label{ppthj}
\eea
\begin{table}[th]
\begin{center}
\begin{tabular}{|c|c|c|c|c|}
\hline
&$ s_1$ (pb)& $s_2$ (pb)& $s_3$ (pb)& $s_4$ (pb)\\ \hline
14 TeV &$0.108$  & $0.034$ &$-0.203$& $0.131$ \\ \hline
\hline
33 TeV& $0.553$ & $0.195$& $-1.04$ & $0.702$  \\
$R^{33}$ & $5.12$ & $5.71$ & $5.14$& $5.37$ \\\hline \hline
100 TeV &$2.83$  & $1.09$ & $-5.31$ & $3.72$ \\ 
$R^{100}$ & $26.2$ & $31.9$ & $26.2$ & $28.5$ \\ \hline 
\end{tabular}
\end{center}
\caption{The fitting coefficients of the process of single Higgs plus single top $p p \to  t/\bar{t} h j$ for the LHC 14 TeV, 33 TeV and 100 TeV collider are tabulated. $R^{33}$($R^{100}$) is defined as $K^{33}$/$K^{14}$($K^{100}$/$K^{14}$), where K denotes $s_1-s_4$. \label{sthtev}}%
\end{table}
As demonstrated in \cite{Biswas:2012bd,Chang:2014rfa}, LHC can determine the sign of $\textrm{sign}(a\,  d_1)$. The total cross section is mainly determined by the term proportional to $a \, d_1$. As shown in Table \ref{sthtev}, in the SM, there exists a strong cancellation between the term $a \, d_1$ and the rest. When this term switches its sign, the the cross section will be enhanced by one order or so. Therefore, this process is sensitive the sign of $a \, d_1$. The single top and Higgs boson production could also be a signal of the flavour changed decay of top quark  $t \to h + u$ and $t \to h + c$ from the processes $p p \to t {\bar t}$ via the flavour changing neutral current processes, as shown in \cite{Wu:2014dba}.

It is also interesting to examine the process of Higgs pair production associated with a single top, the cross section in the SM has been explored in \cite{Demartin:2015uha}. Here we can parameterise the cross section as
\bea
\sigma(p p \to t  h h j)  & =  &
J_1(s) \, a^2 d_1^2 
+ J_2(s) \,a d_1^3 
+ J_3(s) \, a d_1 d_2 
+ J_4(s) \, d_1^4 
+ J_5(s) \, d_1^2 d_2 
\nonumber \\
&  & 
+ J_6(s) \, d_2^2 + J_7(s) \, a^2 d_1 \lambda_3 
+ J_8(s) \, a d_1^2 \lambda_3 
+ J_9(s) \, d_1^3 \lambda_3 
\nonumber \\
&  & 
+ J_{10}(s) \, a d_1 \lambda_3^2 
+ J_{11}(s) \, d_1^2 \lambda_3^2 
+ \cdots \,,
\label{pp2thhj}
\eea
where we only keep those sizeable terms in this formula and drop those tiny terms. The values of $J_i$, at the LHC 14 TeV, a 33 TeV collider, and a 100 TeV collider, are tabulated in Table \ref{thhjtev}.
\begin{table}[th]
\begin{center}
\begin{tabular}{|c|c|c|c|c|c|c|c|c|c|c|c|}
\hline
&$ J_1$ (fb)& $J_2$ (fb)& $J_3$ (fb)& $J_4$ (fb) & $J_5$ (fb) & $J_6$ (fb) & $J_7$(fb) & $J_8$(fb) & $J_9$(fb)
& $J_{10}$(fb) & $J_{11}$(fb)\\ \hline
14 TeV &2.03&-3.23&1.58&2.88&-3.09&0.890&-0.329&0.627&-0.588&-0.0483&0.0528 \\ \hline
\hline
33 TeV &17.7&-32.9&16.5&33.5&-38.5&12.1&-2.31&4.65&-4.58&-0.332&0.371\\
$R^{33}$ & 8.72 & 10.2 & 10.4 & 11.6 & 12.5 & 13.6 & 7.02& 7.42 & 7.79 & 6.87 & 7.03\\\hline \hline
100 TeV &149&-324&164&381&-471&162&-14.9&31.1&-31.3&-2.04&2.29\\
$R^{100}$ & 73.4 & 100.3 & 103.8 & 132.3 & 152.4 & 182.0& 45.3 & 49.6 & 53.2 & 42.2 & 43.4 \\ \hline 
\end{tabular}
\end{center}
\caption{The fitting coefficients of the process of Higgs pair plus single top $p p \to  t/\bar{t} h h j$ for
the LHC 14 TeV, 33 TeV and 100 TeV collider are tabulated. $R^{33}$($R^{100}$) is defined as
$K^{33}$/$K^{14}$($K^{100}$/$K^{14}$), where K denotes $J_1-J_{11}$. \label{thhjtev}}%
\end{table}
It is noticed that there exists a strong cancellation among terms, like ($J_1-J_3$, $J_4-J_6$, $J_8-J_9$). If there is a new physics which can break this strong cancellation, it is expected that this process can a sensitive probe.

\begin{figure}[htbp]
  \centering
  \subfigure{
  \label{Fig7.sub.1}\thesubfigure
  \includegraphics[width=0.40\textwidth]{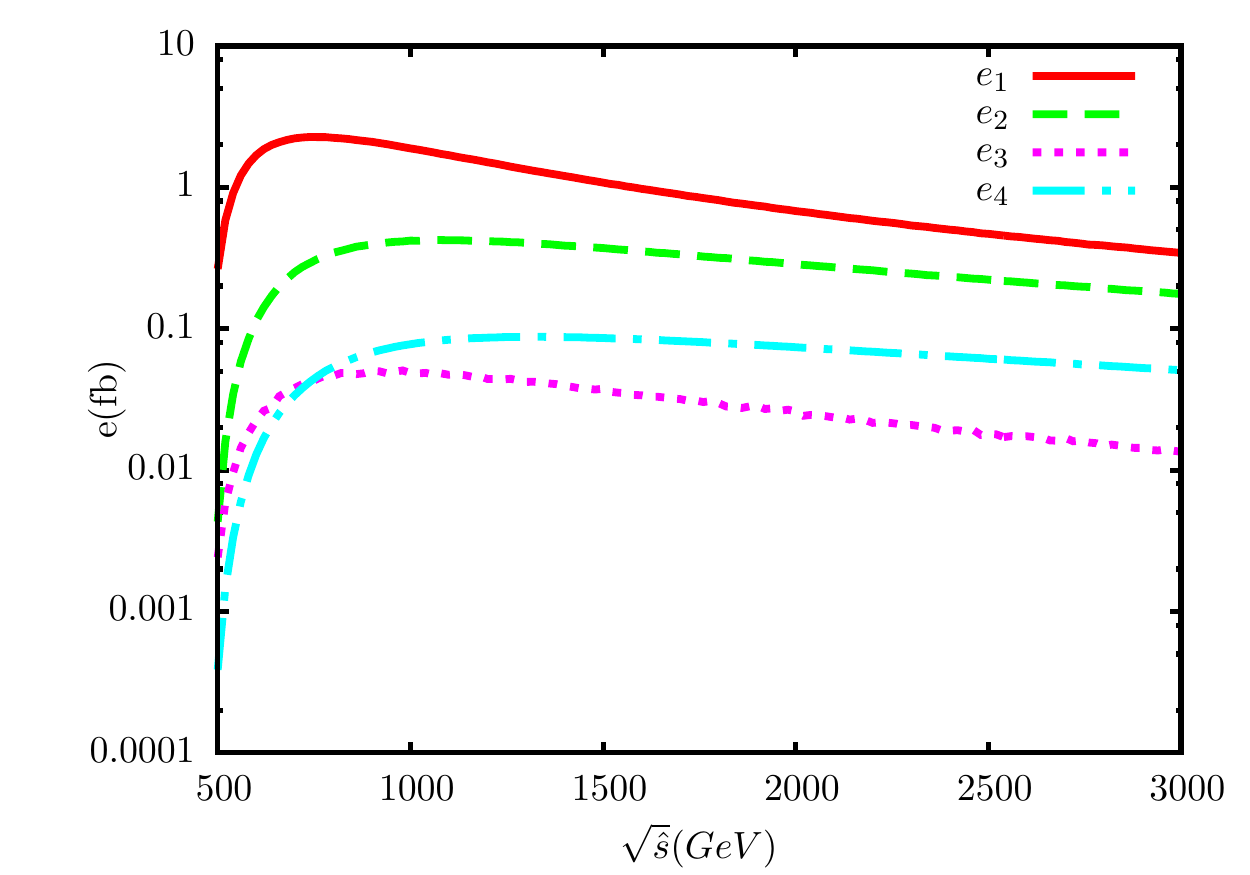}}
  \subfigure{
  \label{Fig7.sub.2}\thesubfigure
  \includegraphics[width=0.40\textwidth]{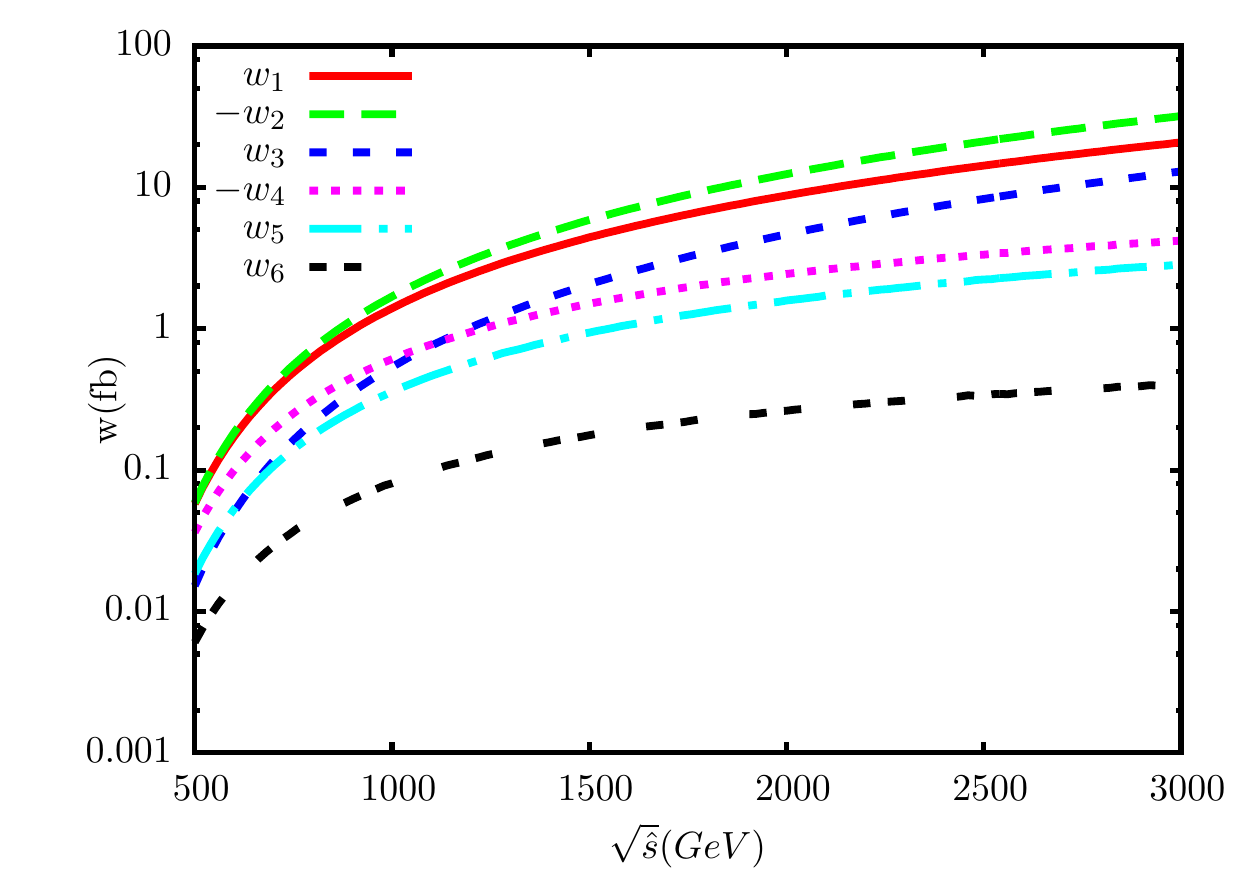}}
  \subfigure{
  \label{Fig7.sub.3}\thesubfigure
  \includegraphics[width=0.40\textwidth]{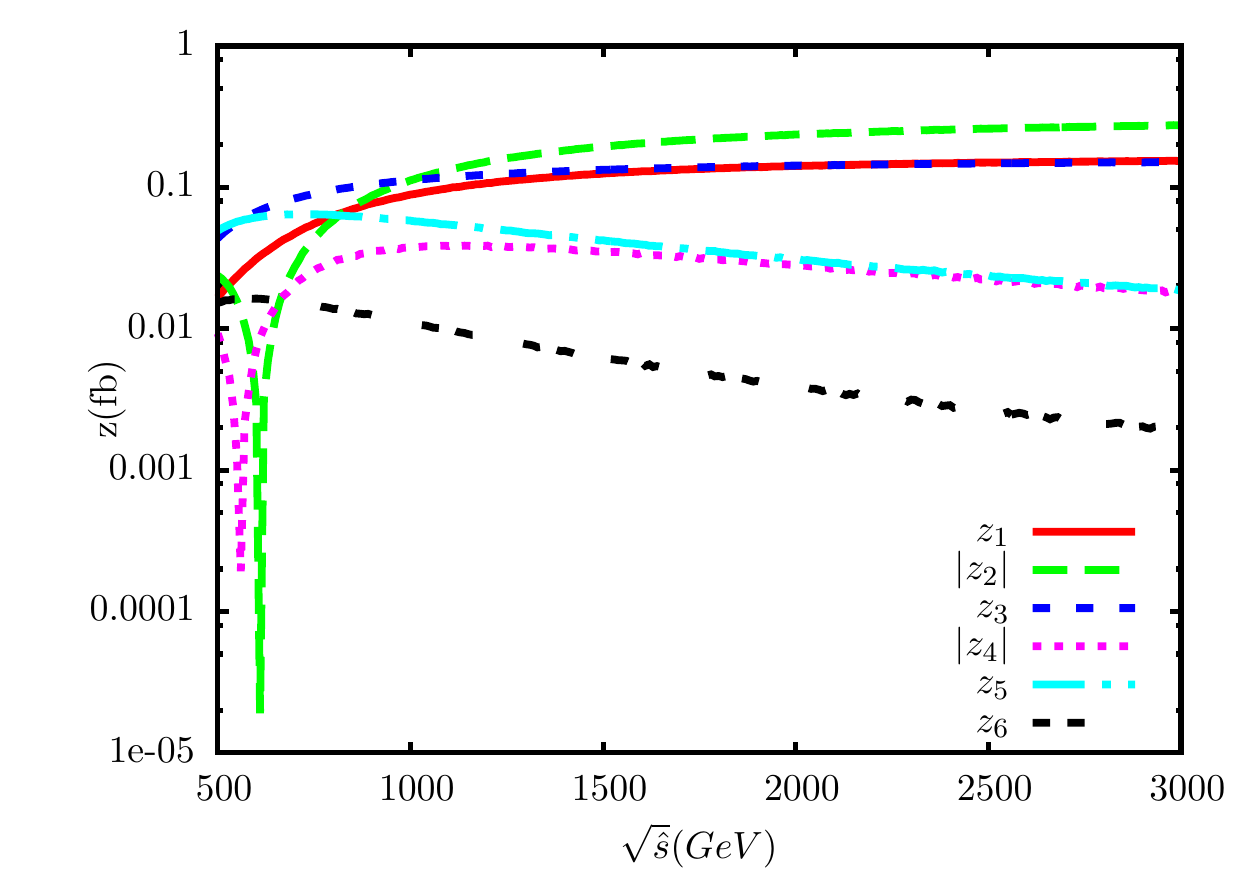}}
  \subfigure{
  \label{Fig7.sub.4}\thesubfigure
  \includegraphics[width=0.40\textwidth]{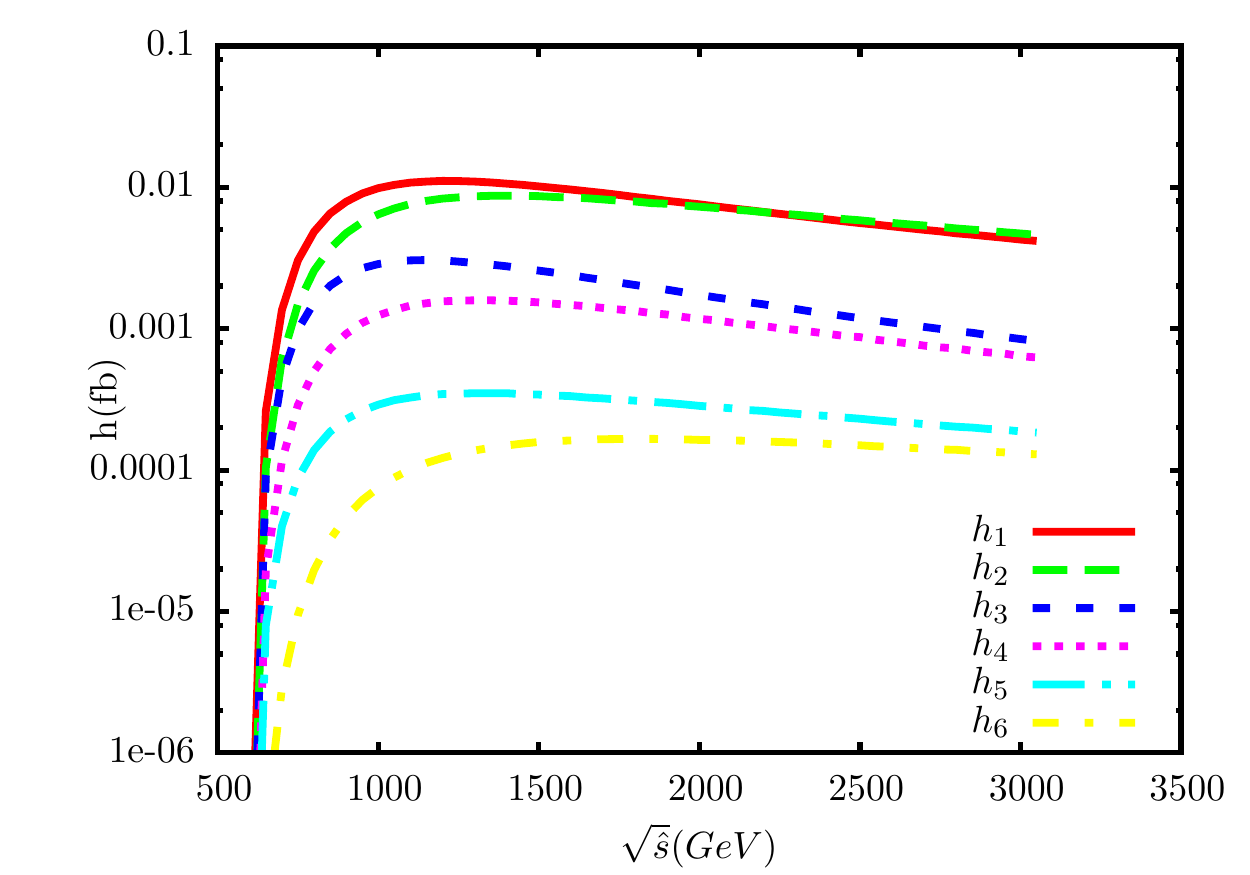}}
  \caption{\label{fig7} The dependence upon the collision energy of form factors given in Eqs. (\ref{eetth}), (\ref{eeww2hh}), (\ref{eezhh}), and (\ref{ee2tthh}) are shown. }
\end{figure}

Below we simply examine the complementarity of $e^+ e^-$ machines to probe trilinear coupling $\lambda_3$. It is noticed that the vertex $t \bar{t} h $ can be probed by measuring the production of $e^+ e^- \to t \bar{t} h $ and its cross section can be parametrised as the following formula 
\bea
\sigma(e^+ e^- \to t {\bar t} h) = e_1(s) a^2 + e_2(s) b^2 + e_3(s) a d_1 + e_4(s) d_1^2 \,,
\label{eetth}
\eea
where $e_i(s)\,,i=1, 2, 3, 4$ are form factors, and their dependence upon the collision energy $s$ can explore numerically. In Fig. \ref{fig7}, we provide the numerical results. The analytic form of these form factors at tree level can be found in \cite{Djouadi:1991tk,Grzadkowski:1999ye}. 

To extract the $t \bar{t} h$ couplings at the ILC, the reference \cite{Yonamine:2011jg} had performed a detailed MC study by considering the $\ell + 6 j$ and 8 j final states. The main background $t \bar{t} W/Z/(gg)$ and $t W b$ are considered. It is observed that b jet veto is crucial to suppress the $tWb$ background. The sensitivity to $a$ is around $10\%$ at the ILC with a integrated luminosity 500 fb$^{-1}$, which is similar to the sensitivity of the LHC high luminosity runs. 

It should be point out that at the hadron colliders, the cross section given in Eq. (\ref{pp2tth}) should have the similar structure as given in Eq. (\ref{eetth}). But due to the large gluon flux, the interference terms proportional to $a$ and the terms independent of $a$ and $b$ are relatively tiny and can be safely neglected.

In contrast, the cross section of the vector boson fusion processes $e^+ e^- \to \nu_e \bar{\nu_e} h h $ (mainly WW fusion processes) increases with the increase of collision energy $\sqrt{s}$ and can be parametrised as
\bea
\sigma(e^+ e^- \to \nu_e \bar{\nu_e} h h) & =  &
w_1(s) \, d_1^4 + w_2(s) \,d_1^2 d_2 + w_3(s) \, d_2^2  \nonumber \\
& + &  w_4(s) \, d_1^3 \lambda_3  + w_5(s) \, d_2 d_1 \lambda_3 +  w_6(s) \, d_1^2 \lambda_3^2  \,,
\label{eeww2hh}
\eea
where form factors $w_1(s)-w_6(s)$ are demonstrated in Fig. \ref{Fig7.sub.2}. It is noticed that $w_2(s)$ has a different sign from $w_1(s)$ and $w_3(s)$, and $w_4(s)$ has a different from $w_5(s)$. Similar to the VBF processes at hadron colliders, the large cancellations among $w_1(s)-w_3(s)$ and $w_4(s)-w_5(s)$ lead to a fact that the total cross section is mainly dependent upon the term proportional to $\lambda_3^2$, which makes the process $e^+ e^- \to \nu_e \bar{\nu_e} h h$ sensitive to the measurement of trilinear coupling $\lambda_3$. A Monte Carlo study on the sensitivity of $\lambda_3$ via  two $jjbbbb$ and $\nu\bar{\nu}bbbb$ final states can be found in \cite{Baur:2009uw}. 

Both the processes $e^+ e^- \to \nu_e \bar{\nu_e} h h $ and the processes $e^+ e^- \to Z h h $ have no dependence upon the coupling of $t \bar{t} h$. The cross section of $e^+ e^- \to Z h h $ can be expressed as
\bea
\sigma(e^+ e^- \to Z h h) & =  &
z_1(s) \, d_1^4 + z_2(s) \,d_1^2 d_2 + z_3(s) \, d_2^2  \nonumber \\
& + &  z_4(s) \, d_1^3 \lambda_3  + z_5(s) \, d_2 d_1 \lambda_3 +  z_6(s) \, d_1^2 \lambda_3^2  \,,
\label{eezhh}
\eea
The form factors are shown in Fig. \ref{Fig7.sub.3}, where it is noticed that $z_2$ and $z_4$ are positive at the very beginning but become negative (the sign is switched in the plot) when $\sqrt{s} > 600$ GeV. In the large $\sqrt{s}$ region, a strong cancellation among $z_1$, $z_2$, and $z_3$ occurs. A similar exact cancellation happens between $z_4$ and $z_5$ when $\sqrt{s}$ is large. Therefore the main contribution  at high energy region is determined by $z_1-z_3$ terms, which is much larger than the terms proportional to $\lambda_3$.

Due to the huge irreducible background from the process of sequential emission of two Higgs bosons and detector effects, even near 500 GeV the uncertainty of $\lambda_3$ is larger than $30 \%$ \cite{Tian:2010np,Tian:2013yda}. With higher collision energy (say $\sqrt{s} \geq 700$ GeV), the production rate proportional to $\lambda_3$ decreases rapidly and the uncertainty to extract $\lambda_3$ increases. Therefore, high energy collisions (say higher than 800 GeV) for the process $e^+ e^- \to Z h h $ could not offer us a better way to extract trilinear Higgs coupling $\lambda_3$.

The structure of the cross section of the processes $e^+ e^- \to t \bar{t} h h$ are more complicated, because all terms in the Lagrangian given by Eq. (\ref{efl}) and Eq. (\ref{hvv}) come into play a role. The leading dominant terms 
\bea
\sigma(e^+ e^- \to t \bar{t} h h) & =  &
h_1(s) \, a^4 + h_2(s) \,a^2 \,\, b^2 + h_3(s) \, a^3 \,\, \lambda_3  \nonumber \\
& + &  h_4(s) \, a \, b^2 \, \lambda_3  + h_5(s)  a^2 \lambda_3^2 +  h_6(s) \, b^2 \,\, \lambda_3^2 + \cdots \,,
\label{ee2tthh}
\eea
where the form factors $h_1-h_6$ are displayed in Fig. \ref{Fig7.sub.4}. Although there are other terms, like terms proportional to $b^2 \lambda_3^2$, $d_1 a^2 \lambda_3$, and so on, but numerically they are small when compared with these 6 form factors so we can simply neglect them. Both this process and $e^+ e^- \to b \bar{b}  h h$ probe trilinear Higgs coupling, as demonstrated in \cite{GutierrezRodriguez:2008nk,GutierrezRodriguez:2009uz}.

It is interesting to notice that the cancellations observed in processes $p p  \to W/Z \,\, h h$, $p p  \to WW/ZZ jj \to  h h j j $, $e^+ e^- \to Z h h $, and $e^+ e^- \to \nu_e \bar{\nu_e} h h$, are not accidental and are tightly related with the unitarity in a theory with spontaneous symmetry breaking \cite{Lee:1977yc,Lee:1977eg}.

In the SM and SUSY, the process $e^+ e^- \to h h$ via loop processes \cite{LopezVillarejo:2008xw,Heinemeyer:2015qbu}, the maximum production rate is around 0.01 fb or so when the collision energy is around 350-450 GeV. In the collision with polarised beans, the production rate can be enhanced by a factor 6-7. For the final state with $4b$,  the main background should be $e^+ e^- \to Z Z \to 4 b$, which is around 36.8 fb when collision energy is set to be 400 GeV. We expect a certain degree of feasibility if the invariant mass of two b jet can be reconstructed. This process can not directly help to probe trilinear Higgs coupling $\lambda_3$. Nevertheless, it might be useful probe to the coupling of $VVhh$. 

In the 2HDM with CP violation, the vertices $Z (h_i \partial h_j  - h_j \partial h_i)$ can be induced at tree level \cite{Mendez:1991gp,Gunion:1997aq,Grzadkowski:1999ye}. If two lightest Higgs bosons of the model are kinematically reachable at a future $e^+ e^-$ collider, this process should be carefully studied. If the $e^+ e^-$ colliders could run in the photon-photon collision mode, then the process $\gamma \gamma \to h h$ \cite{Belusevic:2004pz,Kawada:2012uy,Heng:2013wia} can be useful to directly probe trilinear coupling and the new physics inside the loops.

\section{Discussions and Conclusions}
In this work, we examine the discovery potential of the full leptonic final state of the Higgs pair production at a 100 TeV collider. Based on our provious analysis $gg \to h h \to 3l2j\missE$ , we also explore the correlation between the experimental bounds on the Yukawa couplings $a$ and $b$ and the bounds on $\lambda_3$ at both the LHC and a 100 TeV collider.

Due to the small production of the full leptonic final state in the SM, it is challenging to discover the Higgs pair production via the two modes explored in this work. But if there is new physics which might enhance the Higgs pair production by one or two order in magnitude, like some bench mark points allowed in the 2HDM \cite{Baglio:2014nea}, this mode could be considered. 

It is obviously that if we can combining all final states of Higgs pair production, like $b\bar{b} \gamma \gamma$, $WW^* \gamma \gamma$, $b\bar{b} \tau \tau$, etc, we can achieve a better bound on $\lambda_3$.

Here we would like to briefly address the issues of detector to this type of signal. We show the acceptance efficiencies to the signal events in Table \ref{etaeff}. It is noticed that if the coverage of $\eta$ for lepton can reach to 4, 90 percent of the signal events can be accepted, while in contrast, if a coverage of $\eta$ for lepton can only reach to 2, less than half of signal events can be accepted. If the detectors can cover a large $\eta$ region (say $|\eta(\ell)| < 4$), it is good for the signal of Higgs pair.

The second issue is related with the missing energy resolution, which can be  related with HCAL coverage. As we demonstrated, the missing energy reconstruction is quite crucial in order to obtain kinematical observables which can suppress the background events efficiently. Nevertheless, in the 100 TeV collider, if the HCAL detector coverage is less than $6$, the missing energy from forward jets can be larger than 100 GeV, as demonstrated in \cite{Arkani-Hamed:2015vfh}. So if HCAL detector coverage can reach to $\eta(j) <8$, which can be good not only for the vector boson fusion processes but also for the Higgs pair production processes.

\begin{center}
\begin{table}
  \begin{center}
\begin{tabular}{|c|c|}
  \hline
  Eta Coverage & Signal Acceptance Efficiency\\ \hline
  $|\eta|<2$ & 50\% \\ 
  $|\eta|<3$ & 73\%\\ 
  $|\eta|<4$ & 91\%\\ 
  $|\eta|<5$ & 98\%\\ 
  $|\eta|<6$ & $\approx$100\%\\
  \hline
\end{tabular}
\end{center}
 \caption{\label{etaeff}The acceptance efficiencies of signal events in $\eta$ are tabulated.}
\end{table}
\end{center}

\begin{acknowledgments} We would like to thank Fapeng Huang and Sichun Sun for useful discussions. This work is
  supported by the Natural Science Foundation of China under the grant NO. 11175251, No.
  11305179, NO. 11475180, and No. 11575005. The work of Q. Li and Q.S. Yan is partially supported by CAS Center for Excellence in Particle Physics (CCEPP).
\end{acknowledgments}

\appendix 
\section{QMC, Reweighting}
In this appendix, we explain the QMC and a new reweighting algorithm based on QMC.

Considering a $d$-dimensional integral
\begin{align}
  I(f)=\int_{[0,1]^d}\mathrm{d}^d{x}f(\vec{x}),
\end{align}
a way to estimate it is firstly sampling the integrand at a predefined set of $n$ points
$\{\vec{x}_i |~ \vec{x}_i \in [0,1]^d;~ i = 0,\dots,n-1 \}$, 
then taking the average:
\begin{align}
  Q_{n}(f)=\frac{1}{n}\sum_{i=0}^{n-1}f(\vec{x}_i) \approx I(f).
\end{align}
This method is called quasi-Monte Carlo method, and the point set is called quasi-Monte Carlo
rule\cite{ANU:8877392}.

Comparing to Monte Carlo integration, which using independently distributed random points
and can only achieve a convergence speed proportional to $\mathcal{O}(N^{-0.5})$ ,
QMC integration can achieve a convergence speed near $\mathcal{O}(N^{-1})$,
if a proper QMC rule is adopted.
Two families of QMC rules attract most of interest: one is consist of digital nets and digital sequences; the
other is lattice rules.
In this work, we adopt the rank-1 lattice rule (R1LR):
\begin{align}
  \vec{x}_i=\left \{\frac{i\vec{z}}{n}\right\},~ i=0,\dots,n-1,
  \label{rankonerule}
\end{align}
where $\vec{z}$, known as the generating vector, is required that all its components should be integer and relatively prime to $n$. The braces around the vector in the Eq. (\ref{rankonerule}) means only the fractional part of each component is taken.

The previous algorithm is fully deterministic, and will result in a biased estimation. To achieve an unbiased
result, proper randomization must be introduced. For R1LR, the simplest form of randomization called shifting
can be used.

The algorithm for random shifted R1LR is explained as below\cite{ANU:8877392}:
\begin{enumerate}
  \item Generate $m$ independent random vectors called shifts, $\{ \vec{\Delta}_1, \cdots,
        \vec{\Delta}_m \}$, from uniform distribution in $[0,1]^d$.
  \item For each shift, calculate the integrand at correspondent lattice points and estimate the integral:
        \begin{align}
          Q_{n,k}(f)=\frac{1}{n}\sum_{i=0}^{n-1}f\left(\left\{\frac{i\vec{z}}{n}+\vec{\Delta}_k\right\}\right),
          \quad k = 1,\dots,m.
        \end{align}
  \item The average of these $m$ integral estimations
        \begin{align}
          \bar{Q}_{n,m}(f)=\frac{1}{m}\sum_{k=1}^{m}Q_{n,k}(f)
        \end{align}
        is finally taken as the estimation of the integral $I(f)$ and it can be proved to be
        unbiased\cite{ANU:8877392}.
  \item An unbiased estimation of the mean-square error $\bar{Q}_{n,m}(f)$ can also be obtained by
        \begin{align}
          \frac{1}{m(m-1)}\sum_{k=1}^{m}(Q_{n,k}(f)-\bar{Q}_{n,m}(f))^2.
        \end{align}
\end{enumerate}

This algorithm has been realised and validated in Feynman diagram evaluation by using the sector decomposition method \cite{Li:2015foa}. For more details about QMC and this algorithm, such as the construction of generating vector, the convergence rate, please see the references\cite{ANU:8877392,Kuo2003301,Dick2004493}. 

Besides doing numerical integration to estimate the cross section, generating unweighted events is also
required for an event generator.
Considering a multi-dimensional non-negative function $f(\vec{x})$,
the most widely used algorithm is the rejection algorithm, which is described below:
\begin{enumerate}
  \item Construct a non-negative function $g(\vec{x})$, and it is called sampling function.
  \item Find a value $M$ satisfying
        \begin{align}
          f(\vec{x})\le M g(\vec{x})
        \end{align}
  \item Generate a point $\vec{x}$ with probability density function $\frac{g(\vec{x})}{\int\mathrm{d}x
        g(\vec{x})}$, and a random number $r$ with uniform distribution in $[0,1)$.
        If
        \begin{align}
          r < \frac{1}{M}\frac{f(\vec{x})}{g(\vec{x})},
    \end{align}
        this point is accepted, otherwise this point is rejected.
		This step can be repeated many times to obtain desired number of points.
\end{enumerate}

The efficiency of this algorithm, which is defined by the ratio of accepted number of points and total number
of points in the last step,
is determined by the sampling function $g(\vec{x})$, and this function can also be used
in MC integration via importance sampling.
The most widely used algorithm for the sampling function in high energy physics is the VEGAS algorithm proposed by Lepage\cite{vegas},
which has a factorized structure as the following
\begin{align}
  g(\vec{x})=\prod_{i=1}^{d}g_i(x_i),
\end{align}
where $g_i(x_i)$ are step functions.
This algorithm starts with $g(\vec{x})=1$, then samples with several points based on $g(\vec{x})$
and optimizes $g(\vec{x})$ based on sampled results.
The sampling and optimizing procedure can be iterated for several times, to further
optimize $g(\vec{x})$.
Although this function allow us to deal with function with large peak,
a general function is unfactorizable and the efficiency of this algorithm 
is limited by this.
Another disadvantage of this algorithm is that the optimization is based on information
gained via MC sampling, which suffers from fluctuation.

As we found that QMC could give a better estimation of integral than MC methods, 
based on the structure of R1LR,
we proposed a new $g(\vec{x})$.
The sampling function is constructed based on QMC points, so the fluctuation is much smaller.
In addition, it do not assume a factorized structure, and the efficiency
of this algorithm could be improved to arbitarily near 1 if a larger point set is adopted.

The sampling function in this algorithm is described in the following:

\begin{enumerate}
  \item Any vector $\vec{v_{ij}}=\vec{x_i}-\vec{x_j}$ is the period of the lattice, and we can choose $d$
linearly independent vectors
from them. Since there are multiple choices, we choose the shortest ones and mark them as $\vec{e_i}(i=1,2,\dots,d)$.
  \item With these linearly independent vector and one origin $\vec{x_k}$, we can build an affine coordinate
        system, and mark it as $S_k$.
	  \item The region $\{\vec{y}|0\le y_i \le 1,i=1,2,\dots,d\}$ is called the $k$-th unit cell, where $y_i$s are the coordinates in affine
        coordinate system $S_k$. It can be shown that the volume of one unit cell is $\frac{1}{n}$, and the
        original domain of integration is divided into $n$ unit cells. Any point can only belong into one
		unit cell.
	  \item If a point $\vec{x}$ belong in the $k$-th unit cell, then $g(\vec{x})$ is defined as the following:
\begin{align}
  g(\vec{x})=h_k(\vec{y})
  &=\sum_{a_1,a_2,\dots,a_d\in\{0,1\}}h_k(a_1,a_2,\dots,a_d)\prod_{i=1}^{d}(a_iy_i+(1-a_i)(1-y_i))
\end{align}
Where $\vec{y}$ are the coordinates in $S_k$, corresponding to $\vec{x}$ in original coordinate system.
$h(a_1,a_2,\dots,a_d)$ is the values of function $f(\vec{x})$ at vertices of this unit cell, as below
\begin{align}
  h_k(a_1,a_2,\dots,a_d)&=f(\vec{x_k}+\sum_{i=1}^{d}a_i\vec{e_i}),
\end{align}
and these values has been calculated in previous QMC integration.
\end{enumerate}

%
An importance fact is that the integration of the $g(\vec{x})$ can be easily obtained:
\begin{align}
  I(g)=\int_0^1\cdots\int_0^1\prod_{i=1}^{d}{\ud
  x_i}g(x)=\sum_{i=1}^{n}f(\vec{x_i})=Q_{n}(f(\vec{x}))
\end{align}

The previously defined $g(\vec{x})$ only depends on the lattice structure, and since random shifts keep this
structure,
it can be used in random shifted lattice rule. In detail, if we use $m$ shifted lattice rule, and denote the
intepolation function for each shift as $g_i(\vec{x}),i=1,2,\dots,m$, then the following probability function
can be used:
\begin{align}
  \label{lri}
  p(\vec{x})=\frac{1}{m}\sum_{i=1}^{m}\frac{g_i(\vec{x})}{I(g_i(\vec{x}))}
\end{align}

An important part of this algorithm is to generate random points in term of the probability function 
Eq. \eqref{lri}, which can be achieved via the following steps:

\begin{enumerate}
  \item choose a shift $s$ in $m$ shifts with probability $\frac{1}{m}$
  \item choose the k-th point in the lattice with probability
	$\frac{f(\vec{x_k})}{\sum_{i=0}^{n}{Q_{n,s}(f)}}$
  \item choose $d$ variables $y_1,y_2,\dots,y_d$ according to the following probability density function
        \begin{align}
          f(y)=\begin{cases}
                0,    &|y|>1\\
                1-|y|,&|y|\le1
          \end{cases}
        \end{align}
  \item then $(y_1,y_2,\dots,y_n)$ is the random point satisfying requested distribution in the affine
        coordinate system $S_k$, i.e. the corresponding coordinates in original coordinates system is
        \begin{align}
		  \vec{x}=\{\vec{x_k}+\sum_{i=1}^{d}y_i\vec{e_i}\},
        \end{align}
	  where the pair of braces means that only the fractional part is taken, as before in Eq. (\ref{rankonerule}).
\end{enumerate}

\begin{table}[th]
\begin{center}
\begin{tabular}{|c|c|c|}
\hline
 & VEGAS & QMC-based \\\hline
integration result(fb) & $19.0\pm0.2$& $19.269\pm0.003$\\\hline
reweighting efficiency & $36.6\%$& $75.8\%$ \\\hline
\end{tabular}
\end{center}
\caption{Comparison of convergence  and reweighting efficiency between the VEGAS and the R1LR algorithm (QMC-based).\label{qmceff}}
\end{table}

In Table \ref{qmceff}, we demonstrate the efficiency of this algorithm in comparison with VEGAS algorithm.
The function used in the test is the differential cross section of gluon fusion into Higgs pair at 14TeV LHC, 
where the dimension of integration is 4.
We adopt the R1LR algorithm and perform an integration with a lattice containing 1024 points and repeat the
computation with 4 shifts. After that we test the reweighting efficiency.
For VEGAS algorithm, we compute 1024 points in each iteration and total number of iteration is 4, the same number of points as in the R1LR algorithm.
It is observed that the QMC-based algorithm is better than VEGAS, both in the speed of convergence and reweighting efficiency.

\bibliography{4w4l}

\end{document}